\documentclass[11pt,english,twoside]{article}
\pdfoutput=1
\usepackage{jheppub}

\usepackage{amssymb, amsmath, amsthm,latexsym}
\usepackage[english]{babel} 
\usepackage[latin1]{inputenc} 
\usepackage{graphicx}
\usepackage[all]{xy}
\usepackage{bbm}
\usepackage{psfrag}

\newcommand{\be}{\begin{equation}}
\newcommand{\ee}{\end{equation}}
\def\bea#1\eea{\begin{align}#1\end{align}}
\newcommand{\nn}{\nonumber}

\def\d {{\rm d}}

\newcommand{\cL}{\mathcal L}

\newcommand{\cM}{{\cal M}}

\usepackage{graphicx, float, array, xspace, amscd, amsmath, amsthm, amssymb, latexsym, bbm}
\usepackage{arydshln}
\usepackage[english]{babel}
\usepackage[all]{xy}

\renewcommand\baselinestretch{1.22}
\gdef\@fpheader{$ $}
\gdef\@journal{$ $}

\newcommand{\varstr}[2]{\vrule height #1 depth #2 width0pt}
\newcommand{\beq}{\begin{equation}}
\newcommand{\eeq}{\end{equation}}

\renewcommand{\a}{\alpha}
\renewcommand{\b}{\beta}
\newcommand{\g}{\gamma}\newcommand{\G}{\Gamma}
\newcommand{\e}{\epsilon}
\newcommand{\z}{\zeta}

\renewcommand{\th}{\theta}

\newcommand{\x}{\xi}

\newcommand{\p}{\pi}\newcommand{\vp}{\varpi}

\renewcommand{\t}{\tau}

\newcommand{\ph}{\phi}

\newcommand{\pd}[2]{\frac{\partial #1}{\partial #2}}

\newcommand{\ii}{\text{i}}
\renewcommand{\ee}{\text{e}}
\newcommand{\smallfrac}[2]{\frac{\scriptstyle #1}{\scriptstyle #2}}

\newcommand{\capt}[3]{\parbox{#1}{\renewcommand{\baselinestretch}{1.0}
                                                           \caption{\label{#2}\small\it #3}}}

\newcommand{\ft}[2]{{\textstyle\frac{#1}{#2}}}

\title{Type IIB flux vacua from G-theory I}
\author[a]{Philip Candelas\footnote{candelas@maths.ox.ac.uk}$\!^,$}
\author[b]{Andrei Constantin\footnote{andrei.constantin@physics.uu.se}$\!^,$}
\author[c]{Cesar Damian\footnote{cesaredas@fisica.ugto.mx}$\!^,$}
\author[b]{Magdalena Larfors\footnote{magdalena.larfors@physics.uu.se}$\!^,$}
\author[d]{\linebreak Jose Francisco Morales\footnote{francisco.morales@roma2.infn.it}$\!^,$}

\affiliation[a]{Mathematical Institute, University of Oxford, Andrew Wiles Building, Radcliffe Observatory Quarter, \linebreak Woodstock Road, Oxford, OX2 6GG, UK}
\affiliation[b]{Department of Physics and Astronomy, Uppsala University, SE-751 20, Uppsala, Sweden}
\affiliation[c]{Departamento de Fisica, DCI, Campus Leon, Universidad de Guanajuato, C.P. 37150, Leon, Guanajuato, Mexico}
\affiliation[d]{I.N.F.N. Sezione di Roma ``TorVergata'', Dipartimento di Fisica, Universita di Roma ``TorVergata'', \linebreak Via della Ricerca Scientica, 00133 Roma, Italy}

\abstract{
We construct non-perturbatively exact four-dimensional Minkowski vacua of type IIB string theory with non-trivial fluxes. These solutions are found by gluing together, consistently with U-duality, local solutions of type IIB supergravity on $T^4\times \mathbb{C}$ with the metric, dilaton and flux potentials varying along $\mathbb{C}$ and the flux potentials oriented along $T^4$. 
We focus on solutions locally related via U-duality to non-compact Ricci-flat geometries.  More general solutions and a complete analysis of the supersymmetry equations  are presented in  the companion paper \cite{paper:local}. We build a precise dictionary between fluxes in the global solutions and the geometry of an auxiliary $K3$ surface fibered over $\mathbb{CP}^1$. 
In the spirit of F-theory, the flux potentials are expressed in terms of locally holomorphic functions that parametrize the complex structure moduli space of the $K3$ fiber in the auxiliary geometry.  The brane content is inferred from the monodromy data around the degeneration points of the fiber. 
}

\preprint{UUITP-18/14}

\begin{document}
\pagestyle{myplain}
\phantom{}
\vspace{-10pt}
\maketitle

\flushbottom 

%\section{Introduction and Summary}%-----------------------------------

%\newpage

%\vspace{-9pt}
\section{Introduction} 

In the presence of fluxes, supersymmetry requires that the internal manifold of a type II string compactification be of generalised complex type. %A generalised Calabi-Yau manifold is characterised by the existence of a globally defined, nowhere vanishing spinor.  For a six-dimensional manifold, this amounts to require the existence of a globally defined and nowhere vanishing real two-form $J$ and a complex decomposable three-form $\Omega$.  Supersymmetry is preserved in the four dimensional theory if the spinor, or equivalently these differential forms, satisfy a system of first order differential equations~\cite{Grana:2004bg}. 
Even though the requirements for supersymmetric flux vacua have been known for a long time \cite{Grana:2004bg}, it is still challenging to find explicit solutions when the four-dimensional spacetime is Minkowski.  In a compact space, the charge and tension associated with fluxes must be balanced by the introduction of branes and O-plane sources \cite{Maldacena:2000mw}. The back reaction of these objects on the compact geometry should be consistently included in the picture. These localised defects act as delta-like sources in the equations of motion for the supergravity fields, that, as a consequence, can seldom be solved in an analytic form\footnote{The best understood examples involve warped Calabi-Yau geometries with anti-self dual 3-form fluxes  \cite{Giddings:2001yu}. The warp factor is governed by a harmonic equation in the compact space sourced by fluxes, branes and O-planes that can be typically solved only locally in the region of weak string coupling.}.

The aim of the present paper is to present explicit examples of non-perturbatively exact supersymmetric four-dimensional Minkowski vacua where all the fields can be   written out in an analytic form even in the presence of fluxes. To achieve this, we follow the strategy used in Refs.~\cite{Martucci:2012jk,Braun:2013yla}, where   
 type IIB supergravity backgrounds describing systems of 3-branes and 7-branes on  $K3$ were described in purely geometric terms.  
 
The strategy is inspired by F-theory \cite{Vafa:1996xn}, where backgrounds with 7-branes are described in terms of elliptic fibrations.  The complex structure parameter of the fiber plays the role of the axio-dilaton field $\tau$ of type IIB theory and the degeneration points of the fiber indicate the presence of brane sources. 
On the other hand, the  $SL(2,\mathbb Z)$ self-duality group of type IIB string theory in ten dimensions is identified 
with the modular group of the elliptic fiber. In compactifications to lower dimensions, the number of scalars is larger and the U-duality group bigger. Remarkably, as shown in~\cite{Martucci:2012jk,Braun:2013yla}, the moduli space of six dimensional solutions can be put in correspondence with the moduli space of complex structures of certain $K3$ surfaces, and the flux solutions can be geometrized in terms of auxiliary $K3$-fibered Calabi-Yau threefolds having the $K3$ in question as a fiber. Similar studies where the U-duality groups of type~II theories compactified to lower dimensions  have been geometrized can be found in 
\cite{Kumar:1996zx,Liu:1997mb,Hellerman:2002ax,Hull:2004in,Flournoy:2004vn,Dabholkar:2005ve,Gray:2005ea,Hull:2007zu,Vegh:2008jn,Pacheco:2008ps, Grana:2008yw,McOrist:2010jw,Andriot:2011uh,Coimbra:2011nw,Berman:2011cg,Berman:2011jh,Coimbra:2011ky,Hohm:2011si,Andriot:2012wx,Andriot:2012an,Blumenhagen:2012nt,Coimbra:2012af, Aldazabal:2013mya, Cederwall:2013naa,Blumenhagen:2013aia,Andriot:2013xca,Cederwall:2014opa}.

The resulting geometrical picture is analogous to the F-theory picture: the torus has simply been replaced by a $K3$. The $SL(2,\mathbb Z)$ modular group of the elliptic fiber is replaced by the U-duality group $SO(2,n,\mathbb Z)$ acting on the space of complex structures of the $K3$ fiber. The structure of singularities is richer, allowing for monodromies associated to  brane charges of various types and dimensions. The subtle question of tadpole cancelation, which is a major impediment in the construction of type II vacua, is automatically solved by holomorphicity. Indeed, the monodromy around a contour enclosing all singularities is by construction trivial showing that all brane charges add up to zero. In particular, this implies that all solutions include exotic brane objects \cite{deBoer:2012ma} that at weak coupling should recombine into O-planes. Like in F-theory, all these subtle features of the background are nicely encoded in the auxiliary  geometry. The complex plane closes up into a two-sphere for configurations involving a number of $24$ branes. The resulting four-dimensional vacua can then be viewed as purely geometric solutions of a gravity theory, dubbed {\it G-theory} in~\cite{Braun:2013yla}, compactified on a $K3$ fibration over $\mathbb C\mathbb P^1$.

In the present work, we focus on compactifications of the type IIB string on compact six-dimensional manifolds that are locally isomorphic to $T^4\times \mathbb{C}$ and consider fluxes of a more general type. We construct explicit solutions falling into the familiar class of warped Calabi-Yau geometries with anti-self-dual three-form fluxes and two classes of solutions that are warped complex but non-K\"ahler.  The three classes, that we will refer to as A, B, and C, describe systems of 3,7-branes and 5-branes of NS and R types respectively. In this paper, we restrict ourselves to solutions related to Calabi-Yau geometries via U-dualities. These solutions are completely characterised in terms of up to three holomorphic functions. In the companion paper  \cite{paper:local}, we will present a general analysis of the supersymmetry equations and their solutions, providing examples that are not related to purely metric backgrounds by means of U-dualities. 
  
 We stress that the solutions found are, like their F-theory analogs, non-perturbatively exact. Indeed, the expansion of the supergravity fields around any of the branching points always exhibits, beside the logarithmic singularity, an infinite tower of instanton-like corrections. Generically, the solutions correspond to non-geometric U-folds from the ten-dimensional perspective, since the metric of the internal space, and other supergravity fields, are in general patched up by non-geometric U-duality transformations (see \cite{Andriot:2011uh} for a recent review on non-geometry in string compactifications). 
 Similar ideas were exploited years ago in \cite{Hellerman:2002ax} to describe solutions with non-trivial $H=d B$ field, but vanishing RR fluxes, in terms of elliptic fibrations. 
  
The paper is organised as follows. In Section \ref{sec:localsolutions}, we derive three classes of local solutions with non-trivial fluxes. We start from a flux-less Ricci flat non-compact background describing a fibration of $T^4$ over $\mathbb{C}$ and then, through a sequence of S and T dualities, we obtain flux solutions. 
In Section~\ref{sec:global} we discuss the global completion of the local solutions. We exploit the observation that the $n$ holomorphic functions characterising the local solutions parametrize a double coset space that turns out to be isomorphic to the moduli space of complex structures of algebraic $K3$ surfaces\footnote{An algebraic $K3$ is a surface that can be holomorphically embedded in $\mathbb C\mathbb P^m$, for some $m$, by polynomial equations.} with Picard number~$20-n$. The flux solutions can then be described in terms of an auxiliary threefold  with $K3$~fiber and $\mathbb{CP}^1$ base (the complex plane plus the point at infinity). 
In Section~\ref{sec:example} we discuss the complex structure moduli space for a simple choice of $K3$ surface with $n=2$ complex structure parameters, which serves as our main example.   In Section~\ref{sec:dictionary} we work out the details of the flux/geometry dictionary for various explicit examples of Calabi-Yau threefolds whose fiber is the algebraic $K3$ surface studied in Section~\ref{sec:example}. We use the language of toric geometry, in which the fibration structure of a Calabi-Yau three-fold is specified by a reflexive 4d polytope containing a 3d reflexive sub-polytope  associated to a $K3$ with $n$ complex deformations. 
The holomorphic functions defining the flux vacua are identified with periods of the holomorphic two-form of the $K3$ fiber. We compute the periods and extract the brane content from their monodromies around the points where the complex structure of the $K3$ fiber degenerates. In Section~\ref{sec:OtherExamples} we present two other examples of $K3$ surfaces with Picard number $18$ and using these, we construct $K3$-fibered Calabi-Yau threefolds and discuss the associated flux solutions. The paper is supplemented by four appendices. In Appendix~\ref{sec:ToricGeometry}, we review the main elements of toric geometry needed for our analysis. Appendix~\ref{app:ModuliSpace} contains the more technical details concerning the analysis of the moduli space for the main example, while Appendix~\ref{app:ellipticfibrations} lists the five different elliptic fibration structures for this $K3$ surface. Finally, in Appendix~\ref{app:Pic18List} we list the $8$ other $K3$ surfaces with Picard number $18$ and $2$ complex structures that are present in the Kreuzer-Skarke list.

\section{Local flux solutions dual to Calabi-Yau geometries}%-----------------------------------
%\label{sec:eom}
 \label{sec:localsolutions}
    
\subsection{The ansatz}

In this section we construct a class of type IIB solutions on space-times $\mathbb{R}^{1,3}\times M_6$,  with $M_6$ a non-compact manifold with topology $T^4\times \mathbb{C}$. These solutions form a sub-class of the more general solutions presented in \cite{paper:local}. The metric of the torus $g_{mn}$, the dilation $\phi$, the Neveu--Schwarz (NSNS) $B$-field and Ramond--Ramond (RR) $C_p$-fields are assumed to vary over $\mathbb{C}$. All the non-trivial fluxes are assumed to be oriented along $T^4$.
      
Let $\{y^1,y^2,y^3,y^4\}$ be real coordinates on $T^4$ and $z$ a complex coordinate on $\mathbb C$. In these coordinates, the metric and the fluxes have the generic form:
 \bea
\d s^2& =\d s_4^2+\d s_6^2=e^{2A}\,  \sum_{\mu=0}^3 \d x_{\mu} \d x^{\mu} +\sum_{m,n=1}^4 g_{mn}\, \d y^m\,  \d y^n+   2 \,e^{2D}\,|h(z)|^2\, \d z\, \d \bar{z} 
 \label{ans}\\
B &=\frac{1}{2} \,b_{mn}\, \d y^m\wedge \d y^n \;, \quad     C_2=\frac{1}{2} \,c_{mn}\, \d y^m\wedge \d y^n \;, \quad C_4=c_4 \, \d y^1\wedge \d y^2\wedge \d y^3 \wedge \d y^4 \;,  \nn
\eea
with $A$, $D$, $g_{mn}, b_{mn}, c_{mn}, c_4, C_0$ and $\phi$ varying only over the complex plane.

\subsection{Non-compact Calabi-Yau geometries}

In the absence of fluxes, supersymmetry requires that the internal six-dimensional manifold be of Calabi-Yau type. A six-dimensional Calabi-Yau manifold is characterised by the existence of a closed two-form $J$ and a closed holomorphic three-form $\Omega_3$, that define a K\"ahler structure and, respectively, a complex structure.  A family of closed forms defining Ricci-flat K\"ahler metrics can be written as  
\beq
\begin{aligned}
\Omega_3 &=h \, \d z\wedge  \,\left[ (\d y^4-\tau \d y^1)\wedge   (\d y^3-\sigma \d y^2)   - \beta\, (\d y^1\wedge \d y^4 -\d y^2 \wedge \d y^3)  - \beta^2\, \d y^1\wedge \d y^2\, \right]   \\
J & = \d y^1\wedge  \d y^4 +\d y^2\wedge  \d y^3   +{i\over 2}\,e^{2D}\, |h|^2\, \d z\wedge \d\bar z 
\label{sold}
\end{aligned}
\eeq
with
\beq
e^{2D}=\sigma_2 \tau_2-\beta_2^2~,
\eeq 
and  $\tau = \tau_1 + \ii\,\tau_2$, $\sigma = \sigma_1+\ii\,\sigma_2$, $\beta = \beta_1 + \ii\,\beta_2$, $h$ arbitrary holomorphic functions of $z$.
 It is easy to see that $\Omega_3$ and $J$ given in (\ref{sold}) define an $SU(3)$ structure\footnote{They satisfy $\Omega_3 \wedge J=0$ and
 $\ft{i}{8}\,\Omega_3 \wedge \bar{\Omega}_3 =\ft16 \, J\wedge J\wedge J$.} and are closed,
\beq
\d\Omega_3=\d J=0~.
\eeq
As such, they define a complex structure and a K\"ahler structure. They also define a metric, which turns out to be Ricci-flat. Explicitly, the metric can be written as  \cite{Hitchin:2000jd} (see also \cite{Larfors:2010wb})
\bea
 g_{MN}= -J_{MP} \,  I^P{}_N   \; .
\label{eq:su3met}
\eea
where $ I$ is a complex structure induced by $\Omega_3$:
\beq
 I^P{}_N =  c\,  
 \epsilon^{P M_1\ldots   M_5}\,   ( {\rm Re}\, \Omega_3)_{ N M_1 M_2}\,  (  {\rm Re}\, \Omega_3)_{ M_3 M_4 M_5}  \; ,
\eeq
and the normalisation  constant $c$ is fixed by the requirement $ I_M{}^P  I_P{}^N = -\delta_M^N$. 

Substituting (\ref{sold}) into (\ref{eq:su3met}), one finds the metric  
as
\beq\label{eq:metric}
\d s^2=g_{MN}\, \d x^M\, \d x^N=g_{mn} \,\d y^m\, \d y^n+ e^{2D} \, |h|^2 \, \d z \, \d\bar{z}  
\eeq
with
\vspace{-8pt}
\bea
g_{mn}=e^{-2D} \left(\begin{array}{ccccccc}
{\sigma}_{2} &~~~ & {\beta}_{2} & ~~~&  \,{\beta}_{2} \, {\sigma}_{1} - {\beta}_{1} \, {\sigma}_{2} &~~~& - {\beta}_{1} \, {\beta}_{2}+{\sigma}_{2} \, {\tau}_{1} \\[4pt]
{\beta}_{2} & & {\tau}_{2} &&  \!\!\!\!- {\beta}_{1} \, {\beta}_{2} + {\sigma}_{1} \, {\tau}_{2} & & ~~{\beta}_{2} \, {\tau}_{1} -{\beta}_{1} \, {\tau}_{2} \\[4pt]
~~{\beta}_{2} \, {\sigma}_{1} - {\beta}_{1} \, {\sigma}_{2} &  & - {\beta}_{1} \, {\beta}_{2} + {\sigma}_{1} \, {\tau}_{2} & &  {\rm Im} (\bar\beta^2 \, \sigma )+|\sigma|^2   \, {\tau}_{2} &&  |{\beta}|^{2} \, {\beta}_{2} +{\rm Im}(\beta \bar \sigma \bar\tau)      \\[4pt]
- {\beta}_{1} \, {\beta}_{2} + {\sigma}_{2} \, {\tau}_{1} && ~ {\beta}_{2} \, {\tau}_{1} - {\beta}_{1} \, {\tau}_{2} && |{\beta}|^{2} \, {\beta}_{2} +{\rm Im}(\beta \bar \sigma \bar\tau)  &   &   {\rm Im} (\bar\beta^2 \, \tau )+|\tau|^2   \, {\sigma}_{2} 
\end{array}\right) \label{gmn4}
 \eea
The metric (\ref{gmn4}) can be shown to be Ricci-flat for any choice of the holomorphic functions $\sigma, \tau,\beta, h$ and therefore it defines a local solution of the equations of motion. The sub-class of solutions corresponding to $\beta = 0$ will be of interest later on. In this case, the Ricci-flat metric takes the simpler form: 
\bea
& \d s_6^2 = {1\over \tau_2} \, |\d y^1+\tau\, \d y^4|^2  +  {1\over \sigma_2}\, | \d y^2+\sigma\, \d y^3|^2 + 2   \,  \sigma_2 \, \tau_2 \, \d z \, \d\bar z \,  |h|^2 \label{metric60}
 \eea
and corresponds to a fibration of $T^2\times T^2$ over $\mathbb C$ where the complex structures $\tau,\sigma$ of the two tori vary holomorphically along the plane.

In Section~\ref{sec:global} we will construct global solutions by extending the definition of the holomorphic functions to the whole complex plane up to non-trivial U-duality monodromies around a finite number of singular points. 
%of these functions associated to degenerations of the $T^4$ fiber.  
Moreover, the function $h$ will be chosen such that the metric along the 2d plane is regular at infinity leading to a compact $\mathbb C\mathbb P^1$ geometry.  The resulting supersymmetric four-dimensional vacuum corresponds to a geometric compactification on a Calabi--Yau threefold, realised as a $T^4$ fibration over $\mathbb C\mathbb P^1$, if these monodromies do not include T-duality transformations. If they do, the compactification is a non-geometric version of a Calabi--Yau three-fold.

\subsection{T and S dualities}

The Ricci-flat metric (\ref{gmn4}) can be mapped to flux backgrounds with the help of T and S dualities.
Under T-duality along a direction $y$, the metric in the string frame and the NSNS/RR fields transform as~\cite{b87,b88,Lunin:2001fv}:
\beq
\begin{aligned}
g'_{yy} &= \frac{1}{g_{yy}}, \quad\quad\quad e^{2 \phi '} = \frac{e^{2 \phi}}{g_{yy}}, \quad\quad\quad g'_{y m} = \frac{B_{y m}}{g_{yy}}, \quad\quad\quad B'_{y m}= \frac{g_{y m}}{g_{yy}} \\[8pt]
g'_{m n} &= g_{m n}-\frac{g_{m y}\,g_{n y} - B_{m y}\,B_{n y}}{g_{yy}}, \quad B'_{m n}= B_{m n}- \frac{B_{m y}\,g_{n y}-g_{m y}\,B_{n y}}{g_{yy}}\\[8pt]
 C'_{m ... n \alpha y} &=C_{m ... n \alpha} - \left( n - 1 \right) \frac{C_{[ m ... n | y}\,g_{y | \alpha ]}}{g_{yy}} \\[8pt]
C'_{m ... n \alpha \beta} &= C_{m... n \alpha \beta y} - n\,C_{[m ... n \alpha }\,B_{\beta ] y} - n(n-1)\frac{C_{[ m...n | y}\,B_{|\alpha | y}\,g_{|\beta ] y}}{g_{yy}}
\end{aligned}
\eeq

 On the other hand, for backgrounds with $C_0=0$, the S-duality transformations are
\beq
 \phi'= -\phi \qquad g'=e^{-\phi } g \qquad C_2' =-B \qquad B' = C_2~.  \label{sdual} 
 \eeq 
  
\subsection{Flux solutions}

Starting from the metric  (\ref{gmn4}) and acting locally with T and S dualities, one can generate local Type IIB solutions with various types of fluxes.  We denote the three main classes of solutions as A, B and C with A for warp, B for $B$-flux and C for $C_2$-flux. They are defined by the maps
  \beq\label{eq:dualities}
%A \quad  \overset{ T_{13}}{\longleftrightarrow}   \quad C   \quad  \overset{  S}{\longleftrightarrow}   \quad B
 %\quad  \overset{ T_{14}}{\longleftrightarrow}   \quad    CY
 \text{CY}  \quad  \overset{ T_{12}}{\longleftrightarrow}   \quad  \text{B}   \quad  \overset{  S}{\longleftrightarrow}   \quad \text{C}
 \quad  \overset{ T_{14}}{\longleftrightarrow}   \quad    \text{A}
 \eeq
 %For convenience of the reader we collect the details of the S and T maps in Appendix  \ref{ap:tsdual}.
 
The resulting solutions are characterised by the three holomorphic functions $\sigma,\tau,\beta$, this time encoding the information about the fluxes rather than the metric.
  
 \begin{itemize}
 
\item{A:   The metric is warped flat
\begin{equation}
g_{mn}=e^{\phi-2A}\, \delta_{mn}  \qquad e^{2D}=e^{-2A}=\sqrt{\sigma_2 \tau_2-\beta_2^2} \qquad e^{-\phi}=\tau_2
\end{equation}
 The non-trivial   fluxes are 
 \bea
 C_0 &=\tau_1   \qquad C_4= \left(-  \sigma_1+{2\, \beta_1 \,\beta_2\over \tau_2} - {\tau_1\, \beta_2^2\over \tau_2^2}  \right) \, \d y^1\wedge \d y^2 \wedge \d y^3 \wedge \d y^4\nn\\[4pt]
B &=- {\beta_2\over \tau_2} \, \left(\d y^1\wedge \d y^2-\d y^3\wedge \d y^4\right)  \qquad   C_2= \left( \beta_1 -{ \tau_1\, \beta_2\over \tau_2}\right)\, \left(\d y^1\wedge \d y^2-\d y^3\wedge \d y^4\right)    
    \eea
 This solution describes general systems of D3- and D7-branes and their U-dual\footnote{The $B $ and $C_2$ fluxes are anti-self-dual forms. As such, they go through two-cycles of zero-volume since $ j  \wedge  \chi_a^-  =0$, for any anti-self-dual two-form $\chi_a^-$, where $j$ is the K\"ahler form on $T^4$. The associated brane sources correspond to a D3-brane coming from D5-branes wrapping a vanishing two-cycle on $T^4$. } .

}

\item{B:    The metric and dilaton depend on the imaginary parts of three holomorphic functions $\tau, \sigma$ and $\beta$:
\beq
g = \left(
\begin{matrix} % or pmatrix or bmatrix or Bmatrix or ...
~\tau_2 &~~-\beta_{2} & ~0 &  ~~0 \\
~-\beta_2  & ~~\sigma_2 & ~~0&   ~0 \\
~0   &  ~~0  & ~~\sigma_2 & ~~-\beta_2  \\
~0  & ~~ 0 &  ~~-\beta_2 & ~~\tau_2 \\
   \end{matrix}
\right)  ;
   \qquad  A=0 ~~~\text{ and }  ~~~ e^{2D}=e^{2 \phi}=\sigma_2 \tau_2-\beta_2^2     \label{solb}
\eeq
The real parts of the holomorphic functions specify the non-trivial $B$-field components:
\begin{equation}
\label{solb2}
B=\tau_1 \, \d y^1 \wedge \d y^4 + \sigma_1\, \d y^2 \wedge \d y^3 - \beta_1 (\d y^2 \wedge \d y^4+\d y^1 \wedge \d y^3 )   
\end{equation}
This solution describes general systems of intersecting NS5-branes and their U-duals.
}

\item{C:   This solution is found from B exchanging  the NS and RR two-forms and preserving the metric in the Einstein frame. One finds 
\beq
g = e^{2A} \,  \left(
\begin{matrix} % or pmatrix or bmatrix or Bmatrix or ...
~\tau_2 &~~-\beta_{2} & ~0 &  ~~0 \\
~-\beta_2  & ~~\sigma_2 & ~~0&   ~0 \\
~0   &  ~~0  & ~~\sigma_2 & ~~-\beta_2  \\
~0  & ~~ 0 &  ~~-\beta_2 & ~~\tau_2 \\
   \end{matrix}
\right) ,
   \qquad e^{2D}=e^{-2A}=e^{- \phi}=\sqrt{\sigma_2 \tau_2-\beta_2^2 }   \label{solc}
\eeq
The real parts of the holomorphic functions specify the non-trivial $C_2$-field components
  \begin{equation}
 C_2=\tau_1 \, \d y^1 \wedge \d y^4+ \sigma_1\, \d y^2 \wedge \d y^3  -\beta_1 (\d y^2 \wedge \d y^4+\d y^1 \wedge \d y^3 ) 
  \end{equation}
  This solution describes general systems of intersecting D5-branes and their U-dual. 
}

\end{itemize}

In \cite{paper:local}, we will present generalisations of solutions A, B and C preserving the same supersymmetry charges, but  involving fluxes of a more general type.  In particular, we will present examples of solutions of  A, B and C characterised by $n>3$ holomorphic solutions that cannot be related to a purely metric background via U-duality.

\section{Global 4d supersymmetric solutions}%-----------------------------------
\label{sec:global}

In the previous section, we solved the supersymmetric vacuum equations  of type IIB supergravity without sources. 
This led us to local solutions characterised by a set of holomorphic functions.   In order to define a global four-dimensional solution with Minkowski vacuum, branes and orientifold planes are needed in order to balance the charge and tension contributions of the fluxes in the internal manifold \cite{Maldacena:2000mw}. In our solutions, we will introduce the branes and O-planes as point-like sources in the $\mathbb{C}$-plane, over which the four-torus is fibered.  The monodromies around these points exhibited by the local holomorphic functions specify the brane content. 

\subsection{BPS solutions and moduli spaces}

The solutions under consideration can be viewed as supersymmetric solutions of ${\cal N}=(2,2)$ maximal supergravity in six dimensions with a number of scalar fields varying over the $z$-plane.  The scalar manifold of the maximal supergravity in six-dimensions is the homogeneous space 
\be
{\cal M}_{\,\text{IIB}~{\rm on}~T^4}= SO(5,5,\mathbb Z)\backslash {SO(5,5,\mathbb{R})\over SO(5,\mathbb R)\times SO(5,\mathbb R) } \;  \label{mt4} 
\eeq
understood as a double coset. This space has dimension $25$: $9$ coming from the symmetric and traceless metric on $T^4$, $2 \times 6$ from NSNS and RR two-forms and $4$ from the dilaton, the RR zero- and four-forms. The set of holomorphic fields entering in the supersymmetric solutions of classes A, B and C found in the previous section and described in their full generality in \cite{paper:local}, span a complex submanifold of (\ref{mt4}). The precise submanifold can be identified from the U-duality transformation properties of the various fields (see \cite{Martucci:2012jk} for a detailed discussion). 
 
The solutions found in the previous section involved a number of $n\leq 3$ locally holomorphic functions parametrizing the $n$-complex dimensional submanifold
\vspace{2pt}

\beq
{\cal M}_{\text{BPS}}=  SO(2,n,\mathbb Z) \backslash{SO(2,n,\mathbb{R} )\over SO(2,\mathbb R)\times SO(n,\mathbb R) } \subset {\cal M}_{\,\text{IIB}~{\rm on}~T^4} 
\label{mbps}
\eeq 
\vspace{2pt}

The U-duality group $SO(2,3,\mathbb Z)$ is generated by
\bea
S: &  ~~ \tau \longrightarrow -{1\over \tau} , ~~ \sigma \longrightarrow  \sigma-{1\over 2 \tau}  \, \beta^2,~~ \beta \longrightarrow {1\over \tau} \beta\\[2pt]
T : & ~~ \tau\longrightarrow \tau+1 \qquad  \\[2pt]
W : &~~ \beta \longrightarrow \beta+1 \qquad\\[2pt]
R: & ~~ \sigma \longleftrightarrow \tau
\eea

Indeed, in the frame of solution A, $S$ and $T$ generate the type IIB $SL(2,\mathbb{Z})$ self-duality, $W$ is the axionic symmetry and $R$ is the T-duality along the four directions of the torus exchanging D3 and D7 branes. Solutions in the B and C classes are related to solutions in the A-class by S- and T-dualities and correspond to different orientations of ${\cal M}_{\text{BPS}}$ inside the coset space ${\cal M}_{\,\text{IIB}~{\rm on}~T^4}$.

The moduli space (\ref{mbps}) has complex dimension $n$ and is isomorphic to the moduli space of complex structures on an algebraic $K3$ surface with Picard number $20-n$, which we denote by ${\cal M}_{K3,n}$ \cite{Aspinwall:1996mn}. 
 The holomorphic functions involved in these solutions are defined on the complex plane up to non-trivial actions of the $SO(2,n,\mathbb{Z})$ U-duality group. 

%In the examples discussed in the following sections, we will focus on the simplest case of solutions involving $n=2$ locally holomorphic functions. 
In the case $n=2$, i.e. $\beta=0$,   the variables $ \sigma$ and $\tau$ parametrize the double coset
\beq
{\cal M}_{K3,2}= {O}(\Gamma_{2,2})\backslash { {SO}(2,2,\mathbb{R})\over{SO}(2,\mathbb{R}) \times {O}(2,\mathbb{R})}\cong   {SO}(\Gamma_{2,2})\backslash \left( {SL(2,\mathbb{R} )\over U(1) } \right)^2
\label{moduli22}
\eeq
where ${O}(\Gamma_{2,2})$ is the group of isometries of the transcendental lattice\footnote{For a $K3$ surface $S$, the transcendental lattice is defined as the orthogonal complement of the Picard group (also called the Picard lattice) $\text{Pic(S)} = H^{2}(S,\mathbb Z)\cap H^{1,1}(S,\mathbb Z)$ in $H^2(S,\mathbb Z)$.} of the $K3$ surface in question, or, in the Type IIB language, ${O}(\Gamma_{2,2})$ is the U-duality group:
\beq
 {O}(\Gamma_{2,2})=SO(2,2,\mathbb{Z})\cong  \mathbb{Z}_2 \times SL(2,\mathbb{Z})_{\tau} \times SL(2,\mathbb{Z})_{\sigma}  \label{gamma22}
\eeq
 and $\mathbb{Z}_2$ acts by permuting the two $SL(2,\mathbb R)$ factors. The space (\ref{moduli22}) can be viewed as a $\mathbb Z_2$-quotient of the space of complex structures of a product of two tori.  
 
\vspace{8pt}
In the case $n=3$, the variables $ \sigma, \tau$ and $\beta$ parametrize the double coset
\beq
{\cal M}_{K3,3}= {O}(\Gamma_{2,3})\backslash { {SO}(2,3,\mathbb{R})\over{SO}(2,\mathbb{R}) \times {SO}(3,\mathbb{R})}
\label{moduli3}
\eeq
with
 \beq
 {O}(\Gamma_{2,3})=SO(2,3,\mathbb{Z})\cong  Sp(4,\mathbb{Z}) /\mathbb Z_2
     \label{gamma23}
\eeq
the U-duality group. Notice that $Sp(4,\mathbb{Z})$ is the modular group of genus two surfaces and therefore the solutions in this class can be viewed as a fibration of a genus two-surface over $\mathbb{C}$.

\subsection{From 6d to 4d}\label{sec:from6dto4d}

The class of solutions considered so far, corresponds to solutions of the six-dimensional supergravity equations of motion on an patch of the complex plane where functions $(\sigma,\tau,\dots)$ are holomorphic. These functions span a scalar manifold that is isomorphic to the moduli space $\mathcal M_{K3,n}$ of complex structures of a $K3$ surface with Picard number $20-n$, and can be regarded as coordinates on such a space. 

In order to construct a four-dimensional vacuum, we have to extend these functions to the entire complex plane, including the point at infinity.  However, given that on a compact space there are no holomorphic functions except for constants, the solution must be singular at certain points and multi-valued around paths that encircle those points.  The global consistency of the solution requires that the monodromies  experienced by these functions belong to the U-duality group $SO(2,n,\mathbb Z)$. Indeed, if this is the case, the set of functions $(\sigma,\tau,\dots)$ can be regarded as a single-valued map from the complex plane to the coset space $\mathcal M_{K3,n}$ and as such they make sense as solutions of Type IIB.    Mathematically, we analytically continue the local holomorphic functions to the complex plane with a number of punctures obtained by removing all the singular points.  
  The map 
\begin{equation}
\xymatrix{\overline{\mathbb C}\backslash \{z_1,\ldots,z_m\}~~~ \ar[rr]^{~~~~\sigma,\tau,\ldots} & &~~~\mathcal M_{K3,n}}
\end{equation}
defines a one-dimensional path in the moduli space of complex structures on a $K3$ with Picard number $20-n$, or, put differently, a $K3$ fibration over $\overline{\mathbb C} \backslash \{z_1,\ldots,z_m\}$. The set of points $\{z_1,\ldots,z_m\}$ represents the singular locus and $\overline{\mathbb C} = \mathbb C\cup \{\infty\}$. 

In the Type IIB picture, the locally holomorphic functions entering in the solutions of classes $A, B$ and $C$ are expressed in terms of the dilaton, the metric and the flux potentials. In the case where monodromies involve a shift symmetry of the type $\varphi\to \varphi+q$,
singularities are associated to brane locations. Indeed, the Bianchi identity shows that a monodromy of this type has to be supplemented by the presence of a delta-like source for the supergravity field associated to $\varphi$ in order to compensate for its flux  \cite{Martucci:2012jk}.  For example, for solutions in the A-class with $\beta=0$, $\tau= C_0 + \ii\,e^{-\phi}$ and $\sigma=C_4+ \ii \, {\rm Vol} (T^4)$ a monodromy
%\vspace{-12pt} 
\bea\label{eq:taumonodromy}
\tau &  \longrightarrow\ \tau + q_1    \qquad   \sigma   \longrightarrow\ \sigma + q_2 \qquad \text{with}\qquad q_1,q_2\in\mathbb Z
\eea
around the point $z_0$, indicates the presence of  $q_1$ D7-branes and $q_2$ D3-branes at the point $z_0$. In~general, we will say that a brane of charge $(q_1,q_2)$ is located at $z_0$.  
%
%We will often work with the $SL(2,\mathbb{Z})$ combinations $j(\sigma)$ and $j(\tau)$  rather than with $\sigma$ and $\tau$, so  by $(n_1,n_2)$ charges we will really mean in the  $SL(2,\mathbb{Z})^2$ orbit of   $n_1$ D3-branes and $n_2$ D7-branes. Indeed as in the F-theory case, O-planes and exotic branes will be always realised in terms of elementary branes and their duals.
  
On the other hand, the presence of a brane curves the plane (brane tension) generating a deficit angle of $\pi/ 6$ in the metric \cite{Greene:1989ya}. When a critical number  of 24 branes is reached, the plane curls up into a two sphere. In order to construct four-dimensional solutions, we  will restrict to this critical case, and thus require that the two-dimensional metric satisfies
\beq
\lim_{z\to \infty} \,  e^{2D}\,  |h(z)|^2\, \d z\, \d\bar z \sim  {\d z\, \d\bar z \over |z|^4 }~.    \label{atinf}
\eeq 
In this case, the topology of the six-dimensional compact space can be viewed as a $T^4$ fibration over~$S^2$. The U-duality invariance of the metric requires that  $e^{2D}  |h(z)|^2$ be invariant under  U-duality monodromies. This completely fixes the function $h(z)$. 
For $n=2$, where $e^{2D}=\sigma_2 \tau_2$, an invariant metric with the asymptotic behaviour (\ref{atinf}) is produced by the choice \cite{Martucci:2012jk}
\beq\label{hz1} 
h(z) =   { \eta(\tau)^{2}\,  \eta(\sigma)^{2}  \over \prod_{i=1}^{24} (z- z_i)^{1\over12} }        
\eeq     
with $z_i$ the points where either $\sigma$ or $\tau$ degenerates, corresponding to elementary brane charges $(0,1)$ or $(1,0)$, respectively. Indeed, by going around a path that encircles a brane of charge $(q_1, q_2)$ at~$z_0$, the holomorphic function $h$ returns to its original value, since the phase produced by the factor $(z-z_0)^{(q_1+q_2)/12}$ in the denominator of (\ref{hz1}) cancels against an identical contribution from the transformations of the Dedekind functions in the numerator under (\ref{eq:taumonodromy}).

In the case $n=3$, where  $e^{2D}=\sigma_2 \tau_2-\beta_2^2$, an invariant combination can be written as  
\beq
h(z) = \left( { \chi_{12} (\Pi)  \over \prod_{i=1}^{24} (z- z_i)  } \right)^{1\over 12}     \label{hz2}
\eeq 
with $\chi_{12}(\Pi)$ the cusp form of weight 12 of the genus two surface with period matrix $\displaystyle \Pi=\left(^{\sigma~ \beta}_{\beta ~\tau}\right)$.  
Notice that for $\beta$ small $\chi_{12} (\Pi)   \to   \eta(\tau)^{24}\,  \eta(\sigma)^{24} $ and (\ref{hz2}) reduces to (\ref{hz1}) corresponding to the degeneration of a genus two Riemann surface into two genus one surfaces. The three-parameter solution can then be seen as a deformation of the two-parameter one.

\subsection{The field/geometry dictionary}
%In order to construct global solutions, we noted that the space (\ref{mbps}) is isomorphic to the moduli space ${\cal M}_{K3,n }$ of complex structures for an algebraic $K3$ surface with Picard number $20-n$, i.e. with $n$~complex parameters. 
A global flux solution can be constructed by identifying the holomorphic functions present in the local solution with the complex functions describing the complex structure of certain $K3$ surfaces fibered over a two-sphere. By Torelli's theorem, the moduli space of complex structures on a $K3$ surface is given by the space of possible periods. Thus we will identify the holomorphic functions specifying the flux solutions with integrals of the holomorphic two-form $\Omega$ over a basis $\{e_i\}$ of integral two-cycles spanning the transcendental lattice of the $K3$: 
\beq
{\varpi}_i=\int_{e^i}  \Omega~.
\eeq

In particular, 
a pair $(\sigma,\tau)$ or triplet $(\sigma,\tau, \beta)$ of holomorphic functions parametrizing the double coset spaces (\ref{mbps}) with $n=2,3$ can be associated to auxiliary geometries corresponding to fibrations of $K3$ surfaces with Picard numbers $18$ and, respectively, $17$  over a two-sphere. For convenience, in the examples of the following sections, we will realise such fibration structures using Calabi-Yau threefolds and the language of toric geometry. 
Our working examples are taken from the class of $K3$ surfaces realised as hypersurfaces in toric varieties. Following Batyrev's construction~\cite{Batyrev:1993dm}, the embedding toric variety, as well as the polynomial defining the $K3$ hypersurface, can be encoded in a pair $(\nabla, \Delta)$ of dual three-dimensional reflexive polyhedra. The number of complex structures is given by a simple counting of points in the polytopes (see Eq.~\eqref{eq:complexstructures} in Appendix~\ref{sec:ToricGeometry}, where we review the notions of toric geometry needed for the subsequent discussion).  

The Kreuzer-Skarke list  \cite{Kreuzer:1998vb} contains a complete enumeration of all the $4,319$ reflexive polyhedra in three-dimensions. Among these, $2$ correspond to $K3$ manifolds with Picard number $19$ and $n=1$ complex structure parameter,  $9$ to $K3$ manifolds with Picard number $18$ and $n=2$ complex structure parameters, $24$ to $K3$ manifolds with Picard number $17$ and $n=3$ complex structure parameters and so on. In the following section, we will exploit the well-understood geometry of $K3$ fibrations in order to build up global supersymmetric solutions with non-trivial fluxes. 
Then, in Section~\ref{sec:dictionary}, we will discuss the flux/geometry dictionary for some simple choices of auxiliary Calabi-Yau three-fold geometries that can be realised as $K3$ fibrations over $\mathbb C\mathbb P^1$.

\section{The auxiliary $K3$ surface and its moduli space}\label{sec:example}  
In this section we consider a first example of a $K3$ surface with two complex structure parameters defined by a pair of dual reflexive polyhedra. We explore its complex structure moduli space by  computing the periods of the holomorphic $(2,0)$-form. In the next section, we will embed the $K3$ polyhedron into different four-dimensional reflexive polytopes encoding different fibrations of the $K3$ surface over $\mathbb C\mathbb P^1$ and discuss the associated geometry to flux dictionary.

\subsection{The $K3$ fiber}\label{sec:K3fiber}

Consider the pair of polyhedra $(\nabla,\Delta)$ defined by the data collected in Table~\ref{table:NablaDelta}. Using the conventions of Appendix~\ref{sec:ToricGeometry}, $\nabla$ specifies the embedding toric variety and $\Delta$ the Newton polyhedron that defines the homogeneous polynomial whose zero locus defines a $K3$ surface.

\vspace{12pt}
\begin{table}[H]
\def\str{\varstr{11pt}{5pt}}
\begin{center}
\begin{tabular}{c | >{$~~~} r <{~~~$} || >{$~~~} r <{~~~$} | c}
\cline{2-3}
\varstr{15pt}{9pt}
& \nabla~~~~~~~~~~~~~ & \Delta~~~~~~~~& \\
\cline{2-3}\\[-15pt]
\cline{2-3}
\varstr{14pt}{6pt} $~~~~~~~~~~~~~~~~z_1~~$& w_1 = (\ \ 1,\,\  -1,\  -1) & v_1=(\ \ 0,\ -1,\ -1) & ${z}_{1}^{3} \, {z}_{2}^{3}$\\[3pt]
\str $~~~~~~~~~~~~~~~~z_2~~$&w_2=(-1,\ -1,\ -1) & v_2=(\ \ 0,\,\ \ \ 0,\ \ \ 1) & $ {z}_{3}^{3} \, {z}_{4}^{3}$ \\[3pt]
\varstr{14pt}{6pt} $~~~~~~~~~~~~~~~~z_3~~$& w_3=(\ \ 1,\,\ -1,\ \ \ 2) & v_3=(\ \ 0,\,\ \ \ 1,\ \ \ 0) &${z}_{5}^{3} \, {z}_{6}^{3}$ \\[3pt]
\str $~~~~~~~~~~~~~~~~z_4~~$ & w_4=(-1,\  -1,\ \ \ 2) & v_4=(-1,\ \ \ 0,\ \ \ 0) & ${z}_{2}^{2} \, {z}_{4}^{2} \, {z}_{6}^{2}$ \\[3pt]
\varstr{14pt}{6pt} $~~~~~~~~~~~~~~~~z_5~~$& w_5=(\ \ 1,\,\ \ \ 2,\ -1) & v_5=(\ \ 1,\,\ \ \ 0,\ \ \ 0) & ${z}_{1}^{2} \, {z}_{3}^{2} \, {z}_{5}^{2}$ \\[3pt]
\str $~~~~~~~~~~~~~~~~z_6~~$&w_6=(- 1,\ \ \ 2,\  -1) & v_0=(\ \ 0,\,\ \ \ 0,\ \ \ 0) &${z}_{1} \, {z}_{2} \, {z}_{3} \, {z}_{4} \, {z}_{5} \, {z}_{6}$ \\[3pt]
\cline{2-3}\cline{2-3}
\end{tabular}
\capt{5.7in}{table:NablaDelta}{The list of vertices of $\nabla$ and the list of lattice points in $\Delta$. The left hand side labels $z_1,\ldots,z_6$ represent the homogeneous coordinates associated with the vertices of $\nabla$. The last column shows the monomials associated with the points of $\Delta$.}
\end{center}
\end{table}

Let $z_1,\ldots,z_6$ denote the homogeneous coordinates of the embedding toric variety, associated to the vertices of $\nabla$, as shown in Table~\ref{table:NablaDelta}. The two polyhedra $\nabla$ and $\Delta$, shown in Figure~\ref{fig:dualpair}, define a family of $K3$ hypersurfaces as the zero locus of homogeneous polynomials of the form:  
\bea\label{eq:polynomialSU9SU9}
f _{\rm hom} &=\sum_{a=0}^5  c_a \, \prod_{i=1}^6 z_i^{\langle w^{K3}_i , v^a \rangle+1} \nn\\
&=
-\,{c}_{0} \, {z}_{1} \, {z}_{2} \, {z}_{3} \, {z}_{4} \, {z}_{5} \, {z}_{6}+{c}_{1} \, {z}_{1}^{3} \, {z}_{2}^{3}+{c}_{2} \, {z}_{3}^{3} \, {z}_{4}^{3}+{c}_{3} \, {z}_{5}^{3} \, {z}_{6}^{3}+{c}_{4} \, {z}_{2}^{2} \, {z}_{4}^{2} \, {z}_{6}^{2}+{c}_{5} \, {z}_{1}^{2} \, {z}_{3}^{2} \, {z}_{5}^{2}  \; . 
\eea
Here $v_a$ label the points in $\Delta$ and $w^{K3}_i$ the vertices of $\nabla$. 
The polynomial is homogeneous with respect to three rescalings, given by the weight system presented in Table~\ref{tab:relations1}. The weights have been obtained from the three linear relations 
 $ \sum_i Q_a^i \, w^{K3}_i=0 $  between the vertices of $\nabla$.

\begin{table}[h]
\begin{center}
\begin{tabular}{l|llllllll}
& $z_1$ & $z_2$& $z_3$& $z_4$& $z_5$& $z_6$ \\
\hline
$Q_1$ & 1 & 1 & 1& 1 & 1 & 1\\
$Q_2$ & 2 & 0&1 & 1 & 0 & 2 \\
$Q_3 $& 0& 2 & 2 & 0 & 1 & 1 \\
\end{tabular}
\end{center}
\vspace{2pt}
\caption{ Weight system for the toric variety embedding $S$.}
\label{tab:relations1}
\end{table}

We notice that $f _{\rm hom}$ is invariant under the rescaling $z_i \to \lambda_i \, z_i$ if the coefficients $c_a$ are also properly rescaled. For instance $c_1 \to c_1/(\lambda_1^3 \lambda_2^3)$,  $c_2 \to c_2/(\lambda_3^3 \lambda_4^3)$ and so on. We find the following two invariant combinations:
  \beq
  {\xi\over 27}~=~ { c_1 c_2 c_3 \over c_0^3}   \qquad \text{and}~~~~~~~~      {\eta \over 4}~= ~{ c_4 c_5 \over c_0^2}~.  \label{eq:EtaXiCase3}
  \eeq
The numerical  factors are included here for later convenience. As discussed in Section~\ref{sec:pi00}, $\xi$ and $\eta$ can serve as coordinates on the moduli space. 

 \begin{figure}[h]
\begin{center}
{\hskip0pt
\begin{minipage}[t]{5.4in}
\vspace{30pt}
\raisebox{-1in}{\includegraphics[width=8.2cm]{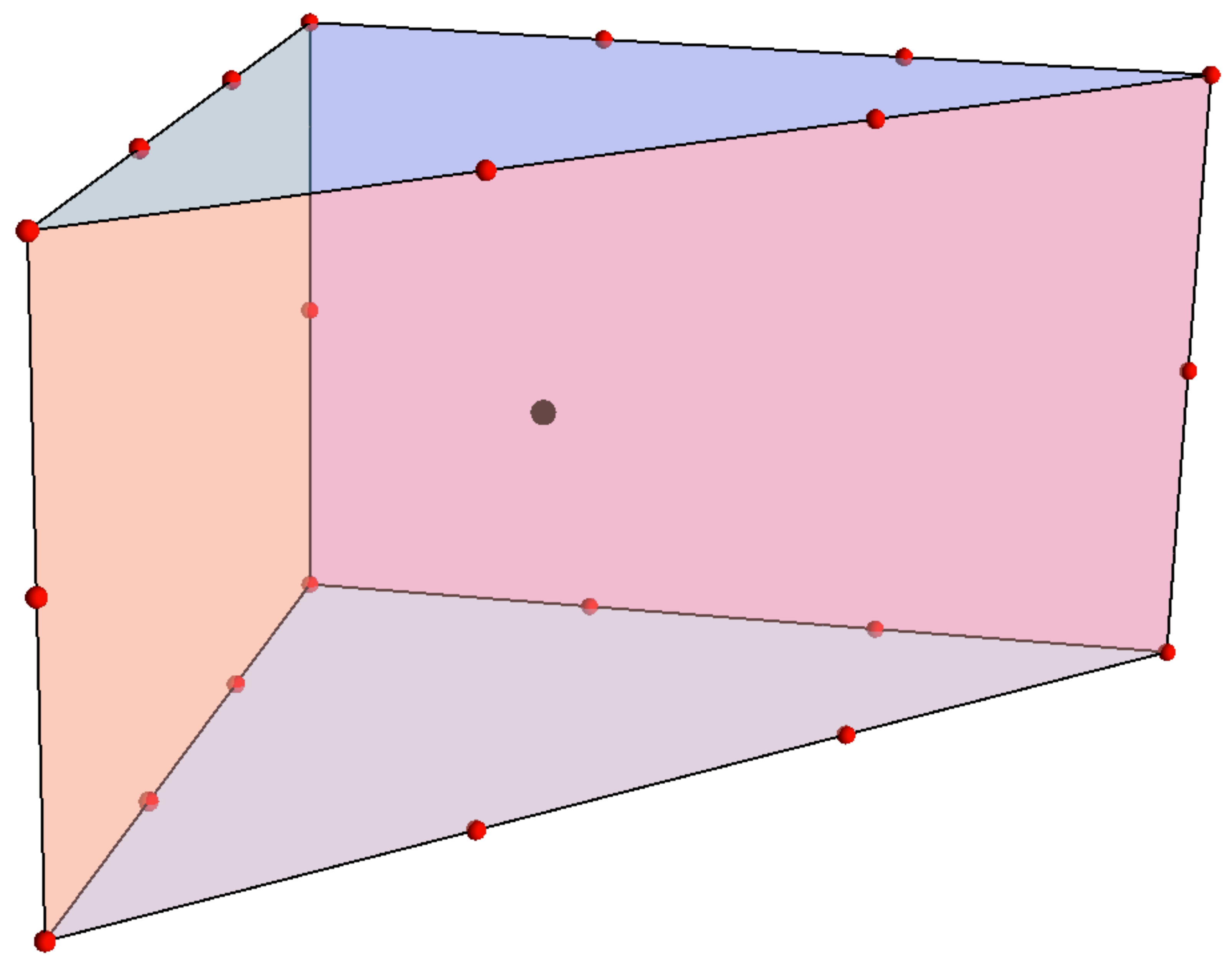}} 
\hfill \hspace*{24pt}
\raisebox{-.37in}{\includegraphics[width=4.1cm]{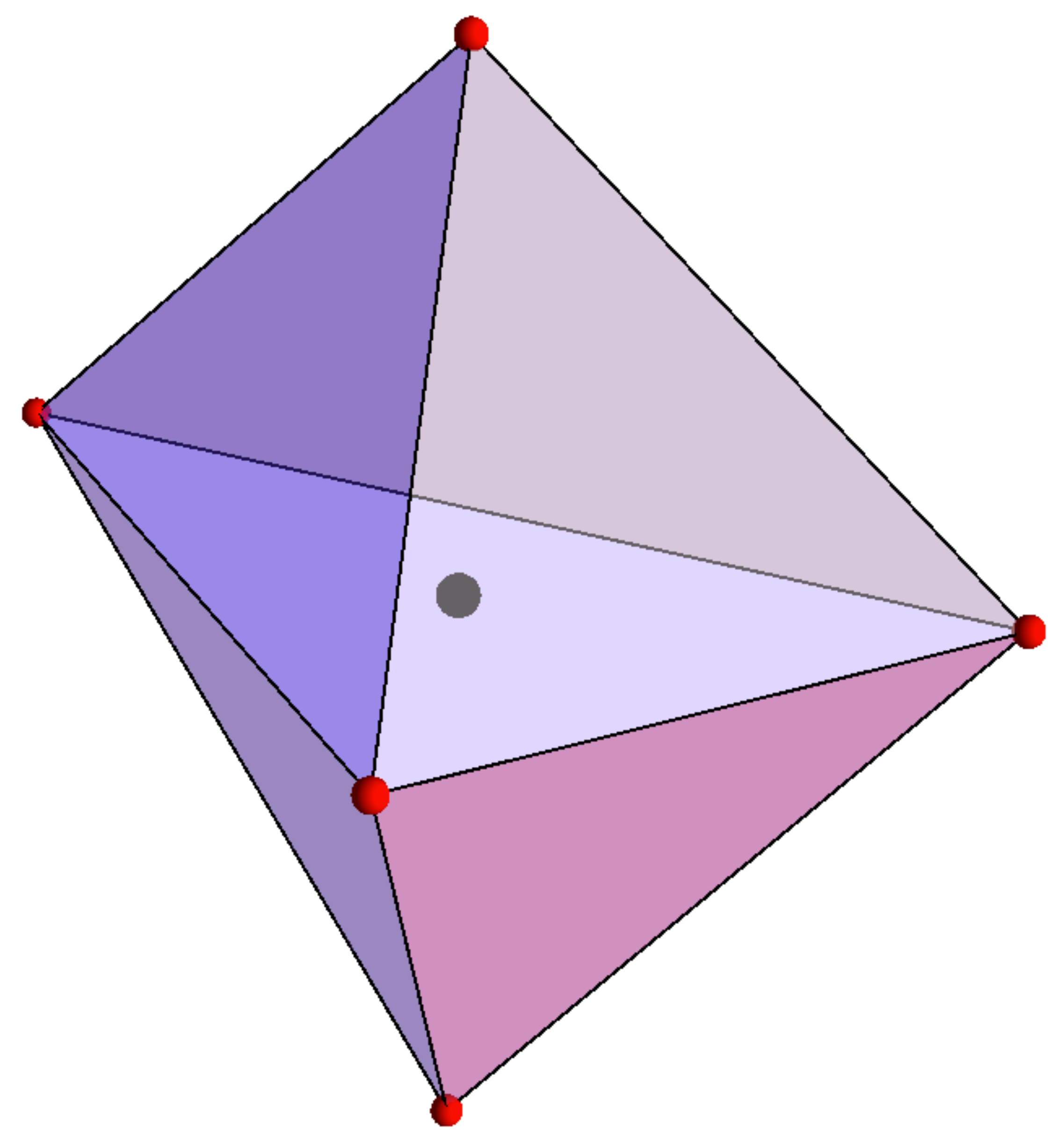}}
\hspace{10pt}
\vspace{10pt}
\end{minipage}}
\capt{6.2in}{fig:dualpair}{The left hand-side polyhedron plays the role of $\nabla$ and defines the ambient toric variety. For the purpose of clarity, the points that are interior to 2-dimensional faces have been omitted from the plot. The right hand-side polyhedron corresponds to the Newton polyhedron~$\Delta$.}
\end{center}
\end{figure}

\subsection{The periods and the $K3$ moduli space}

In this section we outline the computation of the periods of the holomorphic $(2,0)$-form which play a crucial role in the construction of the global solutions. The more technical details of the computation have been transferred to  Appendix \ref{app:ModuliSpace}.

\subsubsection{The fundamental period}\label{sec:pi00}

We are interested in the periods of the holomorphic two-form along the two-cycles spanning the transcendental lattice of the $K3$ surface. First we compute the fundamental period $\varpi_{00}$ of the holomorphic two-form by direct integration, as explained in \cite{Berglund:1993ax} (see also \cite{Griffihs, Morrison:1991cd, Candelas:1993dm, Candelas:1994hw, Hosono:1994ax,Hosono:1993qy}). Including a scale factor of $c_0$, this is given by:
\beq \label{w00}
    \varpi_{00}=-\frac{c_0}{(2\pi i)^6}   \oint_{\mathcal{C}}   \frac{ dz_1\wedge\ldots\wedge dz_6} {f_{\rm hom} }~,
\eeq 
where the cycle  $\mathcal{C}$ is a product of small circles that enclose the hypersurfaces $z_i = 0$. We rewrite the homogeneous polynomial as 
\beq
f _{\rm hom} = -\,c_0\,  z_1\, z_2\, z_3\, z_4\, z_5\, z_6\, (1 - \tilde{f} _{\rm hom}) ~.
\eeq
 
The integral (\ref{w00}) can be evaluated by residues. We find
\beq\label{eq:w00series}
\varpi_{00}(\xi,\eta)~=~\frac{1}{(2 \pi i)^6} ~
    \oint_{\mathcal{C}} \frac{  dz_1\wedge\ldots\wedge dz_6}{z_1\, z_2\, z_3\, z_4\, z_5\, z_6} ~ 
    \sum_{n=0}^{\infty} \tilde{f} _{\rm hom}^n =
    \sum_{k,l=0}^{\infty} a_{k,l} ~
    \Big(\frac{\xi}{27}\Big)^k \Big(\frac{\eta}{4}\Big)^l ,
\eeq
with
\beq\label{eq:akl}
a_{k,l}
%~=~\frac{(3k+2l)!}{(k!)^3 (l!)^2}
~ =~\frac{\Gamma(3k+2l+1)}{\Gamma^3(k+1)\,\Gamma^2(l+1)} %\frac{1}{27^k 4^l}~
\eeq
Only the constant terms in the sum over $\tilde{f} _{\rm hom}^n$ contribute to the residue. As expressed in~(\ref{eq:w00series}), these terms are always powers of $\xi$ and $\eta$. The coefficients in the expansion have been obtained from the multinomial theorem\footnote{More explicitly, the expansion involves terms of the form
 \beq
 \tilde f_{\rm hom}^n =\left( \frac{\sum_{i=1}^5 c_i\,z^{p_i}}{c_0\,z_1\,z_2\,z_3\,z_4\,z_5}\right)^n = \sum_{\stackrel{{n_1+\ldots + n_5 = n}}{n_1\,p_1+\ldots + n_5\,p_5 = n\,p_0}} \frac{n!}{\prod_{i=1}^5 n_i!} \cdot \frac{\prod_{i=1}^5 c_i ^{n_i}}{ c_0^n} ~ + ~\ldots
 \eeq
where $p_1=(3, 3, 0, 0, 0, 0)$, $p_2=(0,0,3,3, 0, 0)$, $p_3=(0, 0, 0, 0, 3, 3)$, $p_4=(2, 2, 2, 0, 0, 0)$, $p_5=(0, 0, 0, 2, 2, 2)$ and $p_0=(1,1,1,1,1)$. The omitted terms are non-constant, hence they do not contribute to the residue. Solving the set of equations 
\beq
\begin{aligned}
n_1+\ldots + n_5 &= n\\
n_1\,p_1+\ldots + n_5\,p_5 &= n\,p_0
\end{aligned}
\eeq
we obtain a class of solutions defined by $n_1 = n_2=n_3$, $n_4=n_5$ and $n = 3\,n_1+2\,n_4$. From these, the invariant combinations \eqref{eq:EtaXiCase3} follow, up to constants.}. 
 
\subsubsection{The periods}
\label{sec:PFmain}
Since the number of algebraic 2-cycles is two for this manifold the set of $\varpi_{00}$ and its derivatives with respect to the parameters $\xi,\eta$, contains four linearly independent elements, that is between any six elements of the set $\{\th_\x^i\,\th_\eta^j\,\vp_{00}\}$,  with $i+j\leq 2$ where 
$\th_\x$ and $\th_\eta$ denote the operators
\beq
\th_\x~=~\x\pd{}{\x}~,~~~\th_\eta~=~\eta\pd{}{\eta}~,
\eeq
there are two linear relations with algebraic, in fact polynomial, coefficients. These linear combinations are the Picard--Fuchs equations.  Explicitly, the fundamental period \eqref{eq:w00series} satisfies the Picard--Fuchs equations $\cL_1 \varpi_{00}=\cL_2\varpi_{00}=0$, with 
\beq\begin{split}
\cL_1~&=~4\,\th_{\eta}^2 - 
\eta\, (3\th_\x+2\th_{\eta} + 2)(3\theta_{\x}+2\th_{\eta} + 1)~,  \\[3pt]
\cL_2~&=~3\,\th_\x(3\th_\x-2\th_\eta) -\x\, (3\th_\x+2\th_\eta+2)(3\th_\x+2\th_\eta+1)
+3\,\eta\,\th_\x(3\th_\x+2\th_\eta+1)~.
\end{split}
\eeq 
The equations $\cL_1 \varpi_{00}=\cL_2\varpi_{00}=0$ can be verified using the series (\ref{eq:w00series}).  

We expect that the differential equations corresponding to $\cL_1$ and $\cL_2$ have four linearly independent common solutions. These can be found using Frobenius' method. Thus we seek four power series that satisfy the differential equations associated to the above Picard-Fuchs operators, of the form 
\beq\label{eq:FrobeniusPeriod0}
\vp~=~\sum_{k,l=0}^\infty a_{k,l}(\epsilon,\delta)~\bigg(\frac{\xi}{27}\bigg)^{k+\e} \bigg(\frac{\eta}{4}\bigg)^{l+\delta} ~=
\sum_{r,s}   \vp_{r,s}\, ( 2\p\ii \e)^r (2\p\ii \delta)^s\, .
\eeq
As explained in detail in Appendix~\ref{sec:PFapp}, requiring  $\cL_1\vp = \cL_2\vp = 0$, determines the expansion coefficients to be   
\beq
a_{k,l} (\epsilon,\delta)~ =~\frac{a_{k+\e,l+\delta}}{a_{\e,\delta}} ~=~ \frac{\Gamma^3(\e+1)\, \Gamma^2(\delta+1)}{\Gamma(3\epsilon+2\delta+1)} ~\frac{\Gamma(3(k+\epsilon)+2(l+\delta)+1)}{\Gamma^3(k+\e+1)\, \Gamma^2(l+\delta+1)}%~ \frac{1}{27^k 4^l}
.
\eeq
The resulting $\vp$ satisfies the Picard-Fuchs equations up to terms proportional to $\delta^2$ and $3\e^2- 2\e \delta$. We say then that $\cL_1\vp = \cL_2\vp = 0$ hold, if $\epsilon$ and $\delta$ satisfy the so-called indicial equations 
\beq
\delta^2=0 ~~~~\qquad~~~~3\e^2- 2\e \delta=0~.
\eeq 
It is an important consequence of the indicial relations that all cubic, and higher, monomials in $\e$~and~$\delta$ vanish. Substituting these relations into (\ref{eq:FrobeniusPeriod0}) one finds the finite expansion
 \beq
\vp~=~\vp_{00} + 2\p\ii\e\, \vp_{10} + 2\p\ii\delta\, \vp_{01} + 
 ( 2\p\ii)^2\e\delta\, \Big( \vp_{11} + \smallfrac13 \vp_{20}\Big)~.
\eeq
with each term defining a solution of the Picard--Fuchs equations. Using these, we form the basis 
 \beq\label{eq:periodbasis0}
\{\vp_0=\vp_{00},~ \vp_1=\vp_{10}, ~\vp_2=3\,\vp_{01}+\vp_{10},~ \vp_3=3\vp_{11}+ \vp_{20}\}~,
\eeq 
which, as explained in Appendix~\ref{sec:PFapp}, is special in that it leads to the factorisation $\vp_0\,\vp_3=\vp_1\,\vp_2$.
The quantities $\vp_{rs}$ are obtained by taking partial derivatives of the Frobenius period \eqref{eq:FrobeniusPeriod0}. Taking partial derivatives produces, on the one hand, logarithms of $\xi$ and $\eta$, and on the other hand, when acting on the $a_{k,l}$ coefficients, quantities of the form
\beq
h_{rs}~=~\frac{1}{(2\p\ii)^{r+s}}\sum_{k,l=0}^\infty     a_{k,l}^{r,s}\, \bigg(\frac{\xi}{27}\bigg)^{k} \bigg(\frac{\eta}{4}\bigg)^{l}\, ,
\eeq
with
\beq
a_{k,l}^{r,s}=\left. \left(\pd{}{\e}\right)^r \! \left(\pd{}{\delta}\right)^s a_{k,l}(\e,\delta)\, \right|_{\e,\delta=0}
\eeq
For the first few terms one finds
\beq
a_{k,l}^{0,0}=a_{k,l}\quad , \quad   a_{k,l}^{0,1}=  {a_{k,l}\over 2 \pi \ii}    (2\,H_{3k+2l}-2\,H_{l} )\quad , \quad   
a_{k,l}^{1,0}=  {a_{k,l}\over 2 \pi \ii}    (3\,H_{3k+2l}-3\,H_{k} )
\eeq
with $H_n=\sum_{m=1}^n {1\over m}$ the harmonic numbers. Note that $h_{00} = \vp_{00}$. The basis of periods~\eqref{eq:periodbasis0} is explicitly given by: 
\beq\begin{split}
\vp_0&=h_{00}\\[8pt]
\vp_1&=\frac{1}{2\p\ii}\,h_{00}\log\left(\frac{\x}{27}\right) + h_{10}\\[8pt]
\vp_2&=\frac{1}{2\p\ii}\,h_{00} \log\left( \frac{\x\,\eta^3}{1728} \right)+ 3\,h_{01}+h_{10}\\[8pt]
\vp_3&=\frac{1}{(2\p\ii)^2}h_{00}
\log\left(\frac{\x}{27}\right)\, \log\left( \frac{\x\,\eta^3}{1728} \right) + 
\frac{1}{2\p\ii} \left(3\,h_{01}+h_{10}\right)\log\left(\frac{\x}{27}\right)\\[8pt]
&\hskip2.32in + \frac{1}{2\p\ii} h_{10}\log\left( \frac{\x\,\eta^3}{1728} \right) +3\,h_{11}+h_{20}~. \\
\end{split}\label{PeriodBasis1}\eeq

The fact that  $\vp_0 \vp_3~=~\vp_1 \vp_2$ can be checked by multiplication of the respective series. Thus, we can define the complex structure variables $\tau^{(i)}$ as the period ratios 
\bea\label{eq:tau1tau20}
\begin{aligned}
\tau^{(1)} &=~\frac{\vp_1}{\vp_0} ~=~\frac{1}{2\p\ii}\log\left(\frac{\x}{27}\right) +\frac{h_{10}}{h_{00}}  \\[8pt]
\tau^{(2)} &=~\frac{\vp_2}{\vp_0}~=~\frac{1}{2\p\ii}\log\left( \frac{\x\,\eta^3}{1728} \right) +\frac{3\,h_{01}+h_{10}}{h_{00}}~.
\end{aligned}
\eea
%with $\tau^{(1)} \tau^{(2)} ~=~\frac{\vp_3}{\vp_0}~$.  

\subsubsection{$j$-invariants}
While $\tau^{(1)}$ and $\tau^{(2)}$ are defined in terms of $(\x,\eta)$ by power series via \eqref{eq:tau1tau20}, we anticipate that the respective $j$-invariants will be algebraically related to $(\x,\eta)$.
%To see this, we pass to the exponentials $q_1=\exp(2\p\ii\t^{(1)})$ and $q_2=\exp(2\p\ii\t^{(2)})$
%\beq
%q_1~=~\frac{\x}{27} \exp\left(2\p\ii\,\frac{h_{10}}{h_{00}}\right)~;~~~~~~~
%q_2~=~\frac{\x\eta^3}{12^3}\exp\left(2\p\ii\,\frac{3\,h_{01}+h_{10}}{h_{00}}\right)~.
%\eeq
We start by computing 
\beq
j_1~=~j(\t^{(1)})~~~\text~~~j_2~=~j(\t^{(2)})
\eeq
as power series in $(\x,\eta)$, with $j$ defined as\footnote{Here $\displaystyle E_{2k}(\t)=1+{ 2\over \zeta(1-2k)} \sum_{n=1}^\infty { n^{2k-1} e^{2n\pi\ii\t} \over  (1-e^{2n\pi\ii\t})}$ with  $\displaystyle{2\over \zeta(-3)} =240$ and $\displaystyle{2\over \zeta(-5)}=-504$. Note also that $\displaystyle j-12^3= {E_6^2 \over \eta^{24}}$.}
\beq
j(\tau)={E_4^3(\t) \over \eta^{24}(\tau)} =\e^{-2\pi\ii\tau}+744+ \ldots  \qquad\qquad q=e^{2\pi\ii \t}  \label{jfuncaa}
\eeq 
 with $E_n$ the Eisenstein series and $\eta$ the Dedekind eta function.

%Note that our notation is that $j_2$ denotes $j(3\t_2)$, not $j(\t_2)$. 
%
Using the small $\xi,\eta$ expansions  one can show that  the U-duality invariant combinations $j_1 \, j_2$ and $j_1+j_2$ are given by rational functions in $\xi$ and $\eta$. Explicitly,
 \beq
 \begin{aligned}
\smallfrac{1}{6^6}\, \left(j_1-12^3\right)\left( j_2-12^3\right) &
= {  \left(8\,\xi^2 + (\eta-1)^3 - 12\,\xi\,\eta+20\,\xi\right)^2 \over \eta^3 \xi^2} ~, \\[5pt]
\smallfrac{1}{6^6}\,  j_1 \, j_2  & ~= ~{  \left(8\, \xi + (\eta -1)^2\right)^3 \over \eta^3 \xi^2  }~.
\label{jjsaa}
\end{aligned}
\eeq
Using these relations, one finds the following remarkably simple expressions
\beq\begin{split}\label{eq:j1j20}
j_1~&=~\frac{6^3}{\x\eta^3}\Big(1-\x -\eta  - \sqrt{D}\Big)^3~,\\[5pt]
j_2~&=~\frac{6^3}{\x\eta^3}\Big(1-\x -\eta + \sqrt{D}\Big)^3~.
\end{split}\eeq
   with $D$ the discriminant that in our case is given by (see Appendix~\ref{app:discriminant}): 
\beq
D=  (\xi-1-3 \eta)^2-\eta (\eta+3)^2 \label{Eq:Discriminant}~.
\eeq
%An immediate consequence of these expressions is that when $D=0$ we have $j_1=j_2$ so the component $C_D$ of the discriminant locus corresponds to the relation $\t^{(1)}=\t^{(2)}$. Moreover, encircling the locus $D=0$ exchanges $j_1$ and $j_2$, which corresponds to the $\mathbb{Z}_2$ factor of the U-duality group. %This suggests that U-duality invariants can be built out of the symmetric combinations $j_1\,  j_2$ and $j_1+j_2$.

The same result can be derived by looking at the loci $\eta=0$ and $\x=0$. The expressions that $j_1$ and $j_2$ take along these curves are collected in Table~\ref{XiEtaLimits}. 

\vspace{8pt}
\begin{table}[H]
\begin{center}
\begin{tabular}{|r|ccc|}
\hline
\vrule height12pt depth8pt width0pt & $\x~=~0$ & \hspace*{10pt} & $\eta~=~0$ \\
\hline\hline
\vrule height23pt depth15pt width0pt $\x\, j_1$~ 
                        & ~$\displaystyle\frac{6^3(1-\eta)^3 (1-\sqrt{1-\eta})^3}{\eta^3} $~ 
                           & & ~$\displaystyle\frac{3^3(1+8\x)^3}{(1-\x)^3}$~ \\
\vrule height20pt depth13pt width0pt $\x\,\eta^3 j_2$~ 
                        & ~$ 6^3 (1-\eta)^3 (1+\sqrt{1-\eta})^3$~
                           & & ~$ 12^3 (1-\x)^3 $~ \\
\hline                           
\end{tabular}
\end{center}
\vspace{-8pt}
\caption{The form of $j_1$ and $j_2$ along the coordinate lines $\x=0$ and $\eta=0$.}
\label{XiEtaLimits}
\end{table}

%Notably, taking the $\eta\to0$ limit in the Picard-Fuchs equations, we observe that $\vp_{00}(\x,0)$ and 
%$\vp_1(\x,0)$ satisfy the corresponding ordinary hypergeometric differential equation in $\x$. From this it is easy to see that
%\beq
%\t^{(1)}(\x,0)~=~\frac{\ii}{\sqrt{3}}\,
%\frac{{}_2F_1\left(\smallfrac13,\smallfrac23;1;1-\x\right)}{{}_2F_1\left(\smallfrac13,\smallfrac23;1;\x\right)}
%\vspace{4pt}
%\eeq
%and it is a classical fact, for $\t^{(1)}$ defined by this relation, that 
%\beq
%j(\t^{(1)})~=~\frac{27(1+8\x)^3}{\x(1-\x)^3}~,
%\eeq
%a fact that can be checked from the series expansions. 

The $\x=0$ column of the table shows that, for general $(\x,\eta)$ the quantities $\x j_1$ and $\x\eta^3 j_2$ cannot be merely rational functions of $(\x,\eta)$. Since, however, all the entries of the table are perfect cubes we can hypothesise that $(\x j_1)^{1/3}$ and $(\x\eta^3 j_2)^{1/3}$ are of the form
\beq
\frac{P + Q\sqrt{D}}{R}
\eeq
for suitable polynomials $P, Q$ and $R$, where $D=0$ corresponds to the discriminant locus, i.e.~the locus in moduli space where the $K3$ is singular. By comparing this form with the series expansions we find again the result (\ref{eq:j1j20}). 
   
The above discussion provided us with a number of coordinates on the moduli space  \eqref{mbps}. We started with the parameters $\xi$ and $\eta$, then we found the period ratios $\tau^{(i)}$  and, finally, we computed the $SL(2,\mathbb Z)$-invariants $j(\tau^{(i)})$.  We will identify the supergravity fields $\sigma$ and $\tau$ entering in the flux solutions with the period ratios $\tau^{(1)}$ and $\tau^{(2)}$, respectively\footnote{A different order in which $\sigma$ and $\tau$ are identified with $\tau^{(1)}$ and $\tau^{(2)}$ corresponds to a different solution.}.  

In the following section, we will consider one-parameter families of $K3$ surfaces obtained by treating the coefficients in \eqref{eq:polynomialSU9SU9}, or alternatively, the parameters $\xi$ and $\eta$, as functions of a single complex parameter $z$. We will obtain such families by explicitly constructing Calabi-Yau threefolds that are $K3$ fibrations over $\mathbb C\mathbb P^1$ using the language of toric geometry.

\section{Geometry to flux dictionary}\label{sec:dictionary}
\subsection{$K3$-fibered Calabi-Yau threefolds and brane solutions} 

There are many ways in which one can fiber the $K3$ surface (\ref{eq:polynomialSU9SU9}) over $\mathbb C\mathbb P^1$ in order to obtain a Calabi-Yau three-fold. Changing the $K3$ fibration leads to a different brane content in the supergravity solutions. Here we consider two different choices of Calabi-Yau threefolds that have the same $K3$ fiber and describe the brane content in each case.

\subsubsection{Calabi-Yau threefolds: a first example}\label{sec:firstCY3}
A three-fold with $K3$ fiber given by (\ref{eq:polynomialSU9SU9}) can be constructed by starting from the polyhedron $\nabla_{K3}$ defined by the data presented in Table~\ref{table:NablaDelta}, and then extending this polyhedron into a fourth dimension by adding points above and below the hyperplane that contains $\nabla_{K3}$. In adding the extra points, one should take care that the resulting four-dimensional polytope $\nabla_{\text{CY}_3}$ is reflexive, and, moreover, that $\nabla_{K3}$ is contained in $\nabla_{\text{CY}_3}$ as a slice, in the sense explained in Section~\ref{sec:ToricGeometry}. The latter condition ensures that the resulting Calabi-Yau three-fold admits a $K3$ fibration structure with a fiber given by $\nabla_{K3}$ and its dual. 

The simplest extension of $\nabla_{K3}$ is to add two points, one above and one below the origin, along the extra direction:
\beq
w_i^{\text{CY}_3}=(0,~w^{K3}_i)   \qquad       w^{\text{CY}_3}_7=(-1,0,0,0) \qquad   w^{\text{CY}_3}_8=(1,0,0,0)  \label{vertone}
\eeq
The resulting Calabi-Yau threefold has Hodge numbers $(h^{1,1},h^{2,1})=(35,11)$, as given by Batyrev's formulae~\cite{Batyrev:1993dm}.
%\vspace{3pt}
%\begin{table}[H]
%\def\str{\varstr{11pt}{5pt}}
%\begin{center}
%\begin{tabular}{c | >{$~~~} r <{~~~$} || >{$~~~} r <{~~~$} | c }
%\cline{2-3}
%\varstr{15pt}{9pt}
%& \nabla_{\text{CY}_3}~~~~~~~~ &  \nabla_{K3}~~~~~~~~&~~~~~~~~~~~~~~~\\
%\cline{2-3}\\[-15pt]
%\cline{2-3}
%\varstr{17pt}{6pt} $~~~~~~~~~~~~~~~~z_1~~$&(\ \ 0,\ \ \ 1,\,\  -1,\  -1)&(\ \ 1,\,\  -1,\  -1) \\[4pt]
%\str $~~~~~~~~~~~~~~~~z_2~~$&(\ \ 0,\ -1, \ -1,\ -1)&(-1, \ -1,\ -1)  \\[4pt]
%\varstr{14pt}{6pt} $~~~~~~~~~~~~~~~~z_3~~$& (\ \ 0,\ \ \ 1,\,\ -1,\ \ \ 2) & ( \ \ 1,\,\ -1,\ \ \ 2)  \\[4pt]
%\str $~~~~~~~~~~~~~~~~z_4~~$ & (\ \ 0,\ -1,\  -1,\ \ \ 2)& ( -1,\  -1,\ \ \ 2) \\[4pt]
%\varstr{14pt}{6pt} $~~~~~~~~~~~~~~~~z_5~~$& (\ \ 0, \, \ \ \, 1,\ \ \ 2,\ -1)& ( \ \ 1,\ \ \ 2,\ -1) \\[4pt]
%\str $~~~~~~~~~~~~~~~~z_6~~$&(\ \  0,\  - 1,\, \ \ 2,\  -1)&(- 1,\, \ \ 2,\  -1)  \\[4pt]
%\str $~~~~~~~~~~~~~~~~z_7~~$&(-1,\ \ \ 0,\ \ \ 0,\ \ \ 0)&  \\[4pt]
%\str $~~~~~~~~~~~~~~~~z_8~~$&(\ \ 1,\ \ \ 0,\ \ \ 0,\ \ \ 0)&  \\[8pt]
%\cline{2-3}\cline{2-3}
%\end{tabular}
%\capt{5.7in}{table:NablaDelta2}{The list of vertices of $\nabla_{\rm{CY}_3}$ and of $\nabla_{K3}$. The polyhedron $\nabla_{K3}$ is contained as a slice in $\nabla_{\rm{CY}_3}$. The left hand side labels $z_1,\ldots,z_8$ represent the homogeneous coordinates associated with the vertices of $\nabla_{\rm{CY}_3}$.}
%\end{center}
%\vskip-10pt
%\end{table}
%
The $8$ vertices of $\nabla_{\text{CY}_3}$ define $8$ vectors, among which $4$ linear relations hold. These lead to the weight system presented in Table~\ref{tab:relations}. 
  \begin{table}[H]
\begin{center}
\begin{tabular}{l|llllllll}
& $z_1$ & $z_2$& $z_3$& $z_4$& $z_5$& $z_6$ & $z_7$ & $z_8 $ \\
\hline
$Q_1$ & 1 & 1 & 1 & 1 & 1 & 1& 0& 0 \\
$Q_2$ & 2 & 0&1 & 1 & 0 & 2& 0& 0 \\
$Q_3 $& 0& 2 & 2 & 0 & 1 & 1& 0& 0 \\
$Q_4 $&  0& 0& 0& 0& 0 & 0 & 1& 1  \\
\end{tabular}
\end{center}
\vspace{-8pt}
\caption{ Weight system for the toric variety embedding the Calabi-Yau three-fold.}
\label{tab:relations}
\end{table}
%\vspace{8pt}

The dual polyhedron $\Delta_{\text{CY}_3}$ contains $18$ points and leads, via~\eqref{fhom}, to the homogeneous polynomial
\begin{equation}\label{eq:CY3}
\begin{aligned}
f _{\text{CY}_3, \,{\rm hom}}
% ~ =&~\sum_{a=1}^{18}  c_a \, \prod_{i=1}^8 z_i^{\langle w^{CY}_i , v^a \rangle+1}
 ~=&~ \left(c_{0,1}\, z_7^2 + c_{0,2}\, z_7\,z_8 + c_{0,3}\, z_8^2 \right) \, z_1\, z_2\, z_3\, z_4\, z_5\, z_6 \\[4pt]
&\! +
\left(c_{1,1}\, z_7^2 + c_{1,2}\, z_7\,z_8 + c_{1,3}\, z_8^2 \right) \,z_1^3\,z_2^3+\left(c_{2,1}\, z_7^2 + c_{2,2}\, z_7\,z_8 + c_{2,3}\, z_8^2 \right) \,z_3^3\,z_4^3 \\[4pt]
&\! +\left(c_{3,1}\, z_7^2 + c_{3,2}\, z_7\,z_8 + c_{3,3}\, z_8^2 \right) \,z_5^3\,z_6^3+\left(c_{4,1}\, z_7^2 + c_{4,2}\, z_7\,z_8 + c_{4,3}\, z_8^2 \right)  \, {z}_{2}^{2} \, {z}_{4}^{2} \, {z}_{6}^{2} \\[4pt]
&\!+\left(c_{5,1}\, z_7^2 + c_{5,2}\, z_7\,z_8 + c_{5,3}\, z_8^2 \right)  \, {z}_{1}^{2} \, {z}_{3}^{2} \, {z}_{5}^{2}  \; . 
\end{aligned}
\end{equation}
%where some of the $c_a$'s have been renamed for convenience.

The form of the polynomial $f _{\text{CY}_3, \,{\rm hom}}$ makes manifest the fibration structure. Indeed, we can write it in the form (\ref{eq:polynomialSU9SU9}) but now with the coefficients $c_a$ replaced by homogenous polynomials of order two in $z=(z_7,z_8)$ parametrizing
% $z=z_{7}/z_{8}$ parametrizing the patch $z_8=1$ of 
the $\mathbb C\mathbb P^1$ base: 
\begin{equation}
\begin{aligned}
f _{\text{CY}_3, \,{\rm hom}} ~~=&~~ - {c}_{0}(z) \, {z}_{1} \, {z}_{2} \, {z}_{3} \, {z}_{4} \, {z}_{5} \, {z}_{6} + {c}_{1}(z) \, {z}_{1}^{3} \, {z}_{2}^{3}+{c}_{2}(z) \, {z}_{3}^{3} \, {z}_{4}^{3}+{c}_{3} (z)\, {z}_{5}^{3} \, {z}_{6}^{3}\\
&~~+{c}_{4} (z)\, {z}_{2}^{2} \, {z}_{4}^{2} \, {z}_{6}^{2}+{c}_{5}(z) \, {z}_{1}^{2} \, {z}_{3}^{2} \, {z}_{5}^{2}  \; 
\end{aligned}
\end{equation}
 
As a result, we can write the parameters $\xi$ and $\eta$ defined by (\ref{eq:EtaXiCase3}) as functions depending on $z$:
\beq
\frac{\xi (z)}{27}   =  { f_6(z)  \over f_2^3(z) }   \qquad~~ \text{and}~~~~~~~~      \frac{\eta(z)}{4} =  {  f_4(z) \over f_2^2(z)} ~,\label{eq:etaxiz2}
\eeq
where the subscripts of the polynomials $f_2(z), f_4(z)$ and $f_6(z)$ indicate their degree in $z$. Explicitly, these polynomials are given by
\beq 
f_2(z)={c}_{0}(z)~,\qquad f_4(z)={c}_{4} (z)\,  {c}_{5}(z) ~,\qquad f_6(z)= {c}_{1}(z) \, {c}_{2}(z)  \,{c}_{3}(z)~.
\eeq

\subsubsection{The brane content}\label{sec:branecontent}
 
A flux solution of class A, B or C is obtained by identifying the holomorphic functions $\sigma(z)$ and $\tau(z)$ characterising the corresponding supergravity solutions with the period ratios $\tau^{(1)}$ and $\tau^{(2)}$.
Here we take
\beq
\sigma(z)=\tau^{(1)}(z)   \quad , \quad \tau(z)=\tau^{(2)}(z)~.
\eeq
The $K3$ fibration discussed in the previous section defines an embedding of the base space $\mathbb C\mathbb P^1$ into the moduli space $\mathcal M_{K3,2}$ (or equivalently the supergravity scalar manifold) through the map $z\mapsto    (\sigma(z),\tau(z))$, defined by the composition of~\eqref{eq:tau1tau20} and~\eqref{eq:etaxiz2}.  

The map $z\mapsto (\xi(z),\eta(z))$ is easier to visualise and, as such, it can give us a more immediate picture of the embedding of the $\mathbb C\mathbb P^1$ base of the Calabi-Yau threefold into the complex structure moduli space of the $K3$ fiber. For the chosen Calabi-Yau threefold, Figure~\ref{fig:Branes} illustrates this embedding in a schematic~way.

\begin{figure}[H]
\begin{center}
{\hskip-8pt
\begin{minipage}[t]{5.2in}
\vspace{-8pt}
\hspace{-30pt}
\raisebox{-1in}{\includegraphics[width=15.4cm]{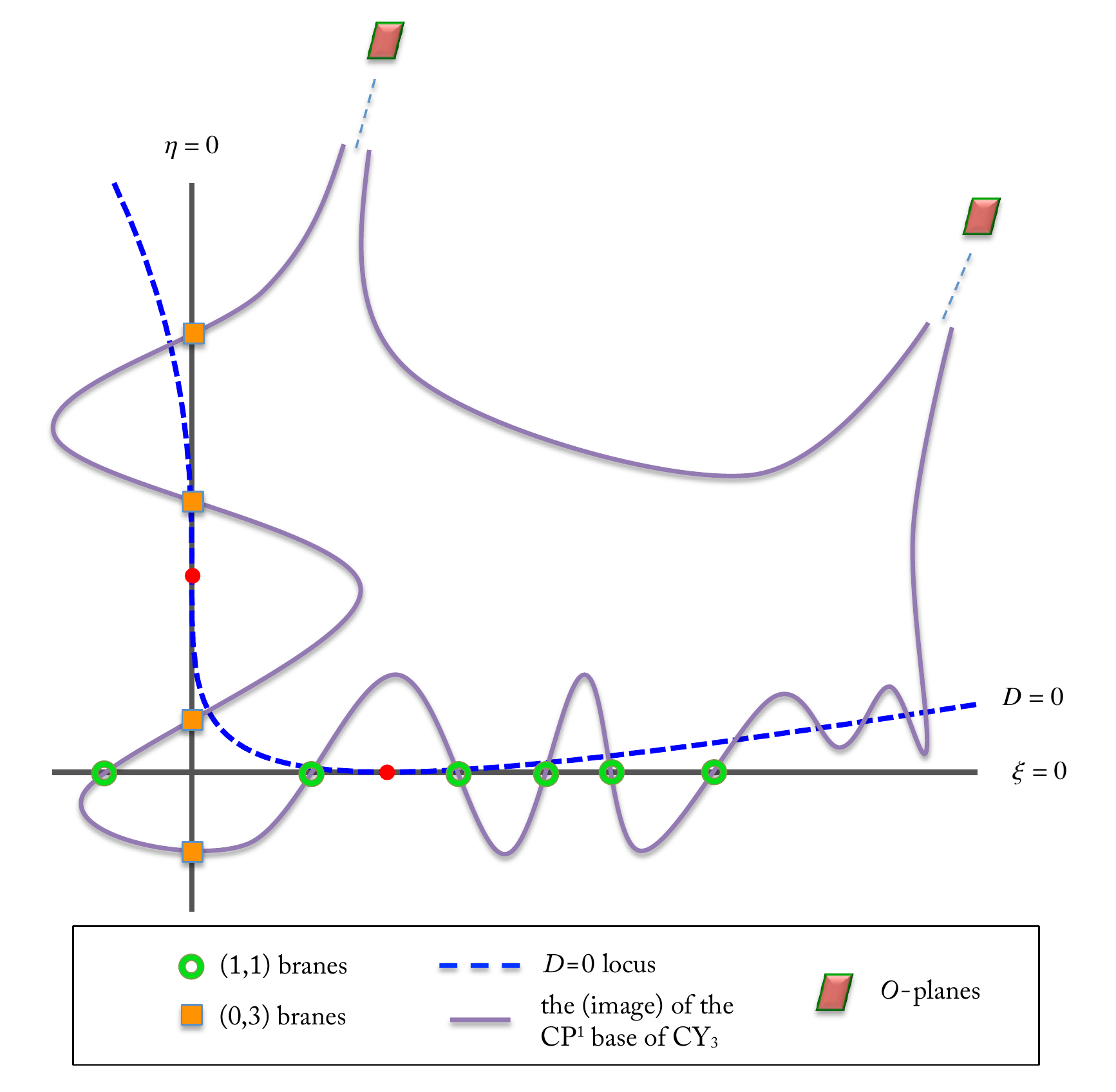}} 
\vspace{-8pt}
\end{minipage}}
\capt{5.7in}{fig:Branes}{The embedding of the (real part of the) $\mathbb C\mathbb P^1$ base (in purple) of the Calabi-Yau threefold into the complex structure moduli space of the $K3$ fiber. 
The green circles and the orange squares correspond to brane locations. The base intersects the $\xi=0$ curve in $6$ points, the $\eta=0$ curve in $4$ points and the $D=0$ curve in $12$ points. At~infinity in the $(\xi,\eta)$-plane we expect to find O-planes -- located at the zeros of $f_2(z)$ --  which cancel the tadpoles associated with the branes.}
\end{center}
\vskip -14pt
\end{figure}

In this section, we will infer the brane content of the flux solution from the geometry of the $K3$~fibration. 
Branes are located at the poles of $j_1$, $j_2$. A simple inspection of (\ref{eq:j1j20}) shows that these poles are located at the zeros of $\xi(z)$ and $\eta(z)$. To find the orders of these poles, we can expand $j_1$ and $j_2$ for small $\xi$ and, separately, for small $\eta$. The exact forms that the $j$-functions take in these limits are given in Table~\ref{XiEtaLimits}. From this we see that both $j_1$ and $j_2$ have a pole of order $1$ along the locus $\x=0$, while $j_2$ has a pole of order $3$ along $\eta=0$. 
 
The same information can be extracted from the monodromies of $\tau$ and $\sigma$. From (\ref{eq:tau1tau20}),  one finds
 \beq
\sigma,\tau ~\sim~ \frac{1}{2\pi\ii}  \log\left(\frac{\xi}{27}\right) ~.     \label{tausigmon}
\eeq
at the points where  $\xi(z)=0$, which according to Eq.\eqref{eq:etaxiz2}, correspond to the roots of $f_6$.
Accordingly, if $\xi(z)\sim (z-z_0)$ has a zero of vanishing order $1$ at $z_0$, then encircling $z_0$ one finds the monodromy   
\beq
\sigma \longrightarrow \sigma+1 \qquad \tau \longrightarrow \tau +1        \label{mon1}
 \eeq
indicating the presence of a brane of charge $(1,1)$, in the sense used in Section~\ref{sec:from6dto4d}. 
For example, for solutions in the A-class, this correspond to a bound state of a D7-brane and a D3-brane,
while for solutions in classes B and C, to the intersection of 5-branes of NS and R type, respectively. Similarly, if $f_6$ has a zero with vanishing order $q$ at $z_0$,  i.e. $f_6(z) = (z-z_0)^q$, one finds that $\sigma$ and $\tau$ exhibit the monodromies associated to a brane of charge $(q,q)$. In a count that includes multiplicities, there are $6$ points that correspond to branes with charge $(1,1)$ in the $z$-plane. 

\vspace{8pt}
Similarly, around the points where  $\eta = 0$, corresponding to the roots of $f_4$, we have
\vspace{-2pt}
\beq
\sigma \longrightarrow \sigma + 3
\eeq
indicating the presence of a brane with charge $(0,3)$. We have a total of $4$ such points. 

The results are summarised in Table~\ref{table:branes1} and illustrated in Figure~\ref{fig:Branes}. In total, we have $6\times 2+4\times 3=24$ branes, corresponding the required number for a compact geometry.

\begin{table}[h]
\begin{center}
\begin{tabular}{c| c c}
& ~~~~$f_6$~~~~ & ~~~~$f_4$~~~~ \\[2pt]
\hline\\[-14pt]
($q_1$, $q_2$)~~ & $(1,1)$ & $(0,3)$
\end{tabular}
\vspace{-8pt}
\capt{6.1in}{table:branes1}{Branes of charge $(q_1,q_2)$ are located at the zeros of $f_6$ and $f_4$.  }
\end{center}
\end{table}

Finally, there are $12$ points where  $D(\xi(z),\eta(z))=0$, see Eq.~\eqref{Eq:Discriminant}. Around these points, there is a $\mathbb Z_2$-monodromy that interchanges $j_1$ and $j_2$, or, equivalently, $\tau$ and $\sigma$ are flipped. The corresponding flux vacuum can therefore be thought of as a non-geometric  $\mathbb{Z}_2$-orbifold by the U-duality element that flips the two fields.\footnote{A non-geometric $\mathbb{Z}_2$-orbifold involving a quotient by T-duality was constructed in \cite{Bianchi:1999uq}.}  

\vspace{8pt}
We can now write down the two-dimensional (holomorphic) metric $h(z)$. Following Eq.~\ref{hz1} and the ensuing remarks, we obtain
\beq
h(z) =  { \eta(\tau)^{2}\,  \eta(\sigma)^{2}  \over  \prod_{i=1}^6 (z-a_i)^{1\over 6} \prod_{i=1}^4  (z-b_i)^{1\over 4}  }  
\eeq
where $\eta(\tau)$ and $\eta(\sigma)$ are Dedekind eta functions, $\{a_i\}$ represent the six roots of $f_6$ and $\{b_i\}$ the roots of $f_4$. The multiplicity of each point is given by the total charge of the corresponding brane -- that is $2$ for the roots of $f_6$ and $3$ for the roots of $f_4$ (see the discussion at the end of Section~\ref{sec:from6dto4d}). This gives a total of $24$ points. It is interesting to note that these points correspond to the degeneration of the product of the two $j$-functions
\beq
j_1\,j_2\sim \frac{\left(8 \xi+ (\eta -1)^2 \right)^3}{\xi ^2\,\eta ^3 }~.
\eeq
Notice that this product is U-duality invariant unlike the individual $j_i$'s. 
The function $h(z)$ is single valued around any of the singular points (brane locations).  Moreover, at large~$z$, $ h(z) \sim z^{-2}$, so after changing coordinates to $1/z$, the metric becomes ${\rm d}z\,{\rm d}\bar z$, and it is regular.  Consequently, the two dimensions parametrized by $z$ are compact \cite{Greene:1989ya}. 

Additionally, the combination 
\beq
   \tau_2 \, \sigma_2\, | h(z)|^2 
\eeq
describing the two-dimensional part of the metric \eqref{metric60}, is invariant under the U-duality group $SO(2,2,\mathbb{Z})\cong  \mathbb{Z}_2 \times SL(2,\mathbb{Z})_{\tau} \times SL(2,\mathbb{Z})_{\sigma}$. 

\vspace{8pt}
Finally, let us point out that taking
\beq
f_4(z)  = f_2(z)^2  ~~~  \qquad \text{and}~~~~~~~~~~~   f_6(z) = f_2(z)^3
\eeq    
leads to constant $\sigma,\tau$ over the whole complex plane. This suggests that the zeros of $f_2(z)$ can be interpreted as the positions of the O-planes that achieve a local tadpole cancelation when branes are exactly located on top of them.

\subsubsection{Calabi-Yau threefolds: a second example}\label{sec:secondCY3}
A different three-fold with the same fiber is obtained by extending the polyhedron $\nabla_{K3}$ defined in Table~\ref{table:NablaDelta} by adding one point ``above" and ``below" each of the six vertices of $\nabla_{K3}$:
\beq
w_i^{\rm CY_3}=(0,~w^{K3}_i)  \qquad      w_{6+i}^{\rm CY_3}=(-1,~w^{K3}_i)  \qquad      w_{12+i}^{\rm CY_3}=(1,~w^{K3}_i)  \quad i=1,\ldots 6 \label{vertsecond}
\eeq
This extension leads to a Calabi-Yau three-fold with Hodge numbers $(h^{1,1},h^{2,1})=(75,3)$. 

The dual polyhedron $\Delta_{\rm CY_3}$ has $8$ points and leads to the following polynomial:
\begin{equation}\label{eq:CY3}
\begin{aligned}
f _{\text{CY}_3, \,{\rm hom}} 
= &~
%\sum_{a=1}^{8}  c_a \, \prod_{i=1}^{18} z_i^{\langle w^{CY}_i , v^a \rangle+1} =
\left(c_{0,0}\, z_7\, z_8\, z_9\, z_{10}\, z_{11}\, z_{12}\, z_{13}\, z_{14}\, z_{15}\, z_{16}\, z_{17}\, z_{18}\right. \\[2pt]
&\left. +~ c_{0,1}\, z_7^2\, z_9^2\, z_{11}^2\, z_{13}^2\, z_{15}^2\, z_{17}^2 +c_{0,2}\, z_8^2\, z_{10}^2\, z_{12}^2\, z_{14}^2\, z_{16}^2\, z_{18}^2 \right) \, z_1\, z_2\, z_3\, z_4\, z_5\, z_6 \\[2pt]
&+c_1\,z_1^3\,z_2^3\,z_7^3\, z_8^3\, z_9^3\, z_{10}^3   + c_2\,z_3^3\, z_4^3\, z_{11}^3\, z_{12}^3\, z_{13}^3\, z_{14}^3 + c_3\,z_5^3\, z_6^3\, z_{15}^3\, z_{16}^3\, z_{17}^3\, z_{18}^3\\[2pt]
&+ c_4\,z_1^2\, z_3^2\, z_5^2\, z_7^2 \,z_8^2 \,z_{11}^2 \,z_{12}^2\, z_{15}^2 \,z_{16}^2 + c_5\,z_2^2 \,z_4^2 \,z_6^2 \,z_9^2 \,z_{10}^2 \,z_{13}^2 \,z_{14}^2\, z_{17}^2 \,z_{18}^2 \; . 
\end{aligned}
\end{equation}
%where some of the $c_a$'s have been renamed for convenience.
We can write the polynomial (\ref{eq:CY3}) in the form of the polynomial defining the $K3$ fiber (\ref{eq:polynomialSU9SU9}), but now with the coefficients $c_a$ replaced by homogenous polynomials of 
 order two in $z=(z_{7},\ldots z_{18})$  parametrizing the $\mathbb C\mathbb P^1$ base of the Calabi-Yau threefold.  Working in the patch $z_9=\ldots = z_{18}=1$, and setting $z=(z_7,z_8)$ we can write again $\xi(z)$ and $\eta(z)$ in the form (\ref{eq:etaxiz2}) with $f_2(z)$ a generic homogeneous polynomial of order 2 and
\beq
 f_4 (z)  \sim   z_7^2 \, z_8^2 \qquad\qquad  f_6 (z)  \sim    z_7^3 \, z_8^3~.
\eeq
Thus for this choice of auxiliary Calabi-Yau three-fold, the branes found in the previous example collide into two groups each of charge: $(3,9)=3\times (1,1)+2\times (0,3)$ located $z_7=0$ and $z_8=0$.

\subsection{The $\eta\to0$ limit}\label{sec:Ftheorylimit}
In order to gain a better understanding of the flux-geometry dictionary, it is instructive to consider the choice $\eta(z) = 0$, that is the situation in which the $\mathbb C\mathbb P^1$ base is embedded along the $\eta=0$ curve. Looking at the Calabi-Yau threefolds discussed in the previous section, this limit corresponds to setting $f_4(z) = c_4(z)\,c_5(z)$ to $0$. 

Along this curve, $j_2\rightarrow\infty$, i.e. $\tau^{(2)} \rightarrow \ii\,\infty$. If we make the identification $\sigma=\tau^{(1)} $ and $\tau=\tau^{(2)}$
for solutions in the A-class, where  $\tau= C_0 + \ii\,e^{-\phi}$,  this limit corresponds to the weak coupling limit with 
$\sigma=C_4+ \ii \, {\rm Vol} (T^4)$ varying over the plane. Alternatively, the same geometry can be used to describe the large volume limit of a varying axio-dilaton dual solution obtained from the identification $\tau=\tau^{(1)} $ and $\sigma=\tau^{(2)}$.
  
We first notice that in the $\eta\to0$ limits, the sums over $k$ defining  $\vp_{00}$ and $\vp_{10}$ can be explicitly performed leading to 
 \beq
 \varpi_{00} 
~  = ~ {}_2 F_1\left(\smallfrac13,\smallfrac23,1,\xi\right)    \qquad    \vp_{10} 
~  = ~ \frac{\ii}{\sqrt{3}}\,{}_2F_1\left(\smallfrac13,\smallfrac23;1;1-\x\right)
 \eeq
%that satisfy the Picard Fuch equation with operator  
%\beq
%\cL_2 ~  
%~  = ~
% 9\,\th_\x^2    -\x\, (3\th_\x +2)(3\th_\x +1)
% ~.
%\eeq
Thus we have
\beq\label{j1eta0}
\tau^{(1)} ~  = ~ \frac{\vp_{10}}{\vp_{00}}
~  = ~\frac{\ii}{\sqrt{3}}\,
\frac{{}_2F_1\left(\smallfrac13,\smallfrac23;1;1-\x\right)}{{}_2F_1\left(\smallfrac13,\smallfrac23;1;\x\right)}   
\eeq
and
 \beq\label{eq:j1LimitEta0}
j_1=j(\tau_1)= \frac{3^3(1+8\xi)^3}{\xi(1-\xi)^3}~.
\eeq 

Remarkably, Eq.~\eqref{j1eta0} provides an explicit expression for $\tau^{(1)}$. On the other hand, in the limit $\eta =0$, $j_1$ develops new poles of order $3$ at $\xi=1$. For the Calabi-Yau three folds discussed in the previous section, $\x(z)$ has degree $6$. As such, we expect to have $18$ branes located, in groups of $3$, at the zeros of $\xi(z)-1$, and $6$ other branes located at the zeros of $\xi(z)$.
\vspace{8pt}

\begin{figure}[t]
\begin{center}
{\hskip12pt
\begin{minipage}[t]{5.2in}
\vspace{10pt}
\raisebox{-1in}{\includegraphics[width=8.cm]{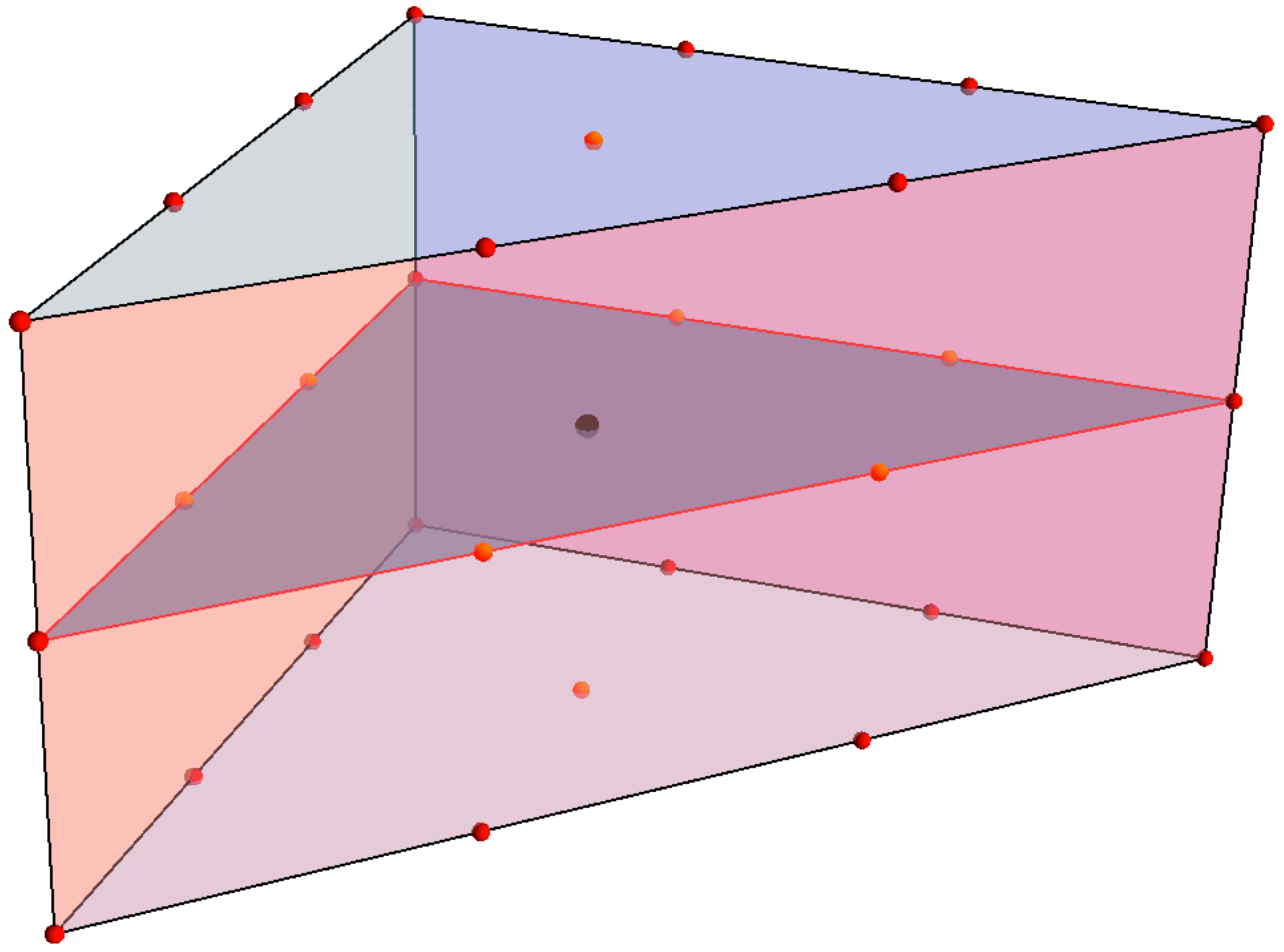}} 
\hfill \hspace*{21pt}
\raisebox{-.43in}{\includegraphics[width=3.4cm]{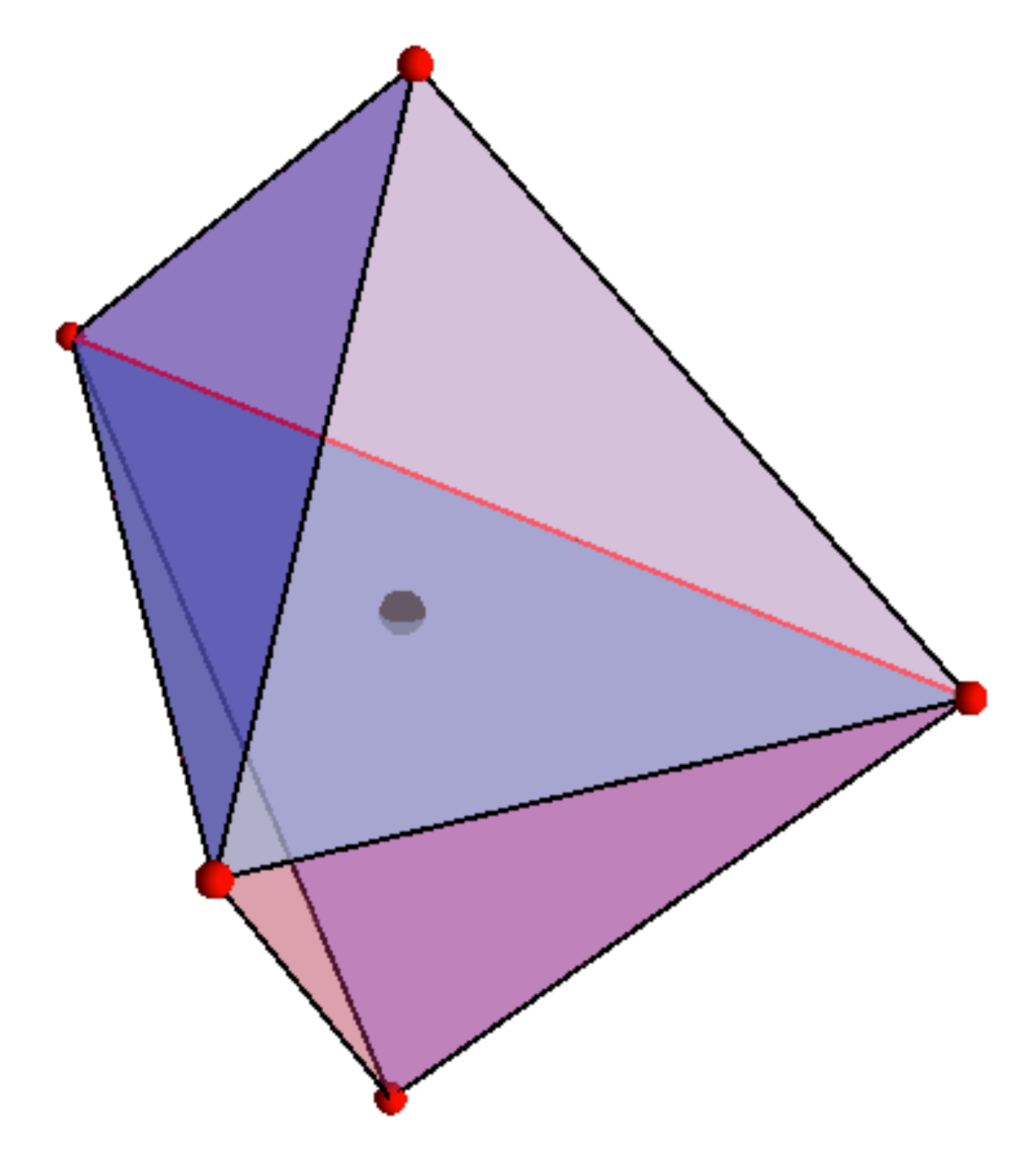}}
\hspace{0pt}
\vspace{2pt}
\end{minipage}}
\capt{5.7in}{fig:NablaDeltaCase3}{The left hand-side polyhedron corresponds to $\nabla$ and defines the ambient toric variety. All the points are included this time. The polygon passing through the origin corresponds to an elliptic curve and shows that the $K3$ surface admits an elliptic fibration structure. The points of~$\nabla$ lying on edges and above the polygon form the extended Dynkin diagram of $SU(9)$. Similarly, the points below the polygon form and $SU(9)$ diagram. As such, this fibration structure is of type $SU(9)\times SU(9)$. Other elliptic fibration structures for this $K3$ are listed in Appendix~\ref{app:ellipticfibrations}. The right hand-side polyhedron corresponds to the Newton polyhedron $\Delta$.}
\end{center}
\vskip -16pt
\end{figure}

Moreover one can identify $j_1$ with the $j$-invariant for an elliptic curve `contained' inside the $K3$ fiber. To this end, consider the pair of polyhedra $(\nabla,\Delta)$ shown in Figure~\ref{fig:NablaDeltaCase3} and associate the homogeneous coordinates $z_1,\ldots,z_6$ to the vertices of $\nabla$ as in Section~\ref{sec:K3fiber}. In addition, associate three further coordinates with the vertices of the polygon that divides $\nabla$ into two halves:
\beq
z_a: (0,-1,-1)\qquad z_b: (0,-1,~2) \qquad z_c: (0,~2,-1)
\eeq

As shown in Figure~\ref{fig:NablaDeltaCase3}, the $K3$ surface in question admits an elliptic fibration structure. In fact, it admits $5$ different fibration structures, as discussed in Appendix~\ref{app:ellipticfibrations}. With the above definitions, the defining polynomial, which now makes manifest this elliptic-fibration structure, becomes: 
\beq\label{eq:polynomialSU9SU9withfiber}
\begin{aligned}
f _{\rm hom} = &~~~{c}_{1} \, {z}_{1}^{3} \, {z}_{2}^{3}\,z_a^3+{c}_{2} \, {z}_{3}^{3} \, {z}_{4}^{3}\,z_b^3+{c}_{3} \, {z}_{5}^{3} \, {z}_{6}^{3}\,z_c^3  
+   \left( {-}{c}_{0} \, {z}_{1} \, {z}_{2} \, {z}_{3} \, {z}_{4} \, {z}_{5} \, {z}_{6} {+}{c}_{4} \, {z}_{2}^{2} \, {z}_{4}^{2} \, {z}_{6}^{2} {+}{c}_{5} \, {z}_{1}^{2} \, {z}_{3}^{2} \, {z}_{5}^{2}\right) \, z_a\,z_b\,z_c \\[2pt]
= &~~~\, a(w)\,z_a^3+b(w)\,z_b^3+c(w)\,z_c^3+d(w)\,z_a\,z_b\,z_c 
\end{aligned}
\eeq
where $w=(z_1,\ldots,z_6)$. The polynomial \eqref{eq:polynomialSU9SU9withfiber} defines the $K3$ surface studied in the previous sections as an elliptic fibration over $\mathbb C\mathbb P^1$. Now we would like to consider the $\eta=0$ direction in the moduli space of this $K3$ surface, which corresponds to setting $c_4=c_5=0$. Along this curve, we have 
\beq\label{eq:xiEllipticFibration}
\frac{\xi}{27}  = -\frac{a(w)\,b(w)\,c(w)}{d(w)^3}= \frac{c_1\,c_2\,c_3}{c_0^3}~.
\eeq

Note that the $w$ dependence drops out from the above expression and, for fixed $c_0,c_1,c_2,c_3$, the parameter $\xi$ is constant along the $\mathbb C\mathbb P^1$ base of the elliptic fibration.
With these identifications, we can proceed to finding the $j$-invariant of the elliptic curve defined by
\beq\label{eq:EllipticCurve}
a\,z_a^3+b\,z_b^3+c\,z_c^3+d\,z_a\,z_b\,z_c=0~.
\eeq
The polynomial~\eqref{eq:EllipticCurve} can be put in the Weierstrass form\footnote{The expressions for $f$ and $g$ were found using Sage ({\tt http://www.sagemath.org}), based on the methods in~\cite{Braun:2011ux}.}
$ z_a^2\,z_c~=~ z_b^3+z_b\,f\,z_c^2+g\,z_c^3$ with
\beq
\begin{aligned}
f &= -\frac{d^4}{48}\left( 8\, \xi + 1 \right)\qquad \qquad
g = \frac{d^6}{32\cdot 3^3 } \left( -8 \,\xi^2 -20\,\xi + 1  \right)~.
\end{aligned}
\eeq
Then the discriminant locus is given by
\beq\label{eq:DiscriminantEC}
\Delta =4\, f^3+27 \, g^2 = \frac{4\, d^{12} }{1728} \, \xi \left(1-\xi   \right)^3
\eeq
 and, hence
\beq
j ~=~ 1728\cdot\frac{4\,f^3}{\Delta} =  \frac{3^3(1+8\xi)^3}{\xi(1-\xi)^3}
\eeq
exactly matching the expression~\eqref{eq:j1LimitEta0}.  Working, e.g.~in the patch $z_2=\ldots = z_6 =0$, we see that $\Delta$ has degree $24$ in $z_1$ and $z_2$, hence we expect $24$ points at which $j$ degenerates.
\vspace{12pt}

\begin{figure}[h]
\begin{center}
{\hskip10pt
\begin{minipage}[t]{5.2in}
\vspace{4pt}
\hspace{-33pt}
\raisebox{-0.4in}{\includegraphics[width=7.cm]{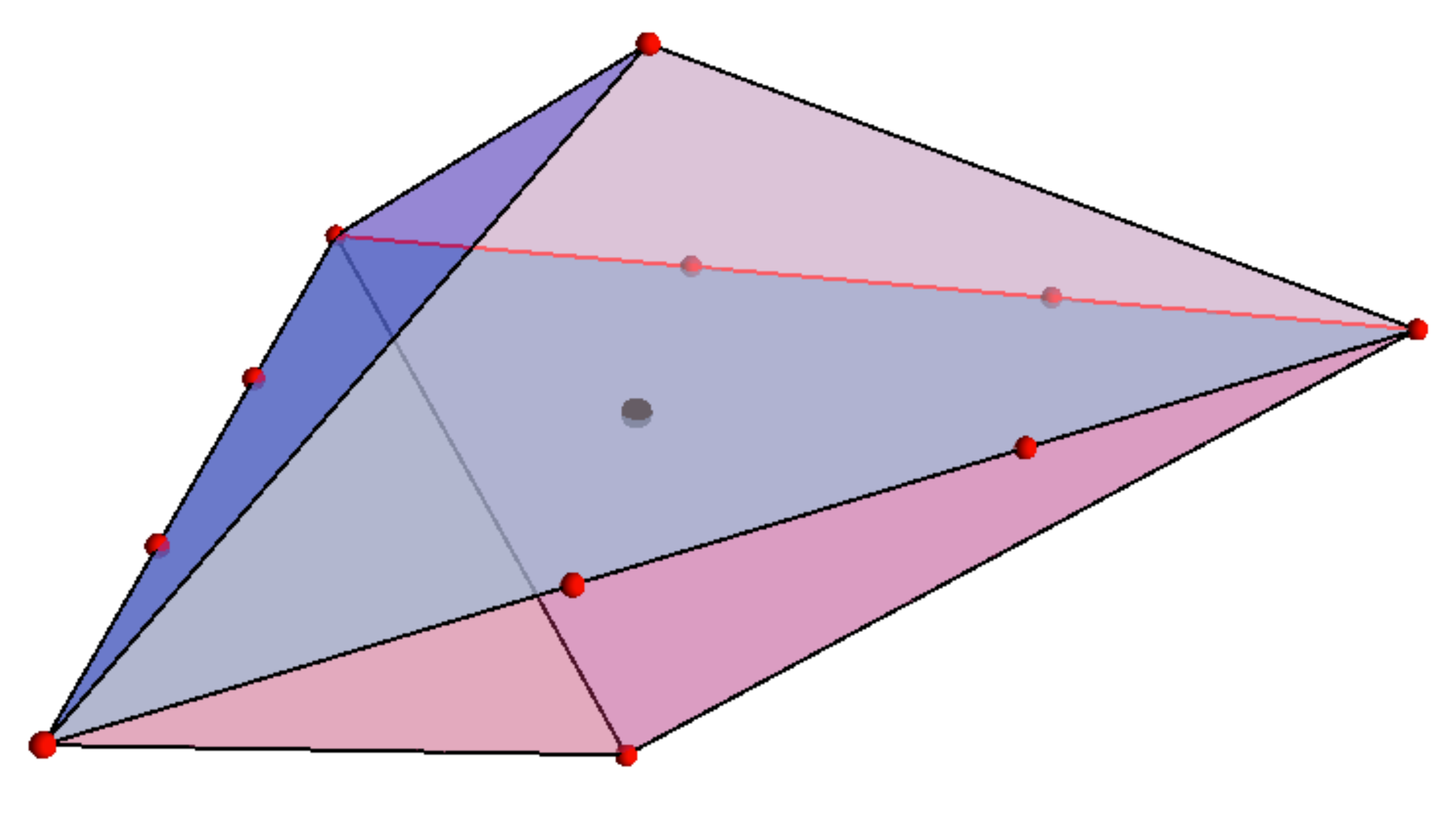}} 
\hfill \hspace*{20pt}
\raisebox{-1in}{\includegraphics[width=7.3cm]{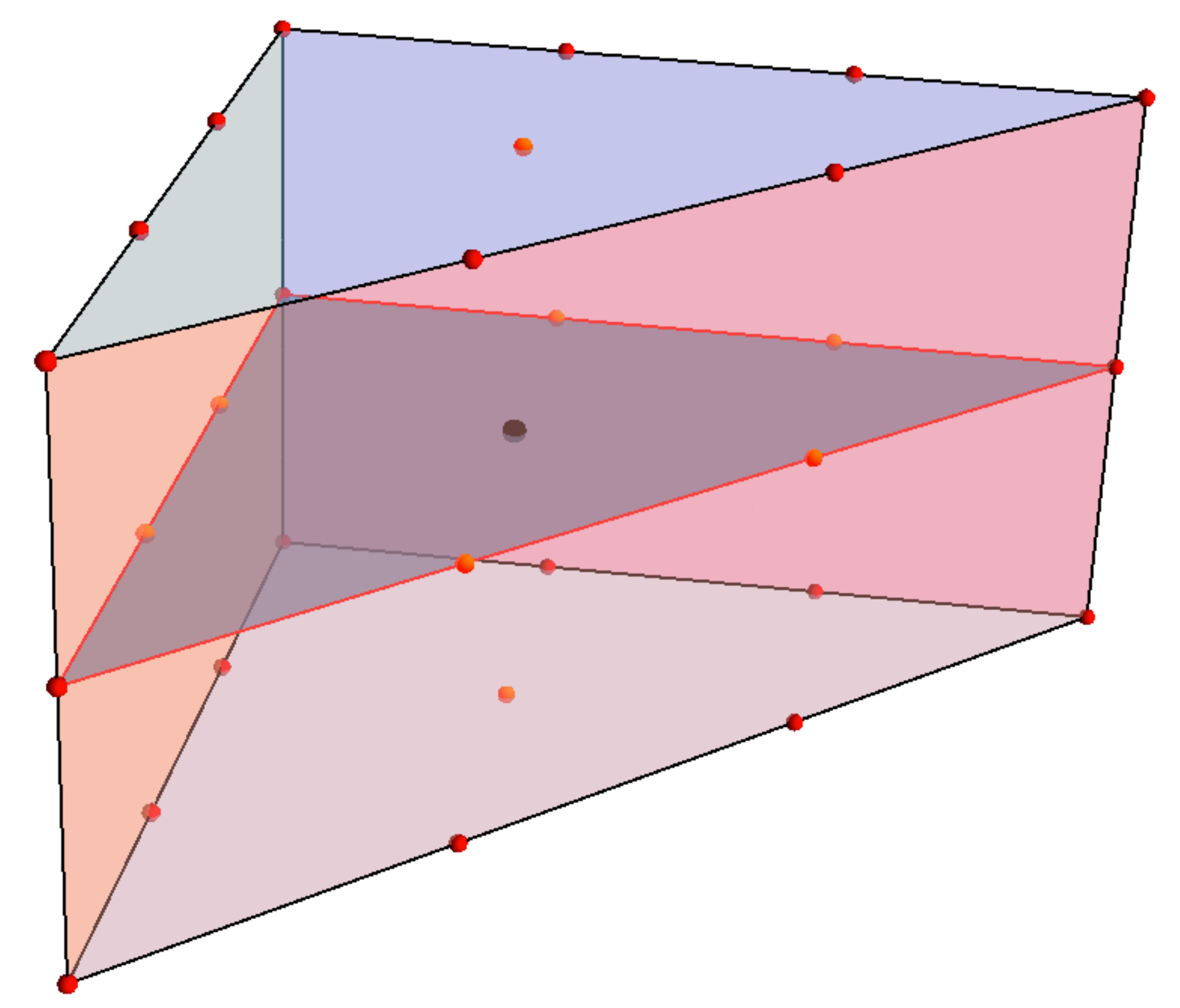}}
\hspace{10pt}
\vspace{-2pt}
\end{minipage}}
\capt{5.7in}{fig:NablaDeltaFibration1}{The polyhedrons $\nabla$ associated  the $K3$ surfaces $\tilde S$ for the two choices of Calabi-Yau three-fold embeddings in sections \ref{sec:firstCY3} (left) and \ref{sec:secondCY3} (right).}
\end{center}
\end{figure}

Further, we can fiber the above $K3$ surface over $\mathbb C\mathbb P^1$ as done in Sections~\ref{sec:firstCY3} and \ref{sec:secondCY3}, obtaining polynomials of the form
\vspace{-8pt}
\beq
f _{\text{CY}_3, \,{\rm hom}}~ =~  a(w,z)\,z_a^3+b(w,z)\,z_b^3+c(w,z)\,z_c^3+d(w,z)\,z_a\,z_b\,z_c 
\eeq
where $z$ parametrizes the $\mathbb C\mathbb P^1$ base over which the $K3$ surface is fibered. This gives an elliptically fibered Calabi-Yau threefold. However, if we drop the $w$ dependence, we obtain a second $K3$ surface, call it~$\tilde S$, that is elliptically fibered with the same fiber as the original $K3$, which we call $S$. 
%Put differently, we look for Calabi-Yau threefolds that admit two different $K3$ fibrations, one with fiber $S$ over the $\mathbb C\mathbb P^1$ parametrized by $z$ and another one with fiber $\tilde S$ over the $\mathbb C\mathbb P^1$ parametrized by $w$.

In Figure~\ref{fig:NablaDeltaFibration1} we have depicted the polyhedra $\nabla$ that defines $\tilde S$ for the Calabi-Yau threefolds considered in Section~sections \ref{sec:firstCY3} and \ref{sec:secondCY3}  respectively.  The two polyhedra are obtained by projecting  $\nabla_{\text{CY}_3}$ given by
(\ref{vertone}) and (\ref{vertsecond})  respectively,
on the first, the third and the fourth coordinates\footnote{$S$ is obtained by projecting $\nabla_{\text{CY}_3}$ on the last three coordinates.}.  
The distribution of branes is very different in the two cases. Indeed, in the first case, the distribution of points above and below the polygon associated to the elliptic fiber gives two Dynkin diagrams that correspond to the trivial~group. This is consistent with the fact that $\xi$ in this case has zeros
 at generic points (i.e.~they do not correspond to the North or the South pole) so no symmetry enhancement is expected.  
  This picture is modified when we look at the Calabi-Yau three-fold considered in Section~\ref{sec:secondCY3}.  The points above and below the polygon are now  associated to two $SU(9)$ extended Dynkin diagrams. This  is consistent with the expected gauge symmetry enhancement in the world volumes of colliding branes. The same results can be found from a careful analysis of the corresponding discriminants  for the two choices of Calabi-Yau threefolds.

\section{ Other examples of auxiliary CY threefolds }\label{sec:OtherExamples}

In Section~\ref{sec:example} we presented a detailed analysis of the complex structure moduli space for a particular $K3$ surface with two complex structure parameters. A similar analysis can be performed for any of the $9$ $K3$ surfaces from the Kreuzer-Skarke list that have Picard number $18$ and two complex structure parameters. These surfaces are defined in terms of polytopes which we present in Appendix~\ref{app:Pic18List}. 

In this section we summarise the results for two other $K3$ surfaces and work out the geometry to flux dictionary derived from fibering these $K3$ surfaces over $S^2$.

\subsection{The second $K3$ surface}
\subsubsection{The moduli space}

The defining pair of dual polyhedra $(\nabla,\Delta)$ for this $K3$ surface is given in  (\ref{eq:verticesSecondK3})  and (\ref{dualcase2}). These polyhedra lead to the following polynomial:
 \beq
f_{\rm hom} =- c_0\,z_1\, z_2\, z_3\, z_4 + c_1\,z_1^{12} + c_2\,z_1^6\, z_2^6 + c_3\,z_2^{12} + c_4\,z_3^3 + c_5\,z_4^2 =0  \label{fe8}
  \eeq
This equation defines a $K3$ hypersurface in $\mathbb C\mathbb P_{[1,1,4,6]}$.

Including for later convenience numerical factors, the relevant combinations of coefficients which appear in the expansion for the fundamental period are
\beq
%\frac{\xi}{3\cdot 12^3} ~=~ \frac{c_1\,c_3\,c_4^4\,c_5^6}{c_0^{12}}\qquad \text{and}\qquad \frac{\eta}{3\cdot 12^2}~=~\frac{c_2\,c_4^2\,c_5^3}{c_0^{6}}~.
\x ~=~ \frac{c_1\,c_3\,c_4^4\,c_5^6}{c_0^{12}}\qquad \text{and}\qquad \eta~=~\frac{c_2\,c_4^2\,c_5^3}{c_0^{6}}~.
\eeq

The Frobenius period is given by
\beq\label{eq:w00seriesSecondK3}
\varpi~=~
    \sum_{k,l=0}^{\infty} \frac{a_{k+\a,l+\b}}{a_{\a,\b}}~ \x^{k+\a}\, \eta^{l+\b} ~,
\eeq
with
\beq
\begin{aligned}
a_{k,l} &~=~ \frac{\Gamma\left(12k+6l+1\right)}{\Gamma^2(k+1) \,\Gamma(l+1)\,\Gamma\left(4k+2l+1\right)\Gamma\left(6k+3l+1\right)}\\[8pt]
&~=~432^{2k+l} \frac{ \G( 2k + l + \frac{1}{6} ) \G( 2k + l  + \frac{5}{6} ) }{  \G^2( k + 1 ) \G( l  + 1 ) 
\G( 2k + l + 1 ) }
\end{aligned}
\eeq
where the last equality has been obtained by using the multiplication formula for the $\G$-function: 
\beq
\G( 1 + m z ) ~=~ (2\p)^{-\frac{1}{2} (m-1)} m^{-\frac{1}{2} + m z} \prod_{r=1}^m \G \left( z + \frac{r}{m} \right)
\eeq
from which it follows immediately that
\beq
\frac{ \G\big( 1 + m (z+\g) \big) }{ \G( 1+ m \g ) } ~=~ 
m^{m z} \prod_{r=1}^m \frac{ \G\big( z + \g + \frac{r}{m} \big) }{ \G\big( \g + \frac{r}{m} \big) }
\eeq

We find that the Picard-Fuchs equations are generated by the second order operators
\beq\begin{split}
\cL_1~&=~\theta_{\eta}\left(2\,\theta_{\xi} + \theta_{\eta}\right) - 432\, \eta \,\left( 2\,\theta_{\xi}+\theta_{\eta} +\smallfrac{5}{6}\right)\,\left(2\,\theta_{\xi}+\theta_{\eta} +\smallfrac{1}{6}\right)~,  \\[3pt]
\cL_2~&=~\eta^2\, \theta_{\xi}^ 2- \xi\, \theta_{\eta}\left( \theta_{\eta}-1\right)
\end{split}
\eeq

The form of the second of these operators is non-standard. Applying Frobenius' method, we find three (instead of two) indicial equations, namely 
$\a (\a-2\b) = 0$, $\b^2 = 0$ and  $\a(\a -1) = 0$.
This is too restrictive. Indeed, the expansion of the Frobenius period in terms of $\a$ and $\b$ has, in this case, only two terms corresponding to $1$ and $\b$. In order to rectify this situation, we perform the change of variables
\beq
u=\frac{\xi}{\eta^2}~ \qquad \qquad v = \eta~.
\eeq

\begin{figure}[h]
\begin{center}
\hskip80pt
\begin{minipage}[t]{3.2in}
%\vspace{10pt}
\raisebox{-1in}{\includegraphics[width=5.cm]{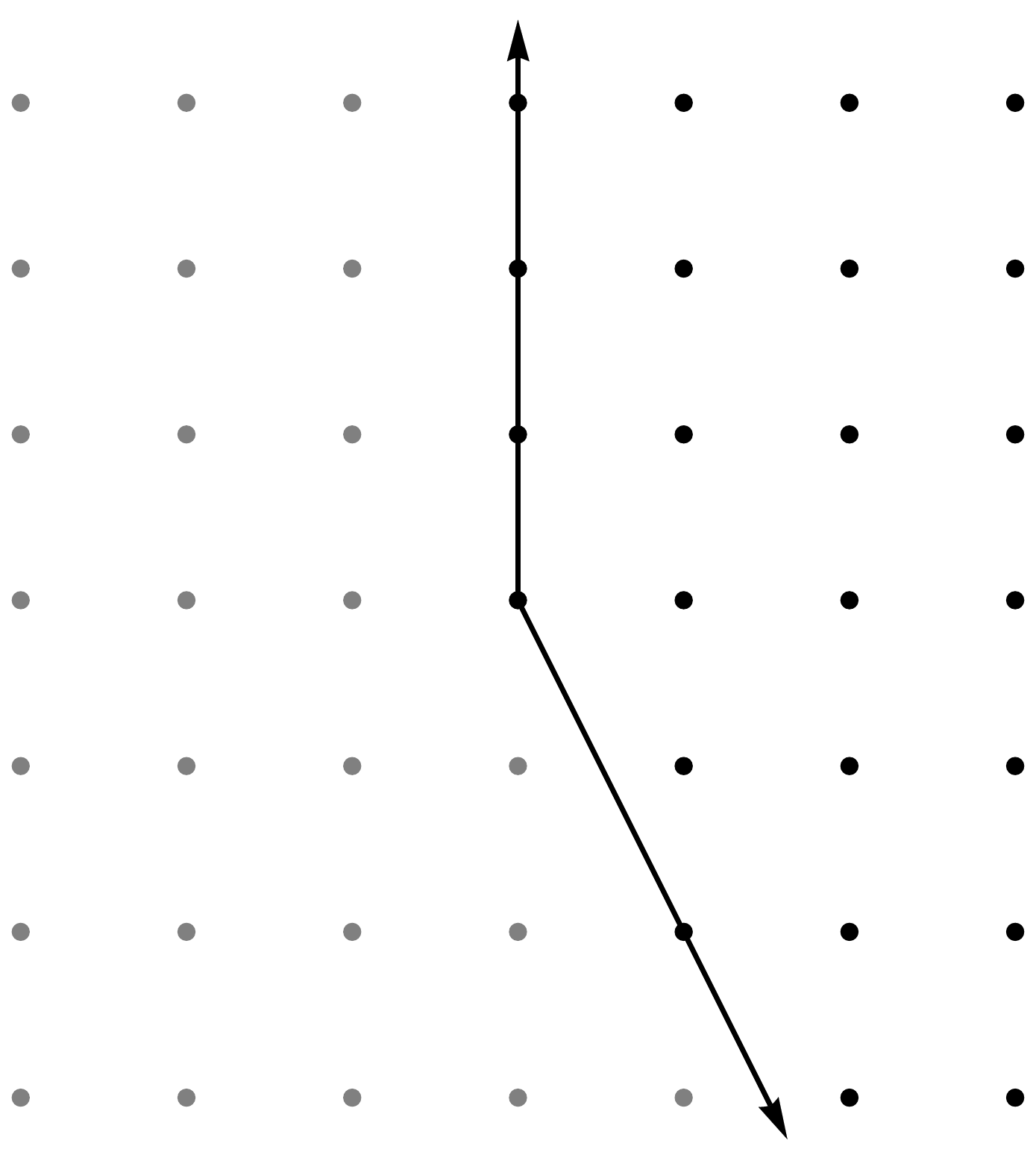}} 
\vspace{-0pt}
\end{minipage}
\capt{6in}{fig:etaxicone}{ The cone of summation in $(k,l)$-space, corresponding to the first quadrant for the integers $(m,n)$. Note that $\vp_{0,0}$ itself restricts to the subcone $k\geq 0,~l\geq 0$ owing to the factor of $\G(n-2m+1)$ in the denominator of the coefficients $b_{m,n}$ (see \eqref{eq:b_mn}). For the other periods, however, the summation corresponds to the points in the larger cone.}
\end{center}
\end{figure}

With this change of variables, the Frobenius period takes the form: 
\beq
\varpi_{00}(u,v) ~=~ \sum_{m,n=0}^{\infty} \frac{b_{m+\e,n+\delta}}{b_{\e,\delta}}   ~   u^{m+\e}\, v^{n+\delta} \; ,
\label{fundperiod2}
\eeq
where
\beq\label{eq:b_mn}
b_{m,n}=\frac{\Gamma(6n+1)}{\Gamma^2(m+1) \,\Gamma(n-2\,m+1)\,\Gamma(2n+1)\,\Gamma(3n+1)}~.
\eeq

\vspace{8pt}
A basis for the Picard-Fuchs operators is given by
\beq\begin{split}
\cL_1~&=~\left(\theta_v - 2\,\theta_u\right) \theta_v - 432\,v\,\left(\theta_v+\smallfrac{1}{6}\right)\,\left(\theta_v+\smallfrac{5}{6}\right)~,  \\[3pt]
\cL_2~&=~ \theta_u^2 - u\,\left(\theta_v-2\,\theta_u\right)(\theta_v-2\,\theta_u-1)
\end{split}
\eeq

%The locus where the $K3$ is singular corresponds to the vanishing of: 
%\beq
%D = 3\cdot 12^5 \xi - (1-3\cdot 12^2 \eta)^2~.
%\eeq

Applying Frobenius' method to the above Picard-Fuchs operators, 
%with $m$ shifted by $\epsilon$ and $n$ shifted by $\delta$, 
we find this time two indicial equations $\epsilon (\epsilon-2\delta) = 0$ and $\delta^2 = 0$, which lead to the following basis of periods 
 \beq\begin{split}
\vp_0&=\vp_{00}=h_{00}\\[4pt]
\vp_1&=\vp_{10}+\vp_{01}=\frac{1}{2\p\ii}\,h_{00} \log\left(uv\right) + h_{10}+ h_{01}\\[4pt]
\vp_2&=\vp_{01}=\frac{1}{2\p\ii}\,h_{00}\log v + h_{01}\\[4pt]
\vp_3&=\vp_{11}+\vp_{02}=\frac{1}{(2\p\ii)^2}\,h_{00}\log\left(uv\right)\log v + h_{11}+ h_{02}
\end{split}\label{PeriodBasis_Case5}\eeq
and $ \vp_{0}\, \vp_{3}~=~\vp_{1}\, \vp_{2}$. 
%It can be checked by multiplication of the series involved, that
%\beq
%\vp_{11}~=~\frac{\vp_{10} \vp_{01}}{\vp_{00}}~~~\text{as}~~~h_{10}\, h_{01}~=~h_{00}\,h_{11}~.
%\eeq
We identify the field combinations $\sigma$ and $\tau$ with the period ratios
\beq
\begin{aligned}
\t^{(1)}~&=~\frac{\vp_{10}}{\vp_{00}} ~=~ {1\over 2\pi i} \log\left(u v\right)  ~+~ \frac{h_{10}+h_{01}}{h_{00}} \\[4pt]
\t^{(2)}~&=~\frac{\vp_{01}}{\vp_{00}} ~=~ {1\over 2\pi i} \log v ~ +~ \frac{h_{01}}{h_{00}}
\end{aligned}
\label{tausigma2_E8E8}
\eeq

\begin{figure}[H]
\begin{center}
\hskip20pt
\begin{minipage}[t]{3.2in}
\vspace{10pt}
\raisebox{-1in}{\includegraphics[width=7.cm]{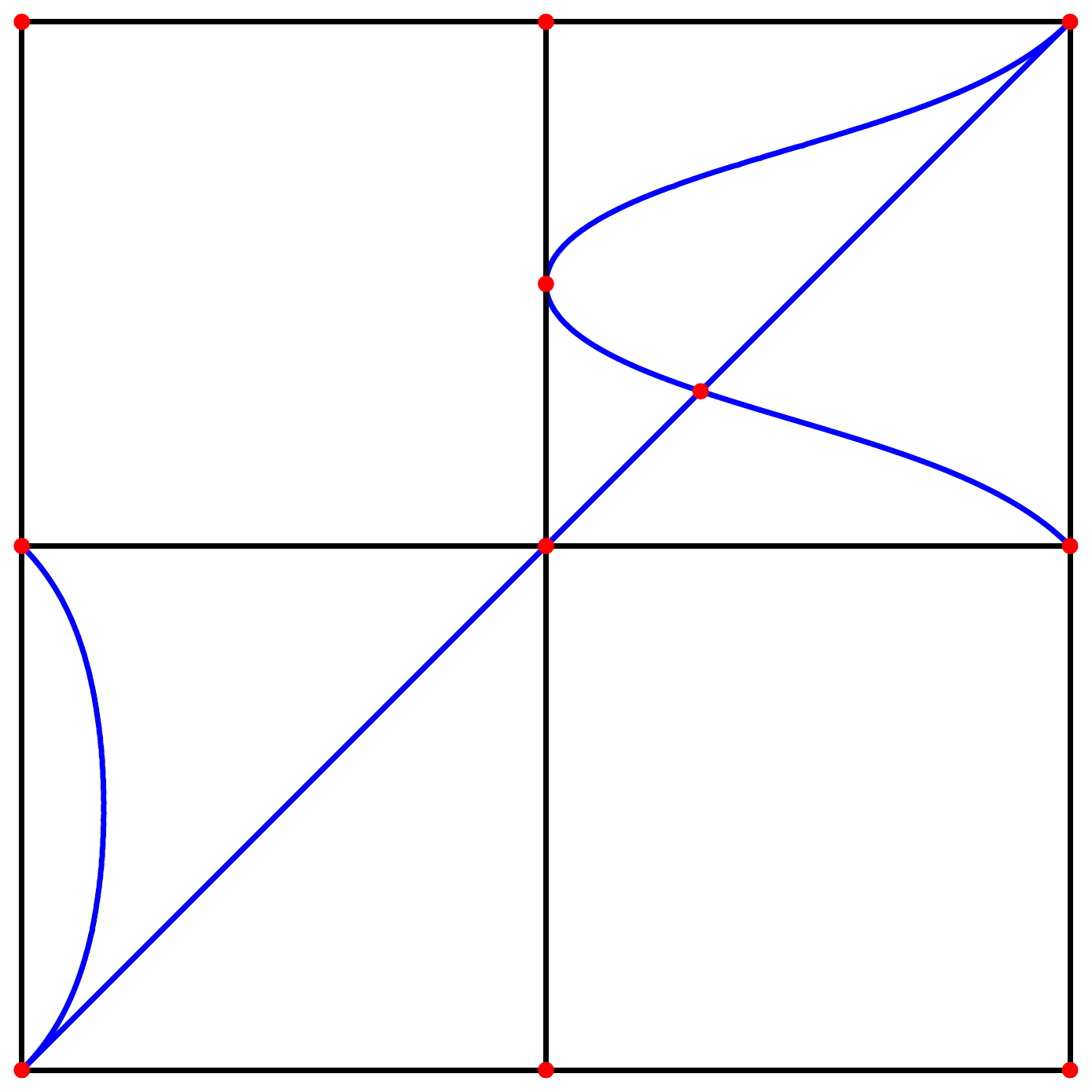}} 
\vspace{-0pt}
\end{minipage}
\capt{6in}{fig:discr_case1}{A sketch of the $(w,v)$-plane, where $w=uv$. The blue curves correspond to $D=0$ and $1-4u=0$. The plane has been compactified to $\mathbb C\mathbb P^1{\times}\mathbb C\mathbb P^1$ so opposite boundaries on the plot, where $w$ and $v$ are infinite, should be identified.}
\end{center}
\end{figure}

The associated $j$-functions satisfy
\beq
\begin{aligned}
j_1+j_2 ~&= ~{ 1 - 3\cdot 12^2 \, v +  12^3 \, uv   \over u\, v} \\[4pt]
j_1 \, j_2 ~&=~{  1 \over u\,  v^2  }
\end{aligned}
\eeq
which, in turn, lead to:
 \beq
 j_{1,2}=\frac{ 1 - 3\cdot 12^2 \, v + 12^3\, v\,u   \pm \sqrt{ (1-4\,u)\,D    }}{2 u\,v }  ~,
 \label{j1j2exact2}
 \vspace{4pt}
 \eeq
 where the discriminant locus, sketched in Figure~\ref{fig:discr_case1}, is given by
 \beq
 D= \left(1- 3\cdot 12^2 v\right)^2 - 3\cdot 12^5 u\, v^2~.
 \eeq

 In Table~\ref{XiEtaLimitsCase1} we collect the expressions that the $j$-functions take along the curves $u=0$ and $v=0$. These limits are important for inferring the brane content of the supergravity solutions discussed below. 
 
 %\vspace{8pt}
\begin{table}[H]
\begin{center}
\begin{tabular}{|r|ccc|}
\hline
\vrule height12pt depth8pt width0pt & $v~=~0$ & \hspace*{10pt} & $u~=~0$ \\
\hline\hline
\vrule height23pt depth15pt width0pt $~u\,v\,j_1$~ 
                        & ~$\displaystyle \frac{1+\sqrt{1-4\,u}}{2} $~ 
                           & & ~$\displaystyle 1- 3\cdot 12^2\,v$~ \\
\vrule height20pt depth13pt width0pt $~v\,j_2$\,~ 
                        & ~$\displaystyle \frac{1-\sqrt{1-4\,u}}{2\,u}$~
                           & & ~  $\displaystyle\frac{1}{ \left(1 - 3\cdot 12^2\,v \right)}$~ \\&&&\\[-14pt]
\hline                           
\end{tabular}
\end{center}
\vspace{-14pt}
\caption{The form of $j_1$ and $j_2$ along the curves $u=0$ and $v=0$.}
\label{XiEtaLimitsCase1}
\end{table}

It is interesting that the last entry of the table corresponds to the following relation:
if $\t$ and $z$ are related by
\beq
\t ~=~ \ii \frac{ \,_{2}F_1(\smallfrac16, \smallfrac56; 1; 1-z) }{ \,_{2}F_1(\smallfrac16, \smallfrac56; 1; z) }~,
\eeq
then
\beq
j(\t) ~=~ \frac{ 432 }{ z(1 - z) }~.
\eeq

%Remarkably, the last entry of the table corresponds to 
%\beq
%\t^{(2)}(0,v) ~=~\ii\, \frac{  \,_{2}F_1(\smallfrac{1}{6}; \smallfrac{5}{6}; 1; 1 - 3\cdot 12^2\,v ) }{  \,_{2}F_1(\smallfrac{1}{6}; \smallfrac{5}{6}; 1; 3\cdot 12^2\,v ) }~.
%\eeq
 
\vspace{12pt} 
\subsubsection{Geometry to flux dictionary}
 
Consider the four-dimensional polytope $\nabla_{\text{CY}_3}$ obtained from the polyhedron \eqref{eq:verticesSecondK3} by adding two points in the fourth dimension, having vertices
\beq
w_i^{\rm CY_3}=(0,~w^{K3}_i)~,   \qquad       w^{\rm CY_3}_5=(-1,0,0,0) ~, \qquad   w^{\rm CY_3}_6=(1,0,0,0)~,   \qquad  i=1,\ldots 4
\eeq
 
The dual polytope contains $18$ points. The defining polynomial is given by (\ref{fe8}), with the coefficients replaced by homogenous polynomials of degree two in $z=(z_5,z_6)$ parametrizing the $\mathbb{CP}^1$~base:
\begin{equation}\label{eq:CY3_Case5}
\begin{aligned}
f _{\text{CY}_3, \,{\rm hom}} 
 ~=&~  - c_0(z) \,z_1\, z_2\, z_3\, z_4 + c_1(z) \,z_1^{12} + c_2(z)\,z_1^6\, z_2^6 + c_3(z)\,z_2^{12} + c_4(z)\,z_3^3 + c_5(z)\,z_4^2 
\end{aligned}
\end{equation}   
 
The period ratios are given by the above formulas with
\beq
u (z)~=~ {c_{1}(z)\, c_{3}(z) \over c_{2}(z)^{2} }~,  \qquad ~~~~~~~~ v (z)~= ~{ c_2(z)\, c_4(z)^2 c_{5}(z)^3 \over c_{0}(z)^6}~.
\label{xizcasesecond}
\eeq

As in Section~\ref{sec:dictionary}, we can find the locations of branes by looking for the poles of $j_1(z)$ and $j_2(z)$, which lie along the curves $u=0$ and $v=0$.
%, or by using the formulas (\ref{tausigma2_E8E8}) for the period ratios and extracting the monodromies around the points where $u(z)=0$ and, separately, $v(z)=0$. 
%To find the location of branes, we look for poles of $j_i$. First we observe that $j_i$ are finite  in the limit where $c_0,c_2$ are small.
%The poles of $j_i$  are located around the remaining  zeros of $u=0$ or $v=0$ so we can used the simpler formulas (\ref{tausigma2}) to read their monodromies at those points.  
We find that both $j_1$ and $j_2$ degenerate at the zeros of $c_4$ and $c_5$ with poles of orders $2$ and $3$, respectively.  Moreover, $j_1$  has single poles at the zeros of $c_1$ and $c_3$. The resulting brane content is summarised in the  table \ref{table:branes2}. Taking into account that each $c_a(z)$ has two zeros, one finds $4\times (0,1)+2\times (3,3)+2 (2,2)$ for a total number of 24 branes, as required.   
 
 \begin{table}[H]
 \begin{center}
\begin{tabular}{c| c c c c}
& ~~~~$   c_0,c_2 $~~~~ & ~~~~$ c_1, c_3 $~~~~ &    ~~~~$ c_4 $~~~~  & ~~~~$ c_5 $~~~~      \\[2pt]
\hline\\[-14pt]
   ($q_1$, $q_2$)~~ & $(0,0)$ & $(1,0) $ & $(3,3)$ & $(2,2)$
\end{tabular}
\vspace{-8pt}
\capt{6.1in}{table:branes2}{Branes of charge $(q_1,q_2)$ are located at the zeros of $c_a$'s.  }
\end{center}
\end{table}

\subsection{The third $K3$}
\subsubsection{The moduli space}

The pair of polyhedra $(\nabla,\Delta)$ defining this $K3$ surface is given in (\ref{eq:nablaCase5}) and (\ref{dualcasethird}). 
These lead to the polynomial:
\beq
f = -c_0\,z_1\, z_2\, z_3\, z_4\, z_5+c_1\,z_1^3\, z_3^3 + c_2\,z_1^3\, z_4^3+c_3\,z_2^3\, z_3^3 + c_4\,z_2^3\, z_4^3 + c_5\,z_5^3   \label{eqdef3}
\eeq
defining a $K3$ hypersurface in the toric variety given by the weight system in Table~\ref{tab:relations_Case5}.

\begin{table}[h]
\begin{center}
\begin{tabular}{l|lllllll}
& $z_1$ & $z_2$& $z_3$& $z_4$& $z_5$&  \\
\hline
$Q_1$ & 1 & 1 & 0& 0 & 1 \\
$Q_2$ & 0 & 0&1 & 1 & 1
\end{tabular}
\end{center}
\caption{Weight system for the toric variety embedding the third $K3$ surface.}
\label{tab:relations_Case5}
\end{table}

The relevant invariant combinations of coefficients for this case are
\beq
\frac{\xi}{27} ~=~ \frac{c_1\,c_4\,c_5}{c_0^{3}}\qquad \text{and}\qquad \frac{\eta}{27}~=~\frac{c_2\,c_3\,c_5}{c_0^{3}}~.
\eeq

The fundamental period is given by
\beq\label{eq:w00seriesSecondK3}
\varpi_{00}(\xi,\eta)~=~
    \sum_{k,l=0}^{\infty} \frac{ \Gamma(3k+3l+1) }{\Gamma(k+1)^2 \,\Gamma(l+1)^2\,\Gamma(k+l+1)} ~
    \Big(\frac{\xi}{27}\Big)^k \Big(\frac{\eta}{27}\Big)^l ~ = ~ \sum_{k,l=0}^{\infty} a_{k,l}\, \Big(\frac{\xi}{27}\Big)^k \Big(\frac{\eta}{27}\Big)^l ~.
\eeq

We find the following second order operators that generate the Picard-Fuchs equations
\beq\begin{split}
\cL_1~&=~9\,\theta_{\xi}^2 - \xi \left(3\,\theta_{\xi}+3\,\theta_{\eta}+1\right)\,\left(3\,\theta_{\xi}+3\,\theta_{\eta}+2\right)~,  \\[3pt]
\cL_2~&=~9\,\theta_{\eta}^2 - \eta \left(3\,\theta_{\xi}+3\,\theta_{\eta}+1\right)\,\left(3\,\theta_{\xi}+3\,\theta_{\eta}+2\right) ~.
\end{split}
\eeq 

%Applying Frobenius' method to the above Picard-Fuchs operators, we find two indicial equations, namely $\epsilon^2=0$ and $\delta^2=0$. With these relations, the Frobenius period can be expanded as
%\beq
%\vp~=~\sum_{k,l=0}^\infty a_{k,l}(\epsilon,\delta)\,\x^{k+\e} \eta^{l+\delta}~ = ~ \vp_{00} + 2\pi\ii\epsilon\, \vp_{10} + 2\pi \ii\delta\, \vp_{01} + \left( 2\pi\ii \right)^2 \epsilon\delta\, \vp_{11}
%\eeq
%where the quantities $\vp_{rs}$ are defined as in \eqref{eq:pi_rs}. As before, the coefficients $a_{k,l}(\e,\delta)$ are 
%defined such that $a_{0,0}(\e,\delta)=1$. Explicitly, 

Applying Frobenius' method to the above Picard-Fuchs operators, we find two indicial equations, namely $\epsilon^2=0$ and $\delta^2=0$
and therefore a basis of periods 
 \beq\begin{split}
\vp_{00}&=h_{00}\\[8pt]
\vp_{10}&=\frac{1}{2\p\ii}\,h_{00}\log\left(\frac{\x}{27}\right) + h_{10}\\[8pt]
\vp_{01}&=\frac{1}{2\p\ii}\,h_{00}\log\left(\frac{\eta}{27}\right) + h_{01}\\[8pt]
\vp_{11}&=\frac{1}{(2\p\ii)^2}\,h_{00}\log\left(\frac{\x}{27}\right)\log\left(\frac{\eta}{27}\right) + h_{11}\\[8pt]
\end{split}\label{PeriodBasis_Case5}\eeq
with $ \vp_{00} \vp_{11}~=~\vp_{10} \vp_{01}$. 
%It can be checked by multiplication of the series involved, that
%\beq
%\vp_{11}~=~\frac{\vp_{10} \vp_{01}}{\vp_{00}}~~~\text{as}~~~h_{10}\, h_{01}~=~h_{00}\,h_{11}~.
%\eeq
We identify the field combinations $\sigma$ and $\tau$ with the period ratios
\beq
\begin{aligned}
\t^{(1)}~&=~\frac{\vp_{10}}{\vp_{00}} ~=~ {1\over 2\pi i} \log \left(\frac{\x}{27}\right)  ~+~ \frac{h_{10}}{h_{00}} \\
\t^{(2)}~&=~\frac{\vp_{01}}{\vp_{00}} ~=~ {1\over 2\pi i} \log \left(\frac{\eta}{27}\right) ~ +~ \frac{h_{01}}{h_{00}}
\end{aligned}
\label{tau1tau2_Case5}
\eeq

The corresponding $j_1=j(\tau^{(1)})$ function takes the form:
\beq\label{eq:j1_Case5}
j_1 = \frac{27}{2 \,\eta ^3 \,\xi}\left[ 2 \eta ^3-486\, \eta  \xi  (-2 \eta -3\xi +3)+ Q \left(\sqrt{D}+\xi -\eta -1\right)  \right]
\eeq
where $Q=(\eta -9 \xi )^3-243 \xi  (-2 \eta -6 \xi +3)$ and $D$ is the discriminant: 
\beq
D = (\eta-\xi)^2 - 2\,(\eta+\xi)+1 ~.
\eeq

The expression for $j_2=j(\t^{(2)})$ is analogous, and can be obtained from \eqref{eq:j1_Case5} by interchanging $\xi$ and $\eta$.

 \vspace{8pt}
\begin{table}[H]
\begin{center}
\begin{tabular}{|r|ccc|}
\hline
\vrule height12pt depth8pt width0pt & $\xi~=~0$ & \hspace*{10pt} & \!\!$\eta~=~0$ \\
\hline\hline
\vrule height23pt depth15pt width0pt $~\x,j_1$~ 
                        & ~$\displaystyle 27\,(1-\eta) $~
                           & & ~$\displaystyle \frac{27\,(1+8\,\xi)^3}{(1-\x)^3}$~~~ \\
\vrule height20pt depth13pt width0pt $~\eta\,j_2$~ 
                        & ~$\displaystyle ~~\frac{27\,(1+8\,\eta)^3}{(1-\eta)^3}$~
                           & & \!\!$\displaystyle 27\,(1-\xi)$~ \\&&&\\[-14pt]
\hline                           
\end{tabular}
\end{center}
\vspace{-8pt}
\caption{The form of $j_1$ and $j_2$ along the curves $\xi=0$ and $\eta=0$.}
\label{XiEtaLimitsCase5}
\end{table}

In Table~\ref{XiEtaLimitsCase5} we have collected the expressions for $j_1$ and $j_2$ in the limits $\xi\rightarrow0$ and $\eta\rightarrow0$. Remarkably, the expressions for $j_1$ in the limit $\eta\rightarrow0$ and for $j_2$ in the limit $\xi\rightarrow 0$ correspond to the $SL(2,\mathbb Z)$-invariants of two different tori. Accordingly, $\tau^{(1)}(\x,0)$ and $\tau^{(2)}(0,\eta)$ are given by 
\beq\label{taulimit_Case5}
\tau^{(1)}(\x,0)~=~ \tau^{(2)}(0,\eta)~=~\frac{\ii}{\sqrt{3}}\,
\frac{{}_2F_1\left(\smallfrac13,\smallfrac23;1;1-\x\right)}{{}_2F_1\left(\smallfrac13,\smallfrac23;1;\x\right)} ~.
\eeq

\vspace{12pt}
\subsubsection{Geometry to flux dictionary}
Consider the four-dimensional polytope $\nabla_{\text{CY}_3}$ obtained from the polyhedron \eqref{eq:nablaCase5} by adding two points in the fourth dimension, above and respectively below the origin. The polytope thus constructed has vertices
\beq
w_i^{\rm CY_3}=(0,~w^{K3}_i)~,   \qquad       w^{\rm CY_3}_6=(-1,0,0,0)~, \qquad   w^{\rm CY_3}_7=(1,0,0,0)~,   \qquad  i=1,\ldots 5
\eeq
%
%
%\beq
%\begin{aligned}
%\{(\,0,-1, -1, ~2),~ (\,0,-1, ~2,& -1),~ (\,0,-1, ~2, ~2), ~(\,0,~2, -1, -1), ~(\,0,-1, -1, -1),\\~&(-1,~0,~0,~0),~(\,1,~0,~0,~0)\}
%\end{aligned}
%\eeq

\vspace{8pt}
The dual polytope contains $18$ points. The defining polynomial is given by (\ref{eqdef3}) but with the $c_a$ coefficients replaced by homogenous polynomials
of order two in the two extra variables $z=(z_6,z_7)$ spanning $\mathbb{CP}^1$:
\begin{equation}\label{eq:CY3_Case5}
\begin{aligned}
f _{\text{CY}_3, \,{\rm hom}} 
%~ =&~ \left(c_{0,1}\, z_6^2 + c_{0,2}\, z_6\,z_7 + c_{0,3}\, z_7^2 \right) \, z_1\, z_2\, z_3\, z_4\, z_5 +
%\left(c_{1,1}\, z_6^2 + c_{1,2}\, z_6\,z_7 + c_{1,3}\, z_7^2 \right) \,z_1^3\,z_3^3 \\
%&\!+\left(c_{2,1}\, z_6^2 + c_{2,2}\, z_6\,z_7 + c_{2,3}\, z_7^2 \right) \,z_1^3\,z_4^3 +\left(c_{3,1}\, z_6^2 + c_{3,2}\, z_6\,z_7 + c_{3,3}\, z_7^2 \right) \,z_2^3\,z_3^3 \\
%&\!+\left(c_{4,1}\, z_6^2 + c_{4,2}\, z_6\,z_7 + c_{4,3}\, z_7^2 \right)  \, {z}_{2}^{3} \, {z}_{4}^{3} +\left(c_{5,1}\, z_6^2 + c_{5,2}\, z_6\,z_7 + c_{5,3}\, z_7^2 \right)  \, {z}_{5}^{3} \nn\\
~=&~  - c_0(z) \, z_1\, z_2\, z_3\, z_4\, z_5 +
 c_1(z)  \,z_1^3\,z_3^3 \ +c_2(z) \,z_1^3\,z_4^3 + c_3(z)  \,z_2^3\,z_3^3 +c_4(z)  \, {z}_{2}^{3} \, {z}_{4}^{3} +c_5(z)  \, {z}_{5}^{3} 
\end{aligned}
\end{equation}   
%with $c_a(z)$ arbitrary homogenous polynomials of order two in the variables $z=(z_6,z_7)$ spanning $\mathbb{CP}^1$.  

The period ratios $\tau^{(i)}$ are given by the formulas in the last section with
\beq
\frac{\xi (z)}{27} ~=~ \frac{c_1(z)\,c_4(z)\,c_5(z)}{c_0(z)^{3}}\qquad \text{and}\qquad \frac{\eta(z)}{27}~=~\frac{c_2(z)\,c_3(z)\,c_5(z)}{c_0(z)^3}~.
\label{xizcasethird}
\eeq

\vspace{8pt}
We notice that $c_5(z)$ appears both in the definitions of $\xi(z)$ and $\eta(z)$, so at the zeros of $c_5$
both $\xi$ and $\eta$ vanish.   To find the location of branes, we look for the poles of $j_1(z)$ and $j_2(z)$. 
As before, the poles of $j_1$ and $j_2$ are located along the curves $\xi=0$ or $\eta=0$ and we can use the simpler formulas (\ref{tau1tau2_Case5}) to read off the monodromies. We find that $\tau^{(1)}$ has a monodromy $\tau^{(1)}\rightarrow \tau^{(1)} +1$ around the zeros of $c_1,c_4,c_5$ while $\tau^{(2)}$  has a similar monodromy around the zeros of $c_2,c_3,c_5$.  The brane content is summarised in the  table \ref{table:branes3}.
 
 \vspace{12pt}
 \begin{table}[h]
\begin{center}
\begin{tabular}{c| c c c}
& ~~~~$   c_1,c_4 $~~~~ & ~~~~$ c_2, c_3 $~~~~ &    ~~~~$ c_5 $~~~~   \\[2pt]
\hline\\[-14pt]
($q_1$, $q_2$)~~ & $(1,0)$ & $(0,1) $ & $(1,1)$
\end{tabular}
\vspace{8pt}
\capt{6.1in}{table:branes3}{Branes of charge $(q_1,q_2)$ are located at the zeros of $c_a$'s.  }
\end{center}
\end{table}

Taking into account that each $c_a(z)$ has two zeros, one finds $4\times (1,0)+4\times (0,1)+2 (1,1) =12$ branes. The intersection of the $(\xi(z),\eta(z))$ curve with the discriminant locus $D=0$ corresponds to $12$ points in this example. Unlike in the previous examples, going around the locus $D=0$ leads to monodromies different from the $\mathbb Z_2$ element that interchanges $j_1$ and $j_2$. This suggests that $12$ additional branes are located at the points of intersection between the curve $(\xi(z),\eta(z))$ and the discriminant locus, leading to a total number of $24$. However, the study of monodromies around these points is more involved.

% The brane locations and charges (up to $SL(2,\mathbb{Z})$ rotations) can be found by expanding $j_{1,2}$ around the zeros of the $c_a$'s with $\sigma$ and $\tau$ charges given by the order of the poles of $j_1$ and $j_2$ respectively.     The results are summarised in the following table
     
%     \bea
%\begin{array}{c|cccccc}
%                     & c_1 & c_2 & c_3 & c_4 & c_5 & c_6\\
%\hline
%  (n_1,n_2) &  (1,1) & (3,0) & (3,0) & (1,1) & (1,1) & (0,0)   
%\end{array}
%\eea
%  Taking into account that each $c_a$ has two zeros, one finds the expected 24 branes.  

\newpage
\section{Conclusions and Outlook}

Based on the ideas presented in \cite{Martucci:2012jk, Braun:2013yla}, in the present paper we studied the possibility of extending the F-theory approach to finding non-perturbative type IIB vacua with non-trivial fluxes. This approach was termed `G-theory'  in \cite{Martucci:2012jk, Braun:2013yla}. While F-theory studies vacua with 7-branes and varying axio-dilaton field, G-theory aims to geometrize several other complex combinations of fluxes.

Concretely, we looked at type IIB solutions on $\mathbb R^{1,3}\times T^4\times S^2$ with the metric, the dilaton and the flux potentials varying over $S^2$ and the flux potentials oriented along $T^4$.
%We started with the local six-dimensional solutions of type IIB supergravity compactified on $T^4$ that were studied in the companion paper \cite{paper:local}.
We started by finding local solutions on $\mathbb R^{1,3}\times T^4\times \mathbb C$ through a sequence of S and T dualities performed on a class of Ricci-flat geometries with trivial fluxes. These solutions are characterised by $n\leq 3$ holomorphic functions that span the moduli space:
\beq
{\cal M}_{\text{BPS}}=  SO(2,n,\mathbb Z) \backslash{SO(2,n,\mathbb{R} )\over SO(2,\mathbb R)\times SO(n,\mathbb R) }
\label{mbps2}
\eeq

The main observation in G-theory is that this moduli space matches the moduli space of complex structures of a $K3$ surface with Picard number $20-n$.
The flux solution can then be viewed as a fibration of an auxiliary $K3$ fibered over $S^2$. The degeneration points of the $K3$ fibration are associated to different types of branes, as indicated by the monodromies of the corresponding holomorphic combinations of flux fields.

In the present paper we focused mainly on the sub-class of solutions corresponding to $n=2,3$. For the $n=2$ case, we worked out in detail the map between the two holomorphic combinations of fluxes and the period ratios of the holomorphic two form describing the complex structure of the auxiliary $K3$ surface. We inferred the brane content of the flux solutions from the $SO(2,n,\mathbb Z)$ monodromies of these functions  around brane locations. In addition, the presence of a brane curves the base space (brane tension) generating a deficit angle of $\pi/6$  \cite{Greene:1989ya}, so a compact $S^2$ arises for a total of $24$ branes.  As a consistency check, we showed that in each example considered this number is reproduced.

 The results obtained here can be easily generalised. In fact, we have worked out the details of the flux-to-period map in a couple of other examples, summarised in Section~\ref{sec:OtherExamples}. We anticipate that a similar analysis can be performed for the local solutions involving $n=3$, $n=4$ and $n=5$ holomorphic functions, a task to which we hope to return in a future publication.

The techniques developed here can also be applied in contexts different from the one under consideration.  Recently, in \cite{Malmendier:2014uka},
non-geometric heterotic backgrounds with $E_8\times E_8$ and $E_8\times E_7$ vector  bundles were studied by relating them to geometries based on $K3$ surfaces admitting $n=2,3$ complex deformations. The flux/geometry dictionary built here provides additional geometries that can be studied from the heterotic
perspective. Conversely, the field/geometry dictionary built in \cite{Malmendier:2014uka} (see also \cite{Martucci:2012jk, Braun:2013yla}) for $K3$ surfaces with $E_8\times E_7$ singularities provides explicit realisations of flux solutions of class A, B and C characterised by $n=3$ holomorphic solutions.
 Finally, it would be nice to study the gauge duals of the supergravity solutions presented here. We observe that near the locations of branes, supergravity
fields exhibit logarithmic divergences and an infinite tower of instanton corrections that can be tested against the dual gauge theory along the lines of \cite{Billo:2012st}.

\section*{Acknowledgements}

The authors would like to thank A.~Braun, V.~Braun, C.~Hull, L.~Martucci, M.~Petrini and D.~Waldram for interesting discussions and valuable comments. In addition, Volker Braun provided much assistance with the use of Sage. CD, ML and JFM would like to thank the Mathematical Institute, University of Oxford and Theoretical Physics group at Imperial College London for their kind hospitality during parts of this project. The work of PC is supported by EPSRC grant BKRWDM00.  AC would like to thank the University of Oxford and the STFC for support during part of the preparation of this paper. The research of ML was supported by the Swedish Research Council (VR) under the contract 623-2011-7205. The work of JFM is supported by EPSRC, grant numbers  EP/I01893X/1 and EP/K034456/1 and the ERC Advanced Grant n.~226455. 
 
%%%%%%%%%%%%%    Appendices   %%%%%%%%%%%%%%%%%%%%%%%%%%%%%%%%%%
 
\begin{appendix}
\newpage

\section{A brief review of toric geometry}  \label{sec:ToricGeometry}

In this section we review the notions of toric geometry needed in Sections~\ref{sec:example},~\ref{sec:dictionary} and \ref{sec:OtherExamples}. 

Let $M,N\cong \mathbb{Z}^{p}$ be two dual lattices. Let $\Delta$ be a lattice polytope in $M$, i.e.~a polytope realised as the convex hull of a finite number of points in $M$. The polytope $\Delta$ is said to be reflexive if it contains the origin as its unique interior point and if the dual polytope $\nabla\subset N\otimes \mathbb R$, defined as
 \beq
 \nabla= \{  w \in N\otimes \mathbb R~ \big| ~\langle v,w \rangle  \geq -1,  \text{ for all } v\in \Delta \}
 \eeq
 is also a lattice polytope. An example of a pair of dual three-dimensional polytopes is given in Figure~\ref{fig:NablaDeltaCase3}. %This pair of dual polyhedra will serve as our main case study in the following section.

Given a pair of reflexive polytopes $(\Delta, \nabla)$, one can construct a $p$-dimensional toric variety from the fan over a triangulation of the surface of $\nabla$, and a Calabi-Yau hypersurface in this toric variety as the zero locus of a polynomial whose monomials are in one-to-one correspondence with the lattice points of $\Delta$. This construction is described in the texts~\cite{0813.14039, 1223.14001, Skarke:1998yk, Avram:1996pj}.

Following Cox's approach, the toric variety can be constructed as an algebraic generalisation of complex weighted projective spaces. Concretely, let $\{w_i\,|\, i=1,\ldots q\}\subseteq (\nabla\backslash \{0\})\cap N$, be a subset of the lattice points of $\nabla$ which includes its vertices and corresponds to a triangulation of its surface. In Cox's approach, one assigns a homogeneous coordinate $z_i\in \mathbb C^q$ to each vertex $w_i$ of $\nabla$. After removing an exceptional set -- in analogy to removing the origin of $\mathbb C^{n+1}$ in the construction of $\mathbb C\mathbb P^n$ (see e.g.~the texts~\cite{0813.14039, 1223.14001, Skarke:1998yk, Avram:1996pj} for details) -- one identifies the points of $\mathbb C^q$ using the certain equivalence relations. The equivalence relations are obtained from the $q-p$ linear relations between the vectors $\{w_i\}$:
 \beq
 \sum_{i=1}^q Q_a^i \, w_i =0     \qquad \text{where } a = 1,\ldots, q-p~.
 \eeq 

The coefficients $Q_a^i$ are called weights and they form a so-called weight system $\{Q_a\}$. 
The equivalence relations between the homogeneous coordinates are, for each $a$, given by
\beq\label{sim}
(z_1,\ldots z_q) \sim (\lambda^{Q_a^1} z_1,\ldots  \lambda^{Q_a^q} z_q ) \qquad \text{for any } \lambda\in  \mathbb{C}^* ~.
\eeq

A Calabi-Yau hypersurface is then defined as the zero locus of a polynomial, homogeneous under any of the relations (\ref{sim}), whose monomials are associated to the lattice points of $\Delta$ 
\beq\label{fhom}
f_{\rm hom}=\sum_{v_a \in \Delta \cap M }  c_a \, \prod_{i=1}^q z_i^{\langle w_i , v^a \rangle+1}~.
\eeq   

The auxiliary geometrical construction presented in the previous section involves a $K3$ fibered over $\mathbb C\mathbb P^1$. Such fibrations can be realised as Calabi-Yau threefolds, described in toric geometry by a pair $(\Delta, \nabla) \subset M{\times}N$ of reflexive four-dimensional polytopes, for which $N$ has a distinguished three-dimensional sub-lattice $N_3$, such that $\nabla_3= \nabla \cap N_3$ is a three-dimensional reflexive polytope. The sub-polytope $\nabla_3$ is associated with the fiber and divides the polytope $\nabla$ into two parts, a top and a bottom~\cite{Candelas:1996su, Candelas:1997pq, Candelas:2012uu}. 
The fan corresponding to the base space, obtained by projecting the fan of the fibration along the linear space spanned by the sub-polytope $\nabla_3$ defines a $\mathbb C\mathbb P^1$ space~\cite{Kreuzer:1997zg}. 
The above description is dual to having a distinguished one-dimensional sub-lattice $M_1 \subset M$, such that the projection of $\Delta$ along $M_1$ is $\Delta_3= (\nabla_3)^{\!^*}$, the dual of $\nabla_3$~\cite{Avram:1996pj}. This dual description is commonly referred to as `a slice is dual to a projection'.

$K3$ surfaces that are elliptic fibrations will also appear in the subsequent discussion. The fibration structure can be seen at the level of polytopes in a manner completely analogous to the Calabi-Yau three-fold case, see Figure~\ref{fig:NablaDeltaCase3} for an example. As before, the two-dimensional polytope $\nabla_2$ corresponding to the elliptic fiber divides the $K3$ polytope $\nabla$ into two parts. It is interesting to note that in this case, the distributions of points above and below the $\nabla_2$ define two affine Dynkin diagrams, which, in the ADE case correspond to the Kodaira degeneration type of the elliptic fiber over two distinguished points in the base space, say the North and the South poles. In F-theory, the corresponding Lie groups appear as gauge groups in the low-energy theory. A similar connection will emerge, in certain limits, in our discussion.

\vspace{4pt}
Finally, let us mention that the number of complex structure parameters of a $K3$ surface can be computed from the combinatorial data of the defining pair of three-dimensional polytopes:
\begin{equation}\label{eq:complexstructures}
n=l(\Delta) - 4~ - \sum_{\substack{2\text{-faces }\theta\subset \Delta}} l^*(\theta )+ \sum_{\substack{\text{edges }\theta\subset \Delta}} l^*(\theta)\,l^*(\theta^*)
\end{equation}
where $l(\Delta)$ denotes the number of integer points of the polyhedron $\Delta$, $l^*(\theta)$ denotes the number of integer points interior to a 2-face or to an edge, and $\theta^*$ is the dual face to $\theta$. Note that edges (1-faces) are dual to edges and vertices (0-faces) are dual to 2-faces.   

%Out of the $4,319$ polyhedra in the Kreuzer-Skarke list, $2$ correspond to $K3$ manifolds with Picard number $19$ and $n=1$ complex structure parameter,  $9$ to $K3$ manifolds with Picard number $18$ and $n=2$ complex structure parameters, $24$ to $K3$ manifolds with Picard number $17$ and $n=3$ complex structure parameters and so on. 
%For example the complete list of K3 with Picard number 18 is displayed in figure 1. 

%For all these cases $\Delta$ has $6$ points and there are no interior points to 2-faces or to edges, hence the above formula gives $n=2$ complex structures.  

%In the next section we will work out the details of the flux/geometry dictionary for a simple choice of a $K3$  fiber with two complex structure parameters. For this choice, we will compute the periods of the holomorphic $(2,0)$-form of the $K3$ and relate them to the coordinates on the complex structure moduli space. We will then embed the $K3$ polytope into a four-dimensional reflexive polytope encoding the $K3$ fibration over $\mathbb C\mathbb P^1$.  
%One can think of the $K3$-fibered Calabi-Yau three-fold as a compact one-parameter curve inside the complex structure moduli space of the $K3$ fiber.  At certain points along this curve, the fiber is singular, or equivalently, the functions $\sigma$ or $\tau$ have logarithmic branch points, which indicates the presence of branes. The type of branes in question can be extracted from the corresponding monodromies of $\sigma$ and $\tau$.

\section{The $K3$ moduli space -- a more detailed presentation}\label{app:ModuliSpace}

\subsection{The discriminant locus}\label{app:discriminant}

The $K3$ surface $S$ is singular at the points where the defining polynomial as well as all its derivatives vanish simultaneously. These conditions are equivalent with the following set of equations\footnote{The equation $f_{\rm him} =0$ is recovered by summing up the LHS of the equations in \eqref{fsing} and equating this with $0$.}
\beq
z_j \, {\partial f_{\rm hom} \over \partial z_j}~=~0~;~~~j=1,\ldots,6~~\text{(no sum)}.   \label{fsing}
\eeq 
Generically, the above system admits no solutions. However, for special choices of coefficients, defining a certain locus in the moduli space (the discriminant locus), the system admits solutions.  
 
If we denote the monomials of $f_{\rm hom}$ (including the $c_a$-coefficients and the signs)  by $m_0, \ldots, m_5$ then the above six conditions become
\beq\begin{split}
m_0 + 3m_1 + 2m_4 ~&=~0~,\qquad m_0 + 3m_1 + 2 m_5~=~0~,\\
m_0 + 3m_2 + 2m_4~&=~0~,\qquad m_0 + 3m_2 + 2 m_5 ~=~0~,\\
m_0 + 3m_3 + 2m_4~&=~0~,\qquad m_0 + 3m_3 + 2 m_5 ~=~0~.
\label{eqdis}
\end{split}\eeq
with the monomials $m_i$ satisfying the relations
 \beq
   m_1 m_2 m_3 = -\,{\xi \over 27} \, m_0^3 \qquad\text{and}\qquad m_4 m_5 = {\eta \over 4} \,   m_0^2 ~.\label{rel}
 \eeq
 From \eqref{eqdis}, it is immediate that $m_1~=~m_2~=~m_3$ and $m_4~=~ m_5$. Then, the equations (\ref{eqdis}) and (\ref{rel}) admit a solution only if $\xi$ and $\eta$  belong to the discriminant locus $D=0$
with
\beq
D=  (\xi-1-3 \eta)^2-\eta (\eta+3)^2 \label{eq:discriminant}~.
\eeq

\vspace{-14pt}
\begin{figure}[H]
\begin{center}
\hskip-50pt
\begin{minipage}[t]{5.2in}
\vspace{10pt}
\raisebox{-1in}{\includegraphics[width=15.cm]{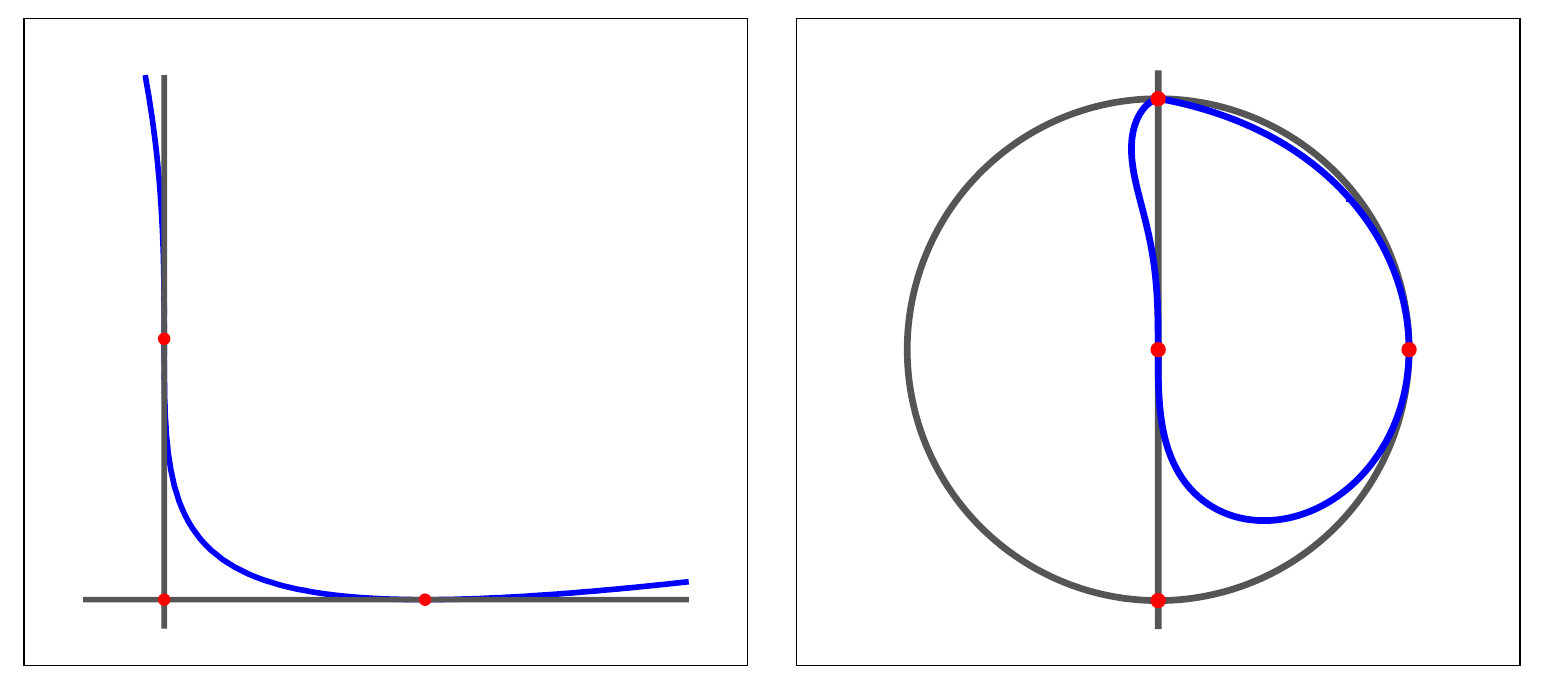}} 
\vspace{-0pt}
\end{minipage}
\capt{6in}{fig:discr_case3}{On the left: the discriminant locus (in blue) showing also the curves $\xi = 0$ and $\eta = 0$. The intersections with $D$ are the points $(1,0)$ and $(0,1)$. The locus $D(\xi,\eta)=0$ contains an extra isolated point in the real plane, $(\xi,\eta)=(-8,-3)$, not included in the plot. On the right, the upper half plane has been mapped to the unit disk in order to show the intersection at infinity.}
\end{center}
\vskip -10pt
\end{figure}

%\bea
% \frac{1}{1-\tilde f_{\rm hom} }= \sum_{n=0}^{\infty} \tilde{f} _{\rm hom}^{\,n} =1+{2 c_4 c_5\over c_0^2} -{ 6 c_1 c_2 c_3 \over c_0^3}+{ 6 c_4^2 c_5^2 \over c_0^4} +\ldots
% \label{tildef}
%\eea
%Plugging (\ref{tildef}) into (\ref{w00}) one finds \footnote{The relevant  $z_i$-independent contributions come from terms 
%$(-\tilde{f} _{\rm hom})^{3k+2l} =  \frac{\Gamma[ 3k+2l+1] }{\Gamma[k+1]^3 \Gamma[l+1]^2}  \left({ c_4 c_5 \over c_6^2 }\right)^{l} 
% \left({- c_1 c_2 c_3 \over c_6^3 }\right)^{k}  +\ldots $. }  
%\beq
%\varpi_{00}= \sum_{k,l=0}^\infty   a_{k,l }    \,\x^{k} \eta^{l}=
%    \sum_{k,l=0}^{\infty} \frac{\Gamma[ 3k+2l+1] }{\Gamma[k+1]^3 \Gamma[l+1]^2}\!\left(\frac{\xi}{27}\right)^k\! \!
%    \left(\frac{\eta}{4}\right)^l \; .
%\label{fundperiod}~
%\eeq

\subsection[The period computation: Picard-Fuchs equations and the method of Frobenius]{The period computation: PF equations and the method of Frobenius}\label{sec:PFapp}

In Section~\ref{sec:PFmain} we found the Picard-Fuchs operators 
\beq\begin{split}
\cL_1~&=~4\,\th_{\eta}^2 - 
\eta\, (3\th_\x+2\th_{\eta} + 2)(3\theta_{\x}+2\th_{\eta} + 1)~,  \\[3pt]
\cL_2~&=~3\,\th_\x(3\th_\x-2\th_\eta) -\x\, (3\th_\x+2\th_\eta+2)(3\th_\x+2\th_\eta+1)
+3\,\eta\,\th_\x(3\th_\x+2\th_\eta+1)~.
\end{split}
\eeq 
by searching for linear relations with polynomial coefficients among the elements of the set 
\beq
\{\th_\x^i\,\th_\eta^j\,\vp_{00}\}_{\,0\,\leq\,i+j\,\leq\,2}~.
\eeq
Alternatively, the first operator $\cL_1$ may also be found from the recurrence relation 
\beq\label{recurrence1}
4 (l+1)^2 a_{k,l+1}=(3k+2l+2)(3k+2l+1) a_{k,l}
\eeq
satisfied  by the expansion coefficients $a_{k,l}$ in  the series (\ref{eq:w00series}) for $\vp_{00}$. A second recurrence relation 
\beq\label{recurrence2}
27(k+1)^3\, a_{k+1,l}~=~(3k+2l+3)(3k+2l+2)(3k+2l+1)\,a_{k,l}
\eeq
leads to the third order operator 
\beq
\widetilde{\cL}_2 =27\,\theta_{\x}^3 - \x (3\theta_{\x}+2\th_{\eta} + 3)(3\theta_{\x}+2\th_{\eta} + 2)(3\theta_{\x}+2\th_{\eta} + 1)~.
\eeq
In this alternative approach (see \cite{Chialva:2007sv} for more details), the second order operator $\cL_2$ can be found by considering combinations of these two operators of the form $\cM\cL_1 + B\widetilde{\cL}_2$, where $M$ is a first order operator and $B$ is a polynomial, and to seek to factor this operator into the product of a first order and a second order operator. 

%\vspace{12pt}
We believe, as indicated above, that the differential equations corresponding to these operators should admit four linearly independent solutions. To find these we have recourse to the method of Frobenius.
Consider a quantity to which we will refer as the Frobenius period
\beq
\vp~=~\sum_{k,l=0}^\infty a_{k,l} (\e,\delta)\,\bigg(\frac{\xi}{27}\bigg)^{k+\e} \bigg(\frac{\eta}{4}\bigg)^{l+\delta}
\eeq
where the coefficients $a_{k,l}$ now depend on $\e$ and $\delta$. We will suppress the arguments $(\e,\delta)$ below. Requiring that the Frobenius period is a solution of the differential equations associated with $\cL_1$ and $\cL_2$ results in certain constraints on the coefficients $a_{k,l}$ and the exponents $\e$ and $\delta$. More precisely, acting with $\cL_1$ and, respectively, $\cL_2$ on $\vp$ produces two series with terms given by: 
\beq
\begin{aligned}
\left( \cL_1\,\vp \right)_{k,l}& {=} \left[4(l+\delta)^2 a_{k,l}  {-} 4\Big(3(k{+}\e){+}2(l{+}\delta{-}1){+}2\Big)\Big(3(k{+}\e){+}2(l{+}\delta{-}1){+}1\Big)a_{k,l-1} \right] \bigg(\frac{\xi}{27}\bigg)^{k+\e} \bigg(\frac{\eta}{4}\bigg)^{l+\delta}\\[8pt]
\left( \cL_2\,\vp \right)_{k,l}& {=} \left[3(k+\e) \Big(3(k{+}\e) - 2(l{+}\delta)\Big)  a_{k,l}  {-} 27\Big(3(k{+}\e-1){+}2(l{+}\delta)+2\Big)\Big(3(k{+}\e-1){+}2(l{+}\delta)+1\Big) a_{k-1,l} \right. \\[8pt] 
&\left.\qquad\qquad+ 12\Big(k+\e\Big)\Big(3(k{+}\e){+}2(l{+}\delta{-}1)+1\Big) a_{k,l-1} \right] \bigg(\frac{\xi}{27}\bigg)^{k+\e} \bigg(\frac{\eta}{4}\bigg)^{l+\delta}
\end{aligned}
\eeq
where the coefficients $a_{k,l}$ are understood to vanish if at least one of the indices is negative.

\vspace{12pt}
If we choose the initial conditions $a_{0,0}=1$ then, $\cL_1\vp=\cL_2\vp=0$ implies
\beq
a_{k,l}~=~\frac{\G^3(\e+1)\G^2(\delta+1)}{\G(3\e+2\delta+1)}\,
\frac{\G(3k+3\e+2l+2\delta+1)}{\G^3(k+\e+1)\G^2(l+\delta+1)}~.
\label{acoeffs}\eeq
The coefficients $a_{k,l}$ reduce to those of \eqref{eq:akl} when $\e=\delta=0$. It is of interest to note also that the coefficients can be expressed simply in terms of Pochhammer symbols.\footnote{
The Pochhammer symbol $(\z)_n$ is defined, for $n=0,1,2,\ldots$, by
$$
(\z)_n~=~\frac{\G(\z+n)}{\G(\z)}~.
$$ 
\vspace{-21pt}

\noindent Thus
$$
(\z)_0~=~1~,~~~(\z)_1~=~\z~~~\text{and}~~~(\z)_n~=~\z(\z+1)\cdots(\z+n-1)~,
$$
so $(\z)_n$ is a polynomial in $\z$ of degree $n$.} In terms of these symbols the coefficients may be written as
\beq
a_{k,l}~=~\frac{(3\e+2\delta+1)_{3k+2l}}{(\e+1)_k^3\, (\delta+1)_l^3}~.
\eeq
This has the consequence that, despite the initial appearance of \eqref{acoeffs}, all the quantities
\beq
a_{k,l}^{(r,s)}~=~\left. \left(\pd{}{\e}\right)^r \! \left(\pd{}{\delta}\right)^s a_{k,l}\, \right|_{\e,\delta=0}~,
\eeq
that we shall require shortly, are rational numbers.
\vspace{8pt}

Furthermore, the $k=l=0$ terms of the equations $\cL_1\vp=\cL_2\vp=0$ have to be treated separately. These imply the following relations
\vspace{-8pt}
\beq
\delta^2~=~0~,~~~~\e(3\e - 2\delta)~=~0.
\label{ideal}\eeq
Multiplying the second of these relations by $\delta$ and $\e$, in turn, we learn that $\e^2\delta=0$ and $\e^3=0$. Thus all the cubic, and higher, monomials in $\e$ and $\delta$ vanish.

We write
\beq\label{eq:pi_rs_app}
\vp_{rs}~=~\frac{1}{(2\p\ii)^{r+s}}\left. \left(\pd{}{\e}\right)^r \! \left(\pd{}{\delta}\right)^s\vp\, \right|_{\e,\delta=0}
\eeq
and expand $\vp$ in terms of $\e$ and $\delta$, taking account of the relations \eqref{ideal},
\be
\vp~=~\vp_{00} + 2\p\ii\e\, \vp_{10} + 2\p\ii\delta\, \vp_{01} + 
 ( 2\p\ii)^2\e\delta\, \Big( \vp_{11} + \smallfrac13 \vp_{20}\Big)~.
\eeq
The four functions identified by this expansion $\{\vp_{00}, \vp_{10},\vp_{01},3\vp_{11}+ \vp_{20}\}$ form a basis for the periods. If we define also the power series $h_{rs}$ by
\beq
h_{rs}~=~\frac{1}{(2\p\ii)^{r+s}}\sum_{k,l=0}^\infty a_{k,l}^{(r,s)}\bigg(\frac{\xi}{27}\bigg)^{k} \bigg(\frac{\eta}{4}\bigg)^{l}\, ,
\eeq
then, writing $\vp_3$  for the combination $3\vp_{11}+ \vp_{20}$, we have the relations
\beq\begin{split}
\vp_{00}&=h_{00}\\[5pt]
\vp_{10}&=\frac{1}{2\p\ii}\,h_{00}\log\left(\frac{\x}{27}\right) + h_{10}\\[5pt]
\vp_{01}&=\frac{1}{2\p\ii}\,h_{00}\log\left(\frac{\eta}{4}\right) + h_{01}\\[5pt]
\vp_{3\phantom{0}}&=\frac{1}{(2\p\ii)^2}h_{00}
\log\left(\frac{\x}{27}\right)\,\log\left( \frac{\x\,\eta^3}{1728} \right) + 
\frac{1}{2\p\ii} \Big(2\,h_{10} + 3\,h_{01}\Big)\log\left(\frac{\x}{27}\right) \\[5pt]
&\hskip2.32in+ \frac{3}{2\p\ii}h_{10}\log\left(\frac{\eta}{4}\right)  + 
\Big(3\,h_{11} + h_{20}\Big)\,.
\end{split}\label{PeriodBasis0}\eeq

\vspace{12pt}
The factors of $2 \pi \ii$ are chosen so that the monodromy matrices, that describe how the periods change as the singular loci $\x=0$ and $\eta=0$ are encircled, are integral:
\beq
T_\x ~=~ \left(
\begin{array}{cccc}
1&~0&~0&~0\\
1&~1&~0&~0\\
0&~0&~1&~0\\
1&~2&~3&~1
\end{array}
\right) ~,
\qquad
T_\eta ~=~ \left(
\begin{array}{cccc}
1&~0&~0&~0\\
0&~1&~0&~0\\
1&~0&~1&~0\\
1&~3&~0&~1
\end{array}
\right).
\label{eq:t_xi0}
\eeq
Note that, if we write $T_\x = {\mathbbm 1} + R_\x$ and $T_\eta = {\mathbbm 1} + R_\eta$, then the matrices $R_\x$ and $R_\eta$ have the same algebra as $\e$ and $\delta$, namely:
\beq
R_\x R_\eta~=~R_\eta R_\x~,~~~R_\eta^2~=~0~,~~~3R_\x^2~=~2R_\x R_\eta~.
\eeq

This $K3$ surface has Picard number 18 and so we expect the periods to factorise and be related to $j$-invariants.
In order to see this factorisation let us return to the generators of the ideal \eqref{ideal}. Modulo $\delta^2$, we may write the first generator as $(\e-\smallfrac13\delta)^2$ so setting $\tilde\e=\e-\smallfrac13\delta$ we may write the generators in the more symmetric form
\beq
\tilde{\e}^2~=~0~~~\text{and}~~~\delta^2~=~0~.
\eeq
Writing $\e$ in terms of $\tilde{\e}$ and $\delta$, we have 
$\x^\e \eta^\delta = \x^{\tilde{\e}} (\eta\x^\frac13)^\delta$, suggesting that we should take as the natural coordinates $\x$ and $\tilde{\eta}=\eta\x^\frac13$.
Note that by setting $\vp_0 = \vp_{00}$, $\vp_1=\vp_{10}$, $\vp_2 =3\, \vp_{01}+\vp_{10}$ and $\vp_3$ as before, we may take a slightly different basis to \eqref{PeriodBasis0} 
\beq\begin{split}
\vp_0&=\ph_0\\[8pt]
\vp_1&=\frac{1}{2\p\ii}\,\ph_0\log\left(\frac{\x}{27}\right) + \ph_1\\[8pt]
\vp_2&=\frac{1}{2\p\ii}\,\ph_0\log\left( \frac{\x\,\eta^3}{1728} \right) + \ph_2\\[8pt]
\vp_3&=\frac{1}{(2\p\ii)^2}\ph_0
\log\left(\frac{\x}{27}\right)\,\log\left( \frac{\x\,\eta^3}{1728} \right) + 
\frac{1}{2\p\ii} \ph_2\log\left(\frac{\x}{27}\right)+ \frac{1}{2\p\ii}\ph_1\log\left( \frac{\x\,\eta^3}{1728} \right) + \ph_3~, \\[12pt]
\end{split}\label{PeriodBasis1_app}\eeq
where
\beq
\ph_0~=~h_{00}~,~~~\ph_1~=~h_{10}~,~~~\ph_2~=~3\,h_{01} +  h_{10}~~~\text{and}~~~
\ph_3~=~3\,h_{11} +\, h_{20}~.
\eeq
Note that 
\beq
\vp_3~=~\frac{\vp_1 \vp_2}{\vp_0}~~~\text{if}~~~\ph_1 \ph_2~=~\ph_0\,\ph_3~.
\eeq
The latter relation does indeed hold as can be checked by multiplication of the respective series. Thus if we write
\beq
\t_1~=~\frac{\vp_1}{\vp_0}~~~\text{and}~~~\t_2~=~\frac{\vp_2}{\vp_0}~~~\text{then}~~~
\t_1\t_2~=~\frac{\vp_3}{\vp_0}~.
\label{tau1tau2_app}\eeq

\newpage
\section{Elliptic Fibration Structures}\label{app:ellipticfibrations}

Elliptic fibration structures for the auxiliary $K3$ surface proved to be important for establishing a link to F-theory, as discussed in Section~\ref{sec:Ftheorylimit}. In this appendix we present the $5$ different fibration structures that can be torically described for the $K3$ surface discussed in Section~\ref{sec:example}. The first fibration structure already appeared in Section~\ref{sec:Ftheorylimit}.

We will present the three-dimensional polytope $\nabla$ and the $5$ different two-dimensional sub-polytopes contained in it as `slices' (see the discussion in Appendix~\ref{sec:ToricGeometry}). In each case, we draw the corresponding  extended Dynkin diagrams, which indicate the degeneration type of the elliptic fiber at $z=0$ and $z=\infty$.

\begin{figure}[h]
\begin{center}
\hskip-35pt
\begin{minipage}[t]{5.2in}
\vspace{10pt}
\raisebox{-1in}{\includegraphics[width=7.cm]{Case3NablaF1.pdf}} 
\hfill \hspace*{10pt}
\raisebox{-.98in}{\includegraphics[width=7.cm]{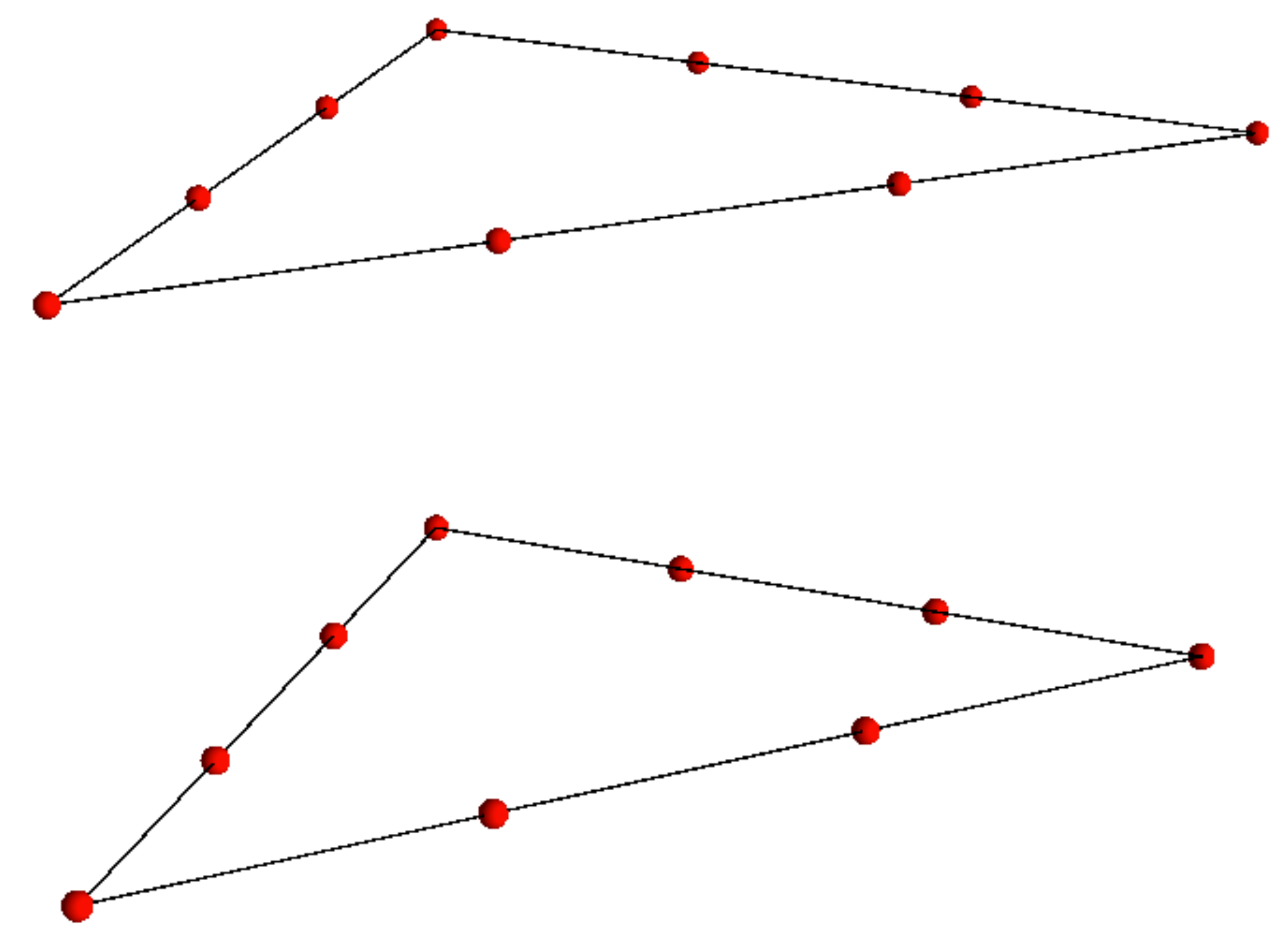}}
\hspace{20pt}
\vspace{10pt}
\end{minipage}
\capt{7in}{fig:fibration1}{The first fibration structure. The Dynkin diagrams correspond to $(SU(9), SU(9))$.}
\end{center}
\vskip -10pt
\end{figure}

\begin{figure}[h]
\begin{center}
\hskip-35pt
\begin{minipage}[t]{5.2in}
\vspace{10pt}
\raisebox{-1in}{\includegraphics[width=7.cm]{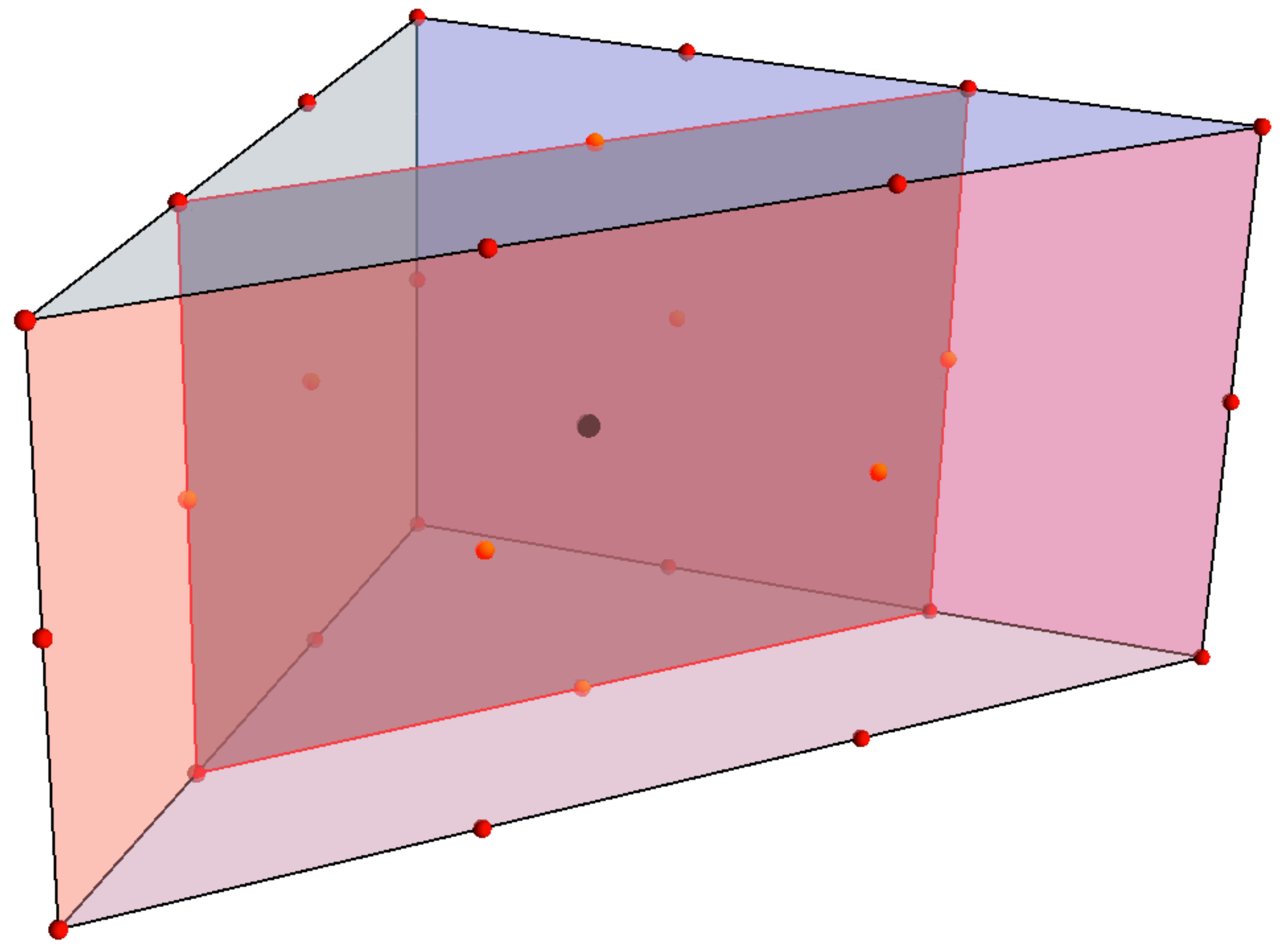}} 
\hfill \hspace*{10pt}
\raisebox{-1.02in}{\includegraphics[width=7.cm]{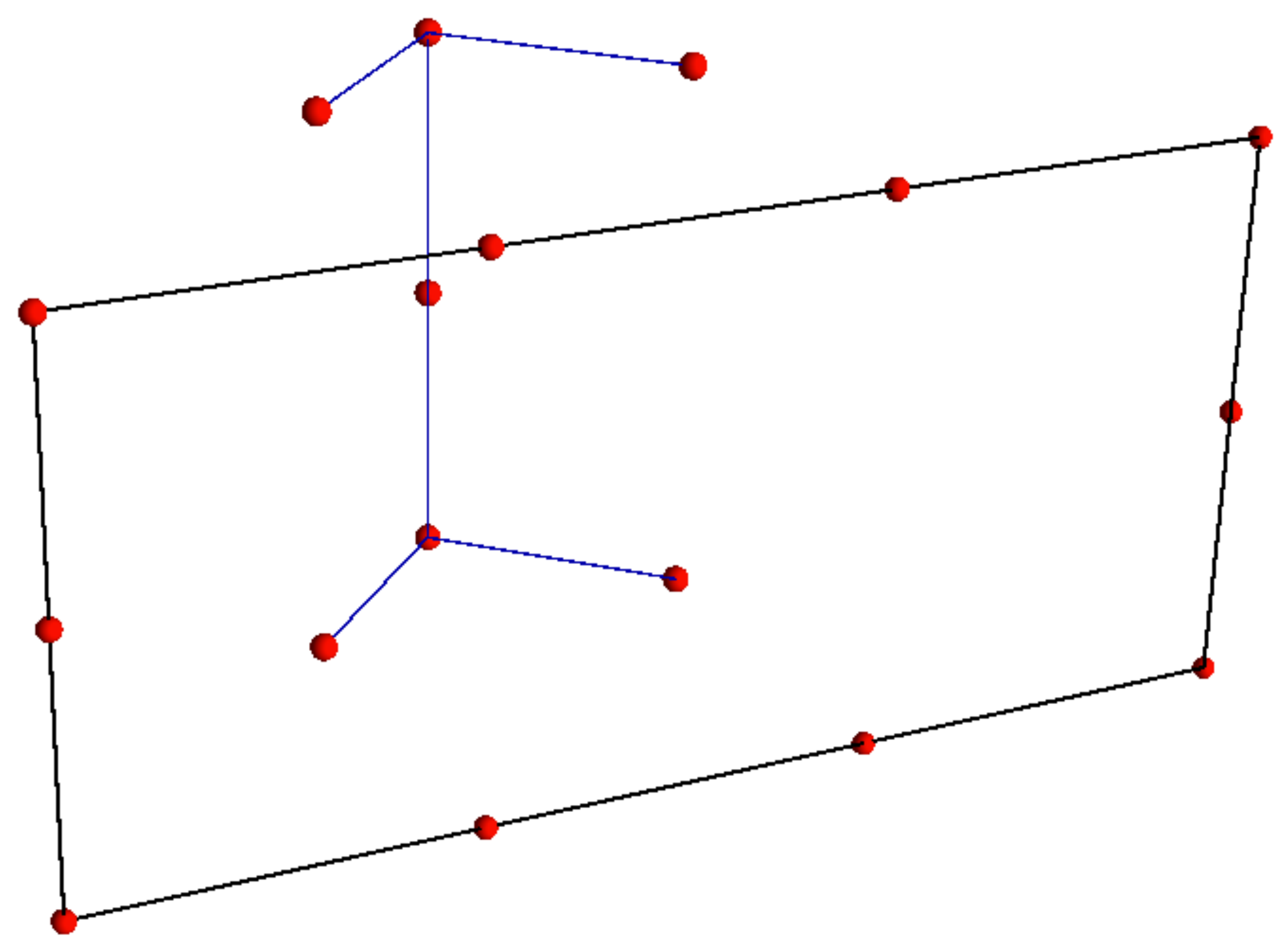}}
\hspace{20pt}
\vspace{10pt}
\end{minipage}
\capt{7in}{fig:fibration2}{The second fibration structure. The Dynkin diagrams correspond to $(SO(12),SU(10))$.}
\end{center}
\vskip -10pt
\end{figure}

\begin{figure}[h]
\begin{center}
\hskip-35pt
\begin{minipage}[t]{5.2in}
\vspace{10pt}
\raisebox{-1in}{\includegraphics[width=7.cm]{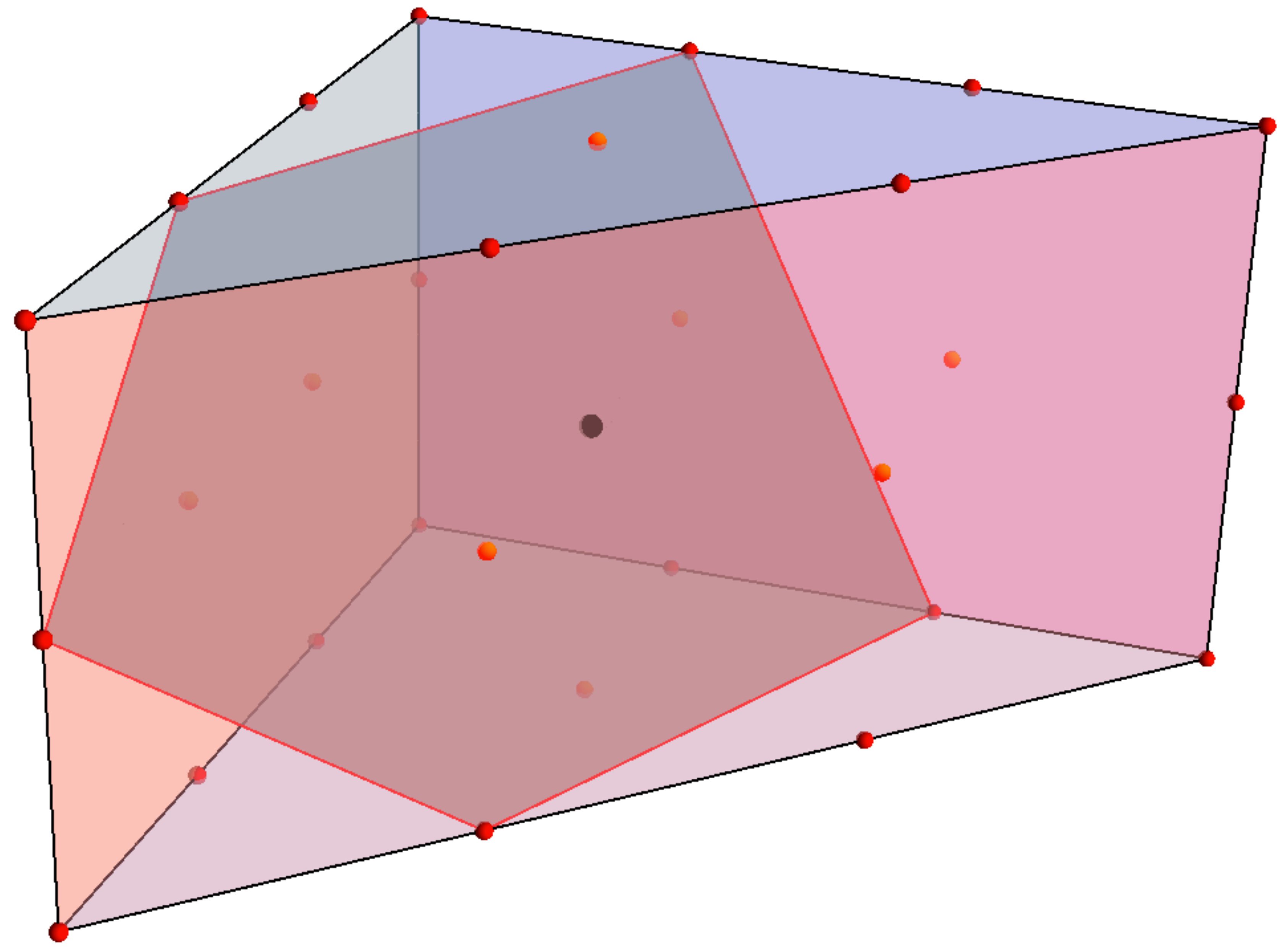}} 
\hfill \hspace*{10pt}
\raisebox{-1.02in}{\includegraphics[width=7.cm]{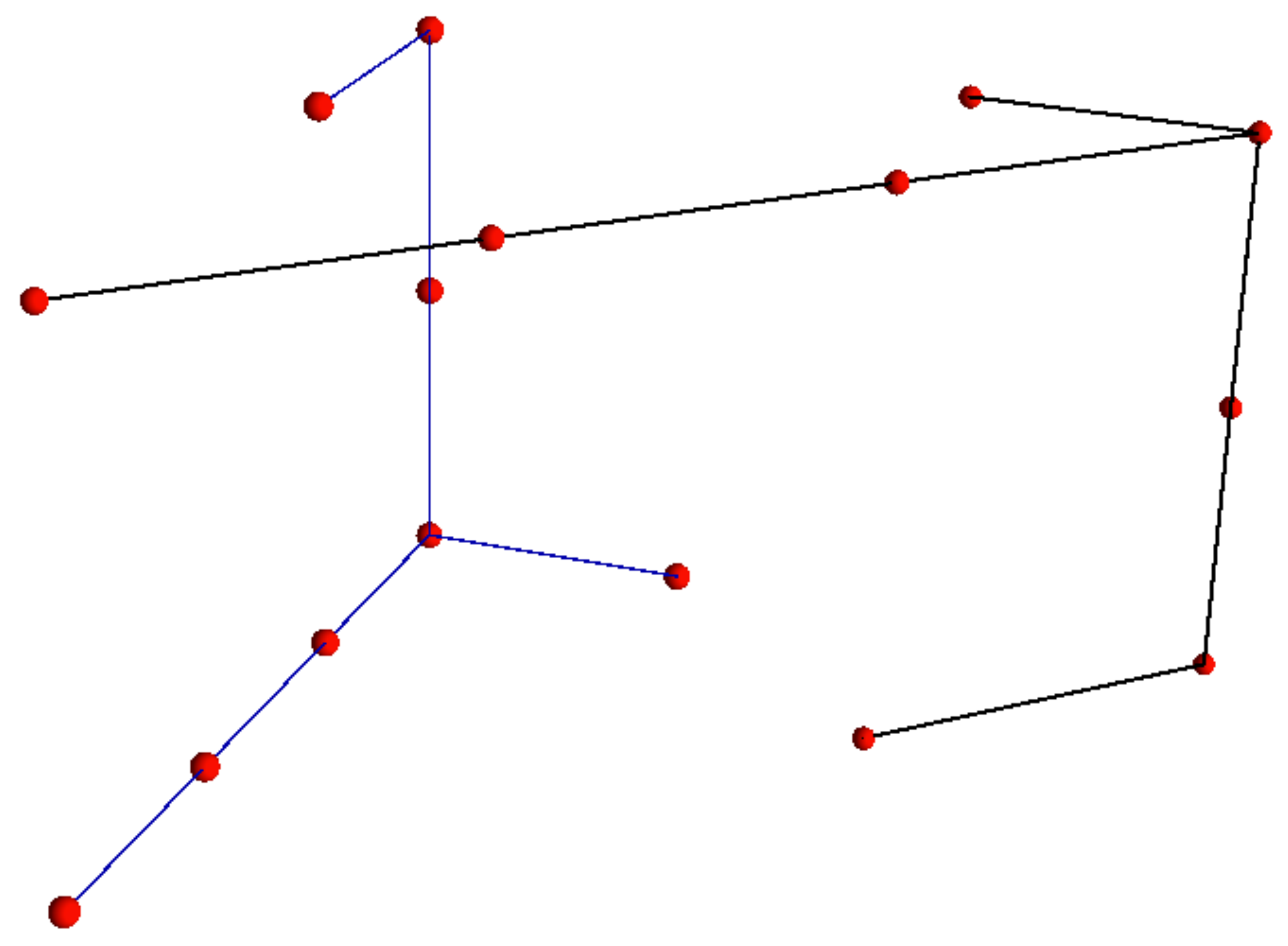}}
\hspace{20pt}
\vspace{10pt}
\end{minipage}
\capt{7in}{fig:fibration3}{The third fibration structure. The Dynkin diagrams correspond to $(E_7,E_7)$.}
\end{center}
\vskip -10pt
\end{figure}

\begin{figure}[h]
\begin{center}
\hskip-35pt
\begin{minipage}[t]{5.2in}
\vspace{10pt}
\raisebox{-1in}{\includegraphics[width=7.cm]{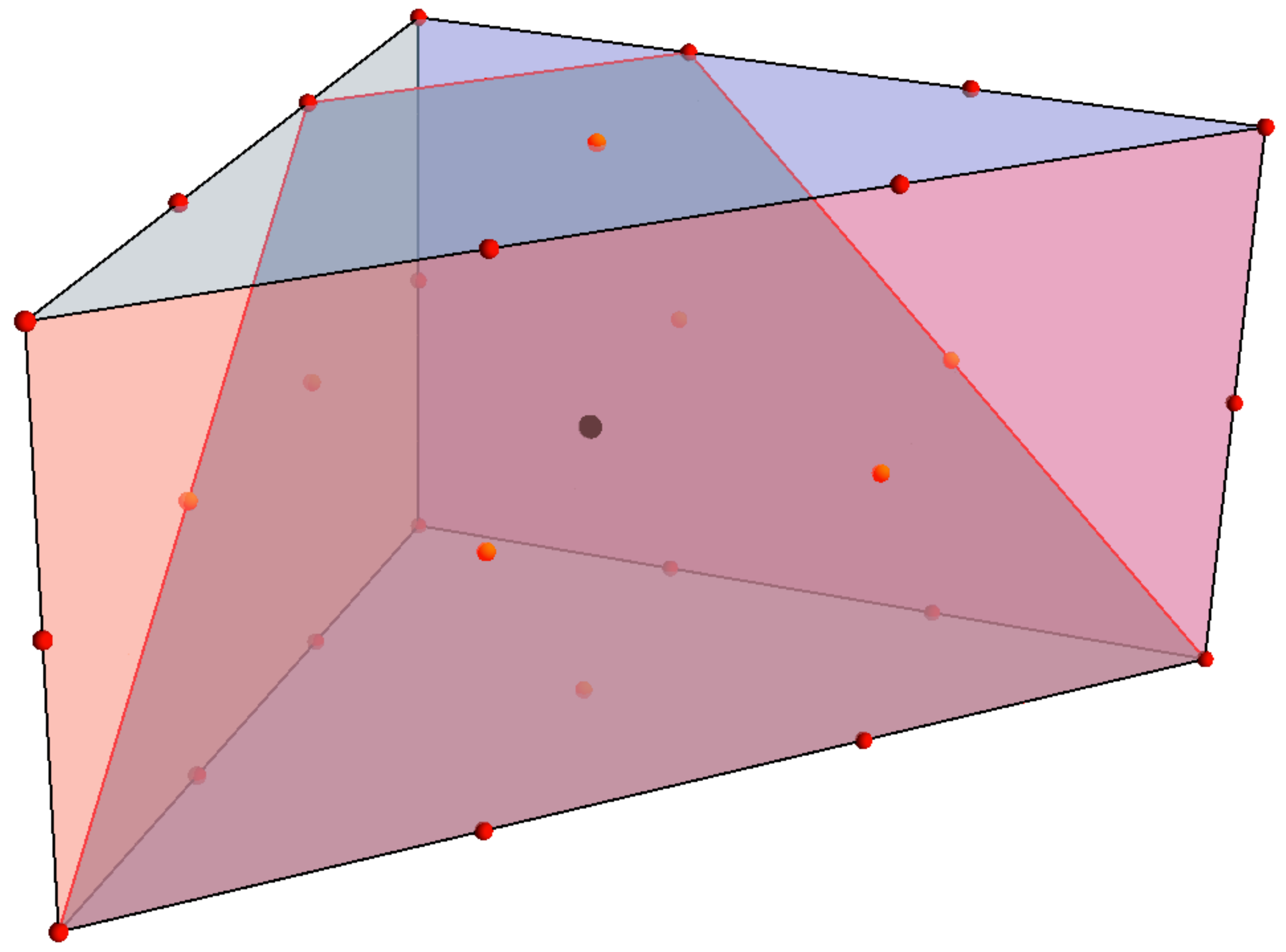}} 
\hfill \hspace*{10pt}
\raisebox{-0.95in}{\includegraphics[width=7.cm]{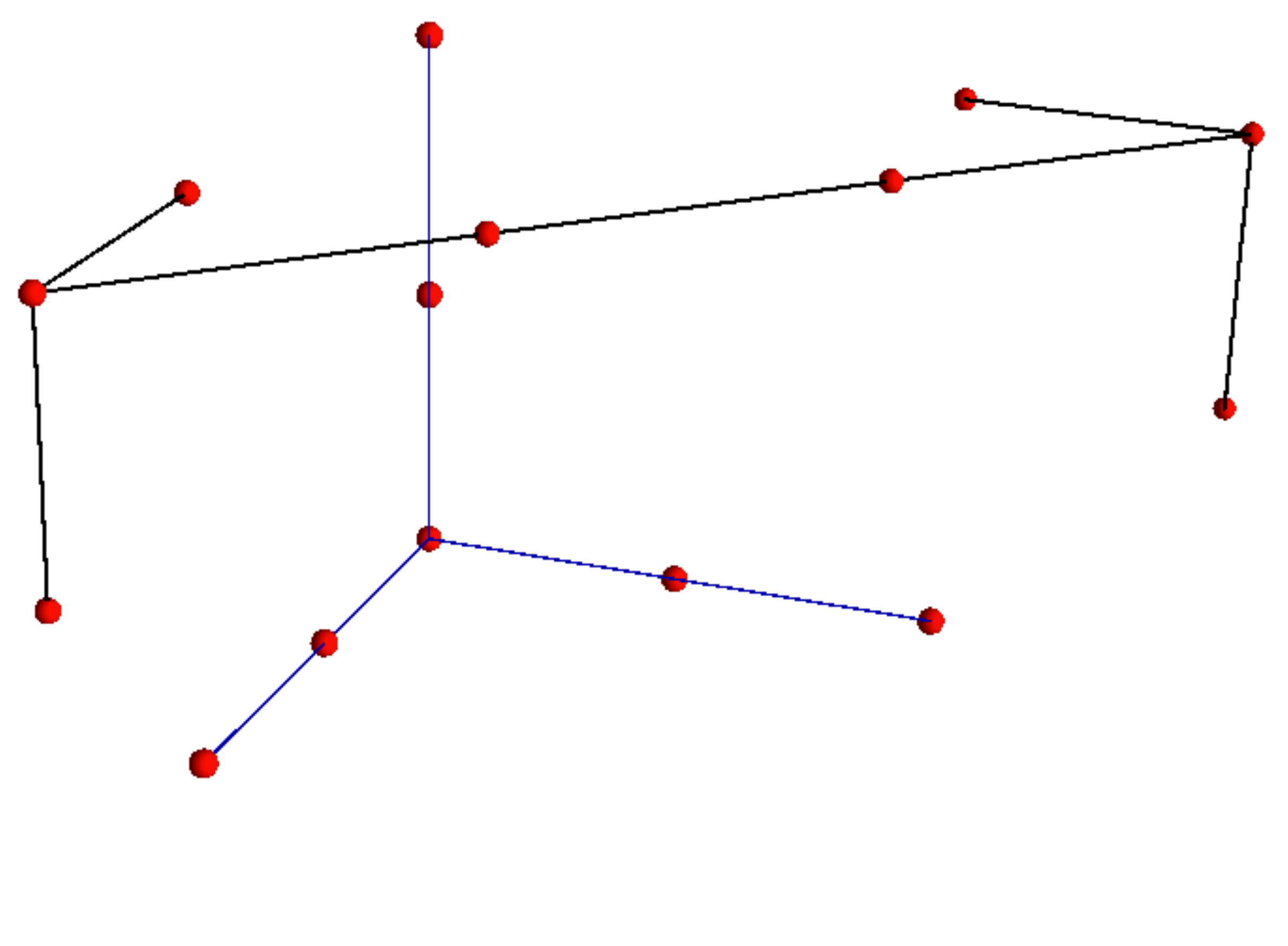}}
\hspace{20pt}
\vspace{10pt}
\end{minipage}
\capt{7in}{fig:fibration4}{The fourth fibration structure. The Dynkin diagrams correspond to $(SO(14),E_6)$.}
\end{center}
\vskip -10pt
\end{figure}

\begin{figure}[h]
\begin{center}
\hskip-35pt
\begin{minipage}[t]{5.2in}
\vspace{10pt}
\raisebox{-1in}{\includegraphics[width=7.cm]{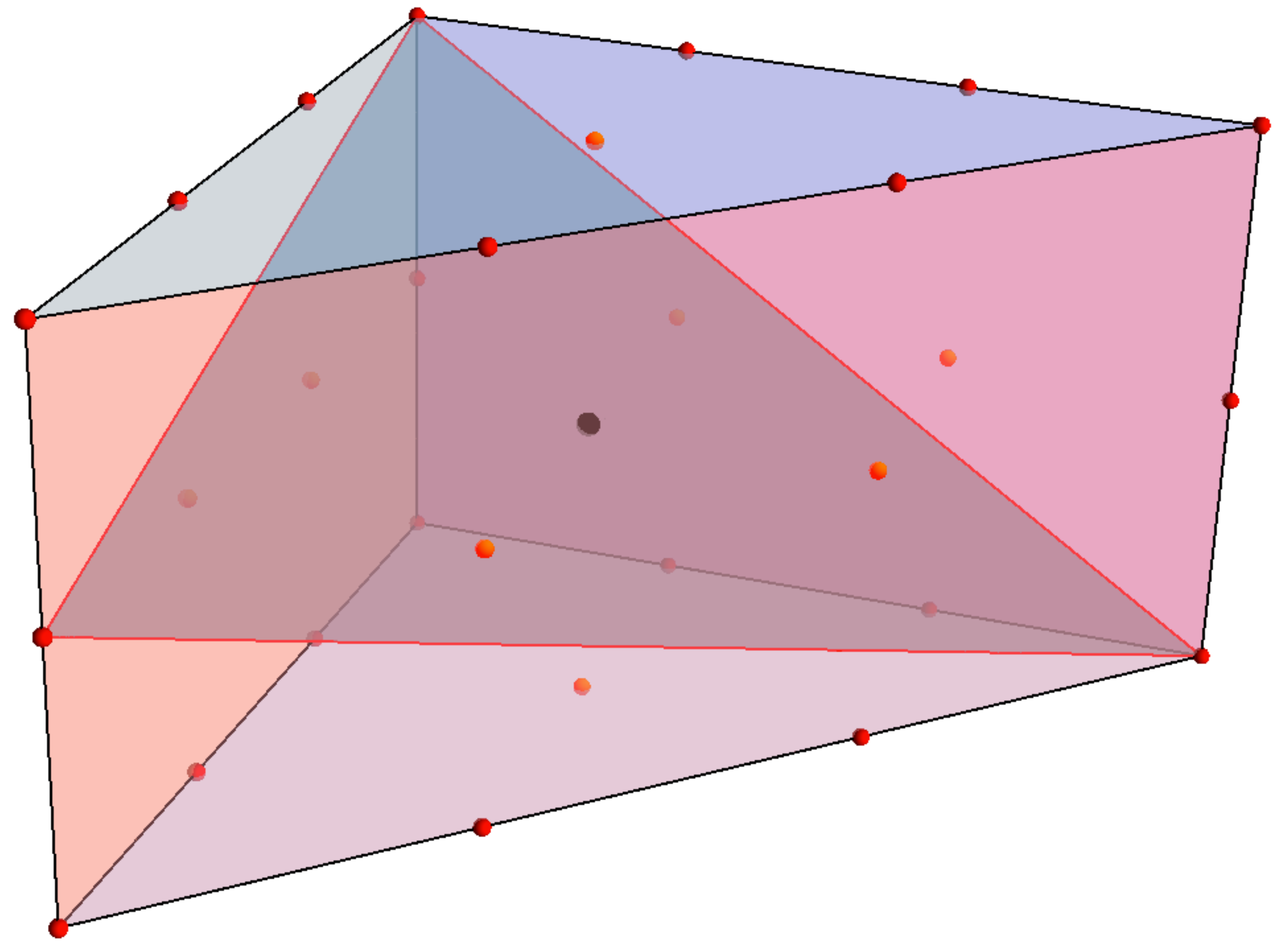}} 
\hfill \hspace*{10pt}
\raisebox{-0.95in}{\includegraphics[width=7.cm]{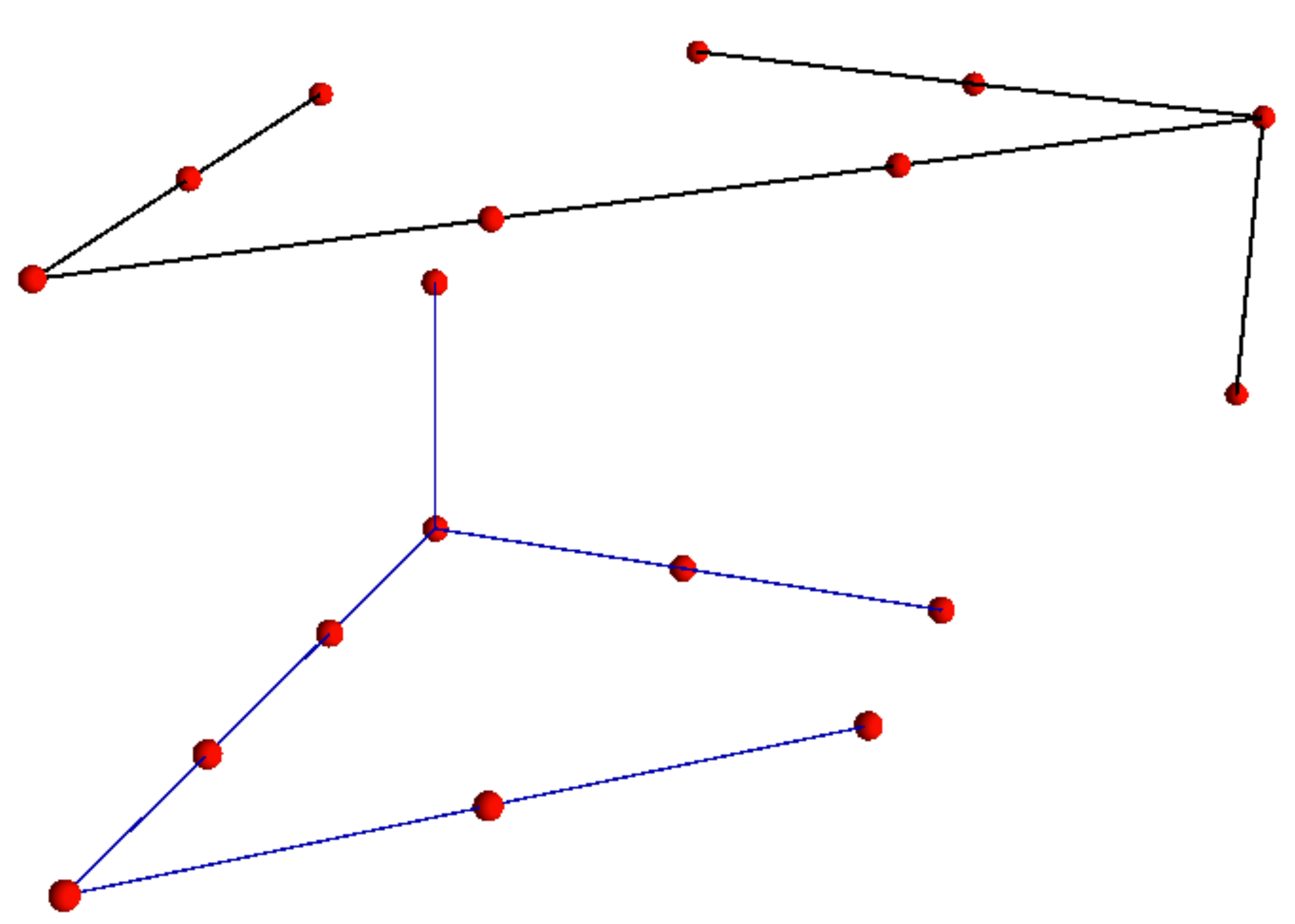}}
\hspace{20pt}
\vspace{10pt}
\end{minipage}
\capt{7in}{fig:fibration5}{The fifth fibration structure. The Dynkin diagrams correspond to $(E_8,E_8)$.}
\end{center}
\vskip -10pt
\end{figure}

\section{$K3$ surfaces with Picard number $18$ and two complex structures}\label{app:Pic18List}
\subsection*{The second $K3$}
The polyhedron $\nabla$ has $4$ vertices:
\beq\label{eq:verticesSecondK3}
w_i=\{ (\,6, ~2, ~3), ~(-6, ~2, ~3),~(\,0, -1, ~0),~ (\,0, ~0, -1)\}~.  
\eeq
This $K3$ admits two fibration structures, as indicated in Figure~\ref{fig:SecondK3}. The first fibration structure is of type $(E_8,E_8)$, while the second one is of type $(SO(32),\{1\})$, where $\{1\}$ denotes the trivial Lie group. This manifold appeared in earlier works, such as \cite{Candelas:1996su, Candelas:1997pq}. 

\begin{figure}[h]
\begin{center}
\hskip-20pt
\begin{minipage}[t]{5.2in}
\vspace{10pt}
\raisebox{-1in}{\includegraphics[width=5.9cm]{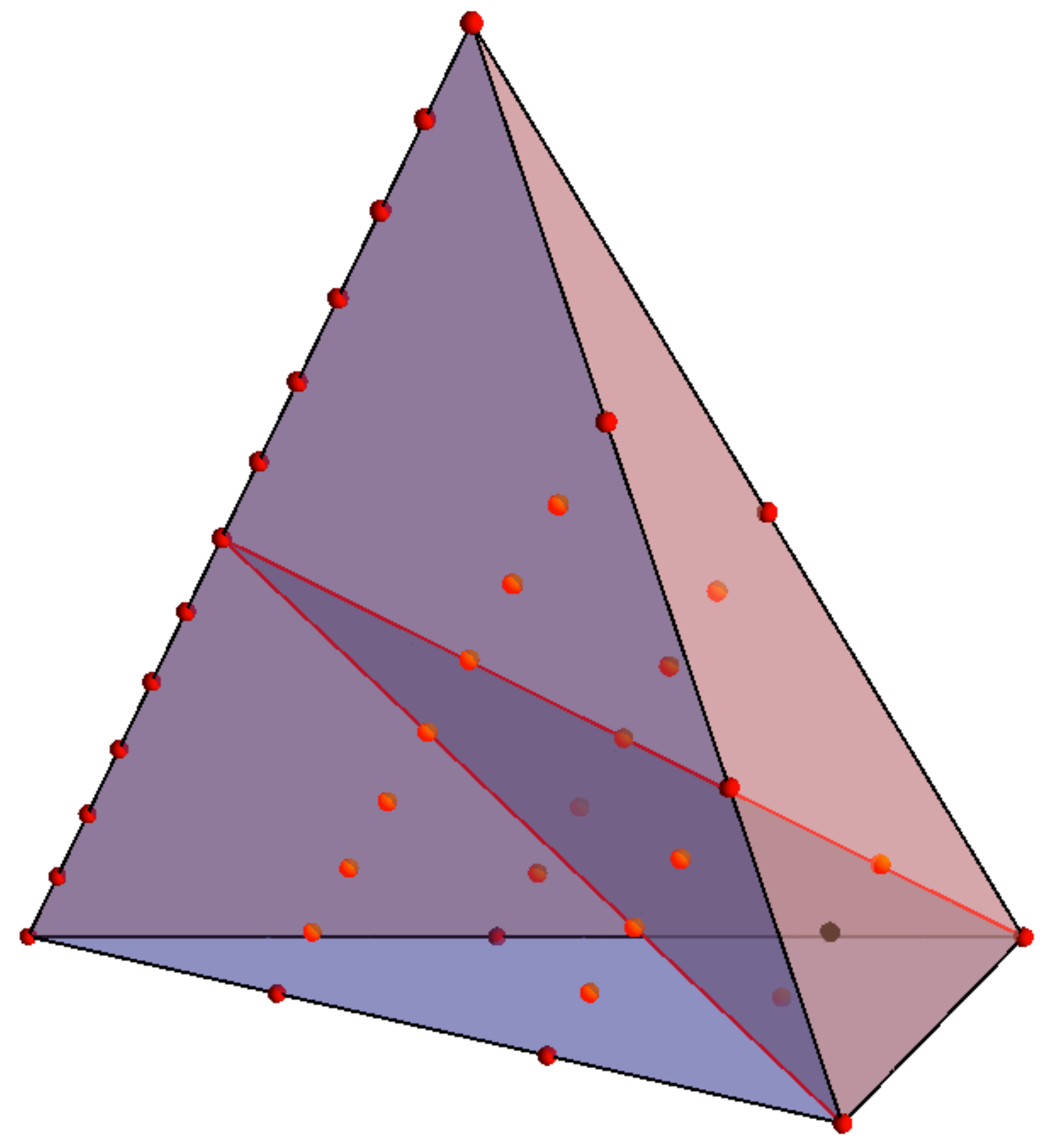}} 
\hfill \hspace*{40pt}
\raisebox{-0.95in}{\includegraphics[width=5.9cm]{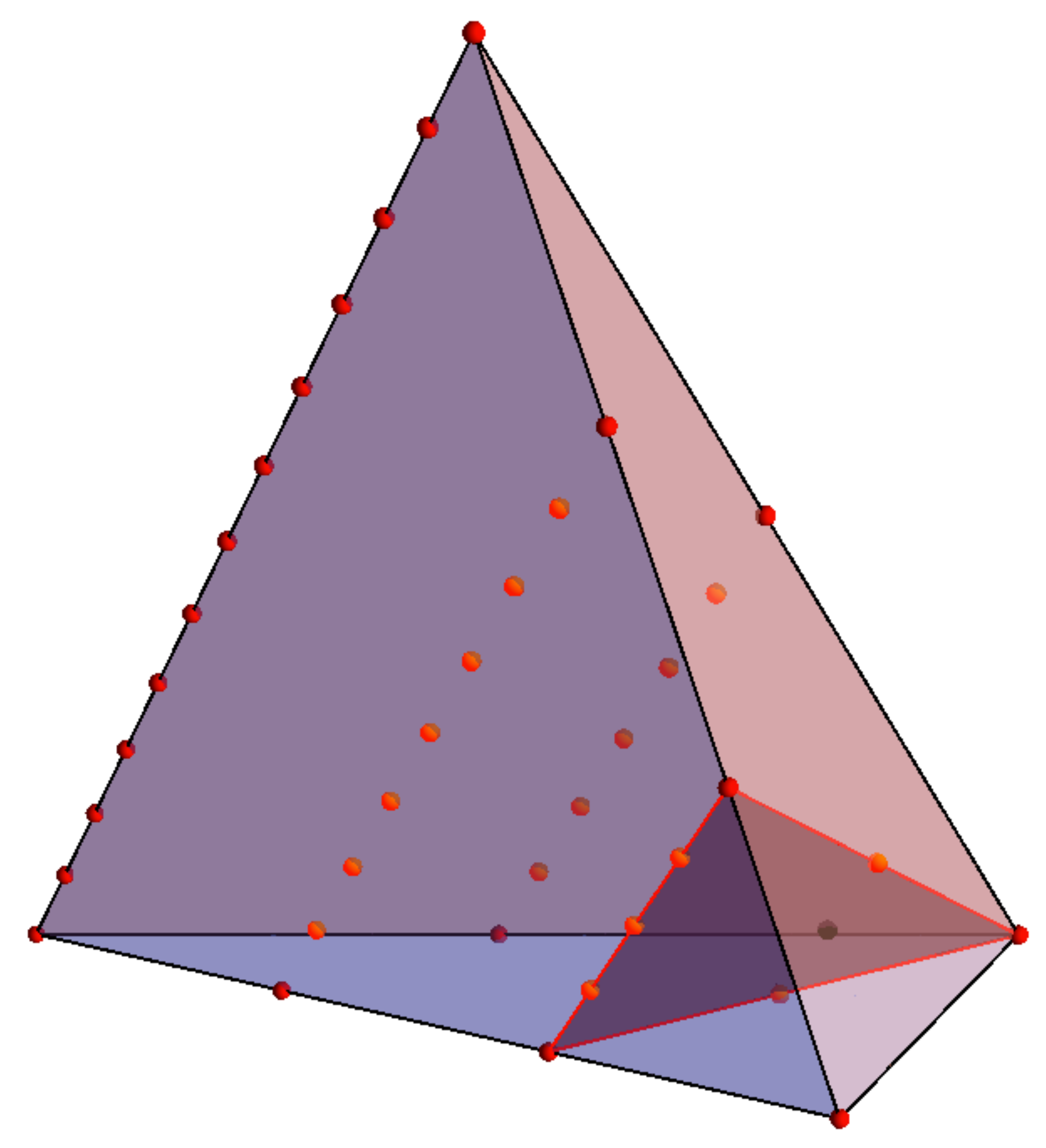}}
\hspace{20pt}
\vspace{10pt}
\end{minipage}
\capt{6.2in}{fig:SecondK3}{The second $K3$ surface. Both polyhedra correspond to $\nabla$ and indicate two different elliptic fibration structures, of types $E_8\times E_8$ and $SO(32)\times \{1\}$}
\end{center}
\vskip -10pt
\end{figure}

The dual polyhedron $\Delta$ has $9$ points in total, and $4$ vertices:
\beq
v_a=\{(-1,~1,~1),~(\,1,~1,~1),~(\,0,~1,-1),~(\,0,-2,~1)\}~. \label{dualcase2}
\eeq

Denoting by $z_1,z_2,z_3,z_4$ the homogeneous coordinates associated with the vertices of $\nabla$, in the order given in Eq.~\eqref{eq:verticesSecondK3}, we obtain the following polynomial:
\beq
f = - c_0\,z_1\, z_2\, z_3\, z_4 + c_1\,z_1^{12} + c_2\,z_1^6\, z_2^6 + c_3\,z_2^{12} + c_4\,z_3^3 + c_5\,z_4^2 + c_6\,z_1^4\, z_2^4\, z_3 + c_7\,z_1^2\, z_2^2\, z_3^2 + c_8\,z_1^3\, z_2^3\, z_4 ~.
\eeq
The monomials associated to points which are interior to the facets of $\Delta$ can be removed by a suitable change of coordinates, such that $c_6=c_7=c_8=0$, leaving a polynomial of generic form
\beq\label{eq:polynomialSecondK3app}
f = -c_0\,z_1\, z_2\, z_3\, z_4 + c_1\,z_1^{12} + c_2\,z_1^6\, z_2^6 + c_3\,z_2^{12} + c_4\,z_3^3 + c_5\,z_4^2 ~.
\eeq

\subsection*{The third $K3$}
The third $K3$ surface is given by the polyhedron $\nabla$ with vertices
\vspace{-4pt}
\beq\label{eq:nablaCase5}
w_i=\{(-1, -1, ~2),~ (-1, ~2, -1),~ (-1, ~2, ~2), ~(-1, -1, -1), ~(\,2, -1, -1)\}
\vspace{-4pt}
\eeq
and its dual $\Delta$, with vertices
\vspace{-4pt}
\beq
v_a=\{(\,0, ~0, ~1),~ (-1, -1, ~0), ~(\,1,~ 0,~ 0),~ (-1, ~0, -1),~ (\,0,~ 1,~ 0)\}~.   \label{dualcasethird}
\vspace{-4pt}
\eeq
%This $K3$ surface admits $6$ different elliptic fibration structures.
%
This pair of polyhedra leads to the following defining polynomial:
\beq
f = -c_0\,z_1\, z_2\, z_3\, z_4\, z_5+c_1\,z_1^3\, z_3^3 + c_2\,z_1^3\, z_4^3+c_3\,z_2^3\, z_3^3 + c_4\,z_2^3\, z_4^3 + c_5\,z_5^3
\eeq
The manifold admits $5$ different elliptic fibration structures, depicted below.

\begin{figure}[h]
\begin{center}
\hskip-24pt
\begin{minipage}[t]{5.2in}
\vspace{10pt}
\raisebox{-1in}{\includegraphics[width=5.2cm]{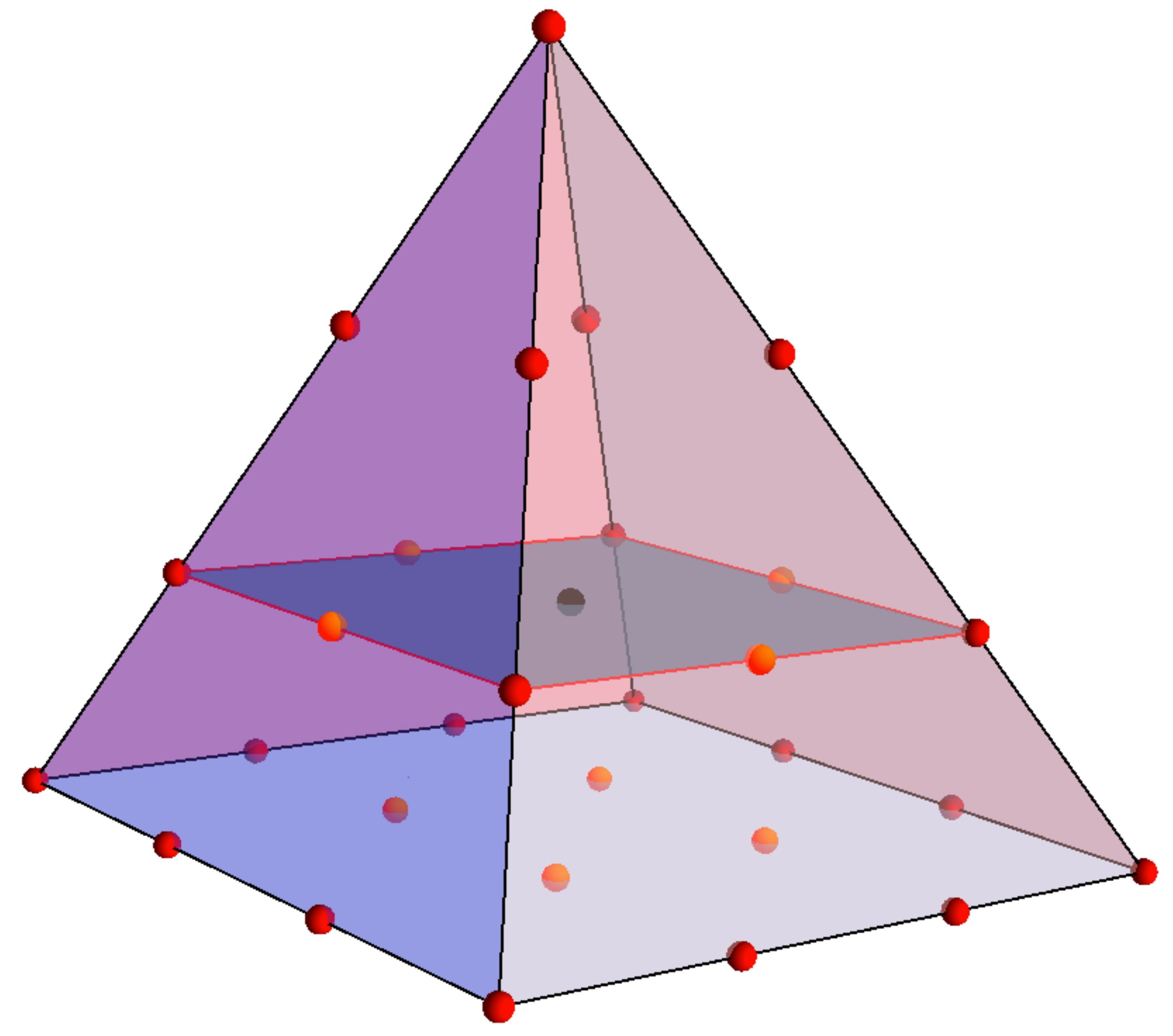}} 
\hfill \hspace*{30pt}
\raisebox{-0.95in}{\includegraphics[width=5.2cm]{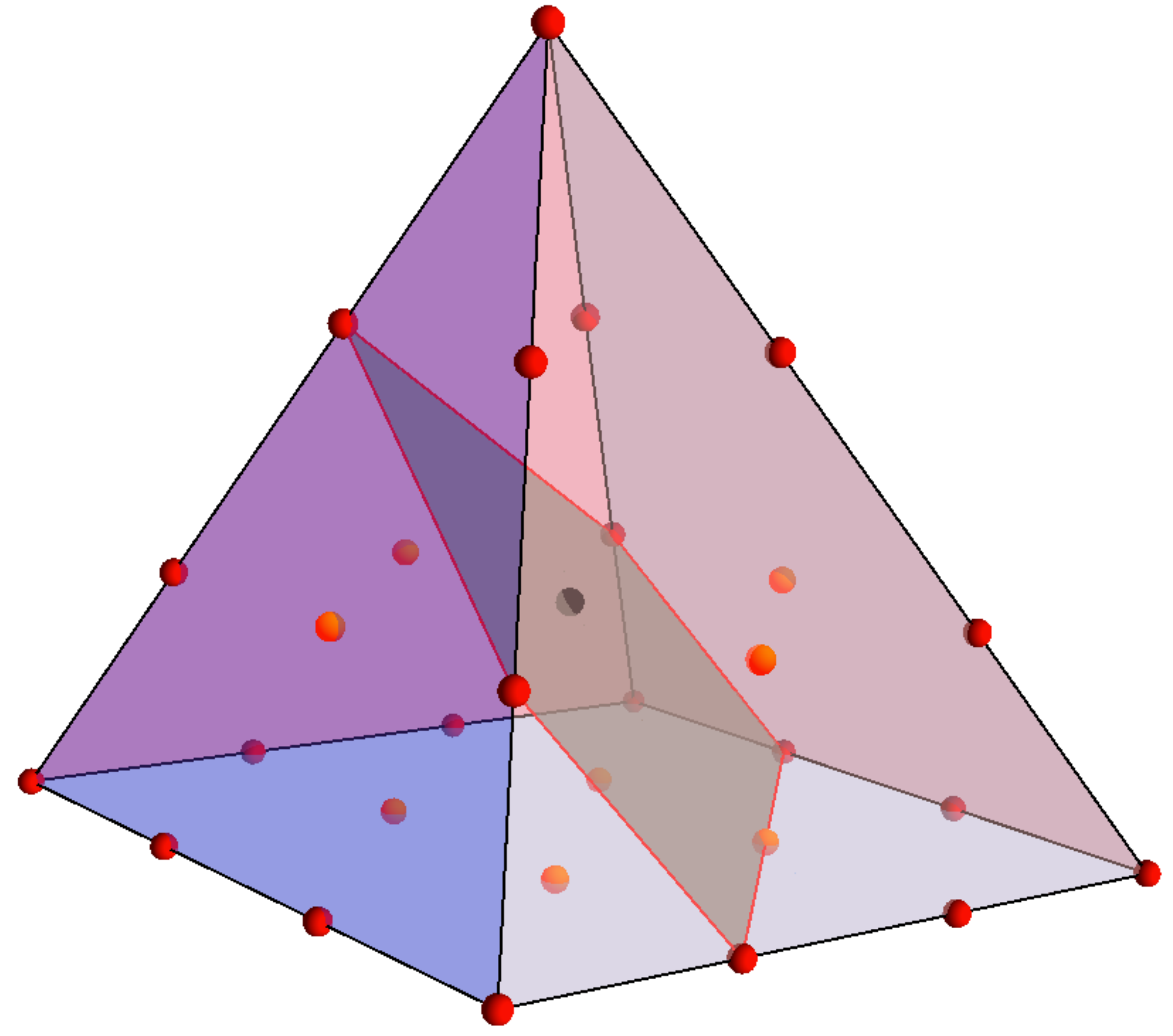}}
\hspace{20pt}
\vspace{10pt}
\end{minipage}
\end{center}
\vskip -20pt
\end{figure}

\begin{figure}[h]
\begin{center}
\hskip-14pt
\begin{minipage}[t]{5.2in}
\vspace{-10pt}
\raisebox{-1in}{\includegraphics[width=5.2cm]{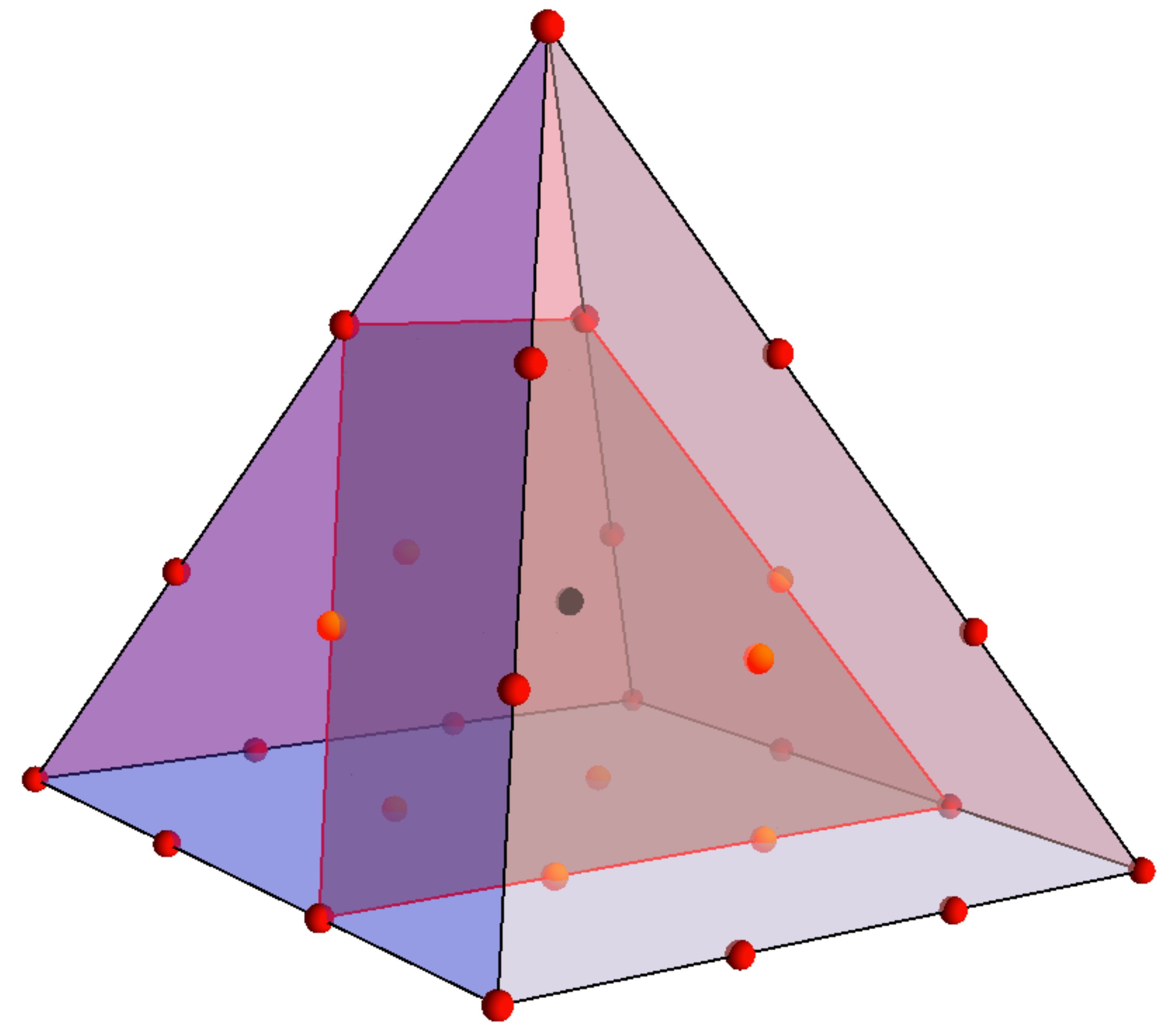}} 
\hfill \hspace*{30pt}
\raisebox{-0.95in}{\includegraphics[width=5.2cm]{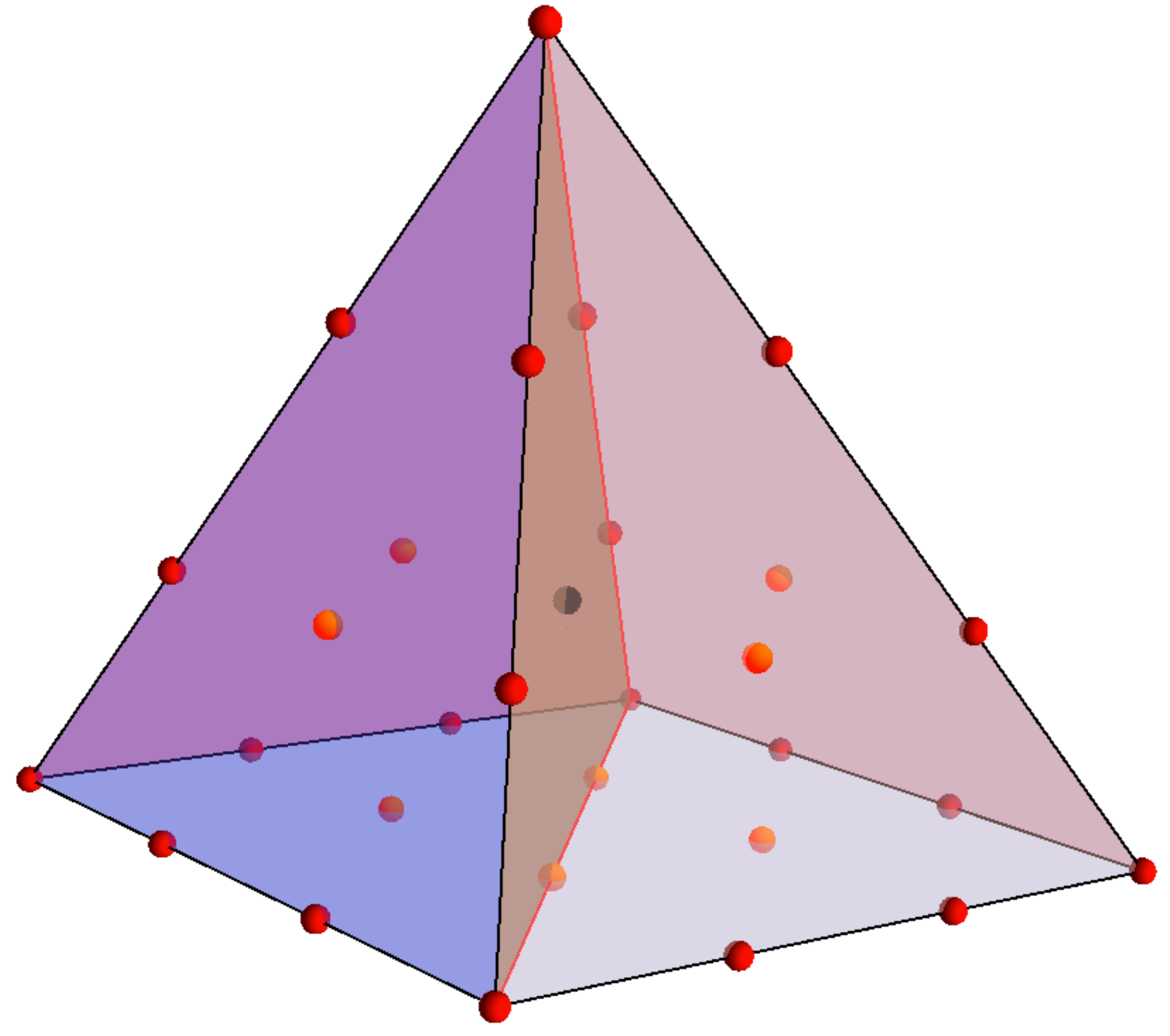}}
\hspace{20pt}
\vspace{10pt}
\end{minipage}
\end{center}
\vskip -50pt
\end{figure}

\begin{figure}[H]
\begin{center}
\hskip70pt
\begin{minipage}[t]{3.4in}
\vspace{-3pt}
\raisebox{-1in}{\includegraphics[width=5.2cm]{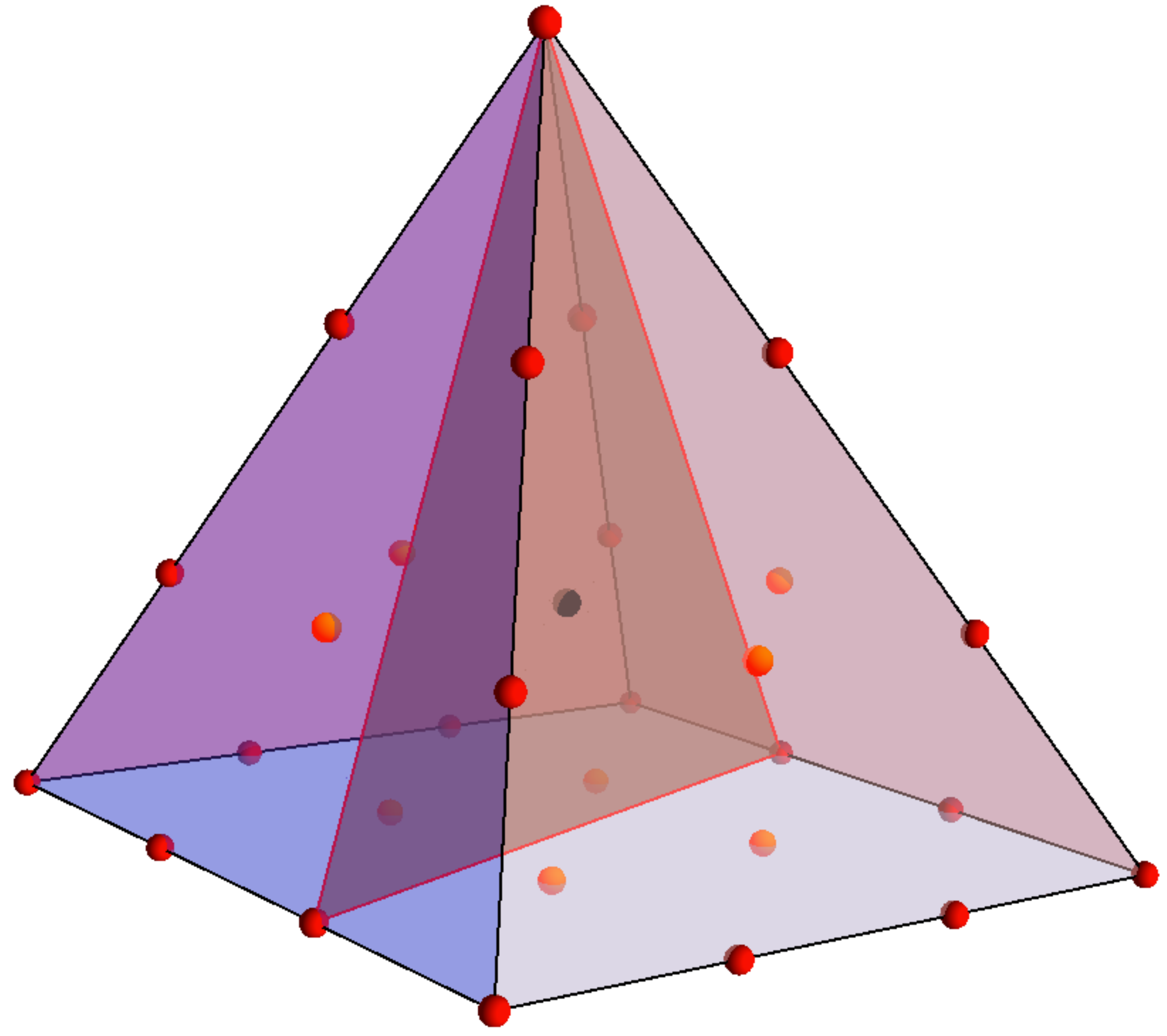}} 
\hfill \hspace*{30pt}
%\raisebox{-0.95in}{\includegraphics[width=7.cm]{Case4NablaFib6.pdf}}
\hspace{20pt}
\vspace{0pt}
\end{minipage}
\capt{6.2in}{fig:FifthK3}{The polyhedron $\nabla$ for the fifth $K3$ surface. This $K3$ admits five different elliptic fibration structures, of types $SO(8)\times SU(12)$, $SO(14)\times E_7$, $SU(10)\times SO(14)$, $E_6\times E_6$ and $E_8\times E_8$, from top left to bottom.}
\end{center}
\end{figure}

\newpage
\subsection*{The fourth $K3$}
The fourth $K3$ surface is given by a polyhedron $\nabla$ with vertices
\beq
w_i=\{(-1, -1, -1), ~(\,0, -1, ~2), ~(-1, ~4, -1), (\,1, ~0, -1),~ (-1, -1,~ 4),~ (\,0, -1, -1)\}
\eeq
and its dual $\Delta$, with vertices
\beq
v_a=\{(-2, -1, -1), ~(-1,~ 1,~ 0),~ (\,0, ~1,~ 0), ~(\,1,~ 0,~ 0), ~(\,0, ~0, ~1)\}~.
\eeq
Disregarding the points interior to the facets of $\Delta$, we obtain the following defining polynomial: 
\beq
f = -c_0\,z_1\, z_2\, z_3\, z_4\, z_5\,z_6+c_1\, z_1^5\, z_6^3 + c_2\,z_2^3\, z_5^5+c_3\,z_3^5\, z_4 + c_4\,z_1 \,z_3^6\, z_5 + c_5\,z_2\, z_4^2\, z_6
\eeq
This $K3$ surface admits $4$ different elliptic fibration structures.

\begin{figure}[h]
\begin{center}
\hskip-90pt
\begin{minipage}[t]{5.2in}
\vspace{20pt}
\raisebox{-1in}{\includegraphics[width=7.7cm]{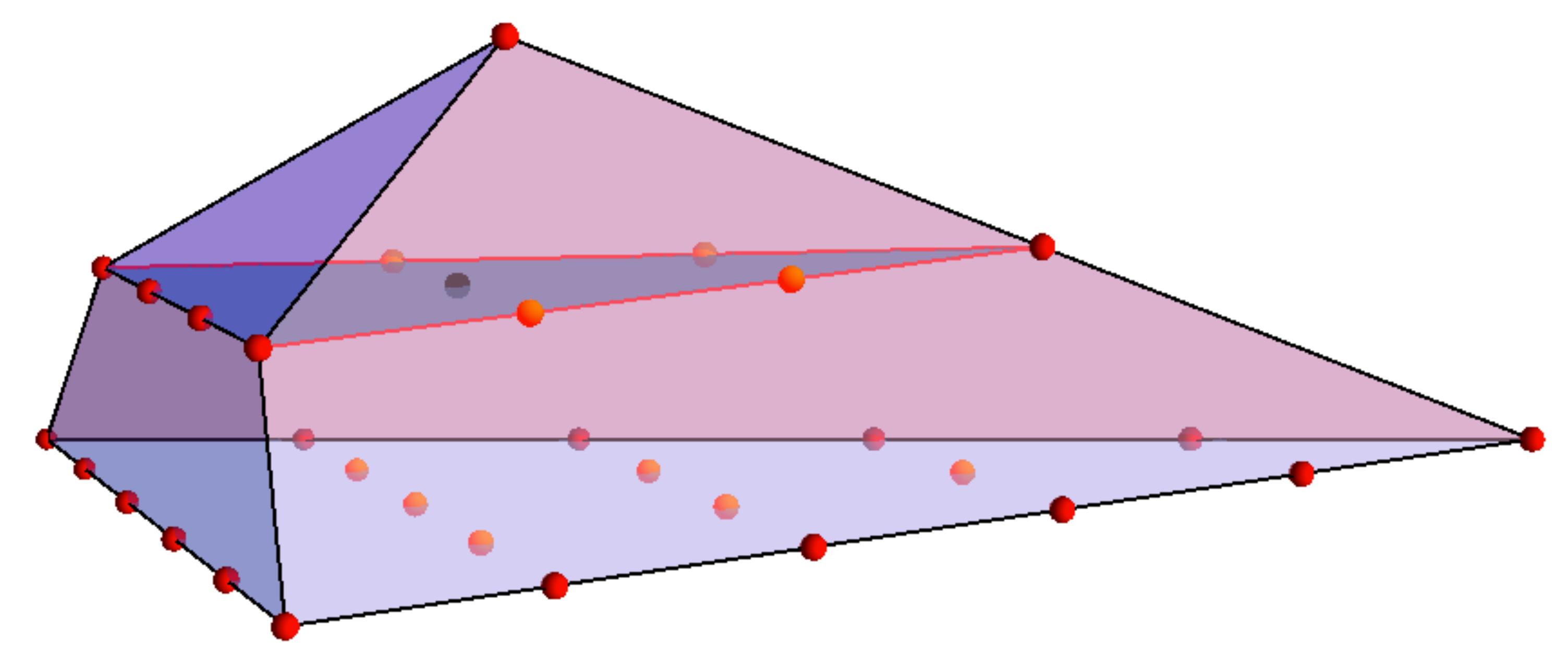}} 
\hfill \hspace*{21pt}
\raisebox{-0.95in}{\includegraphics[width=7.7cm]{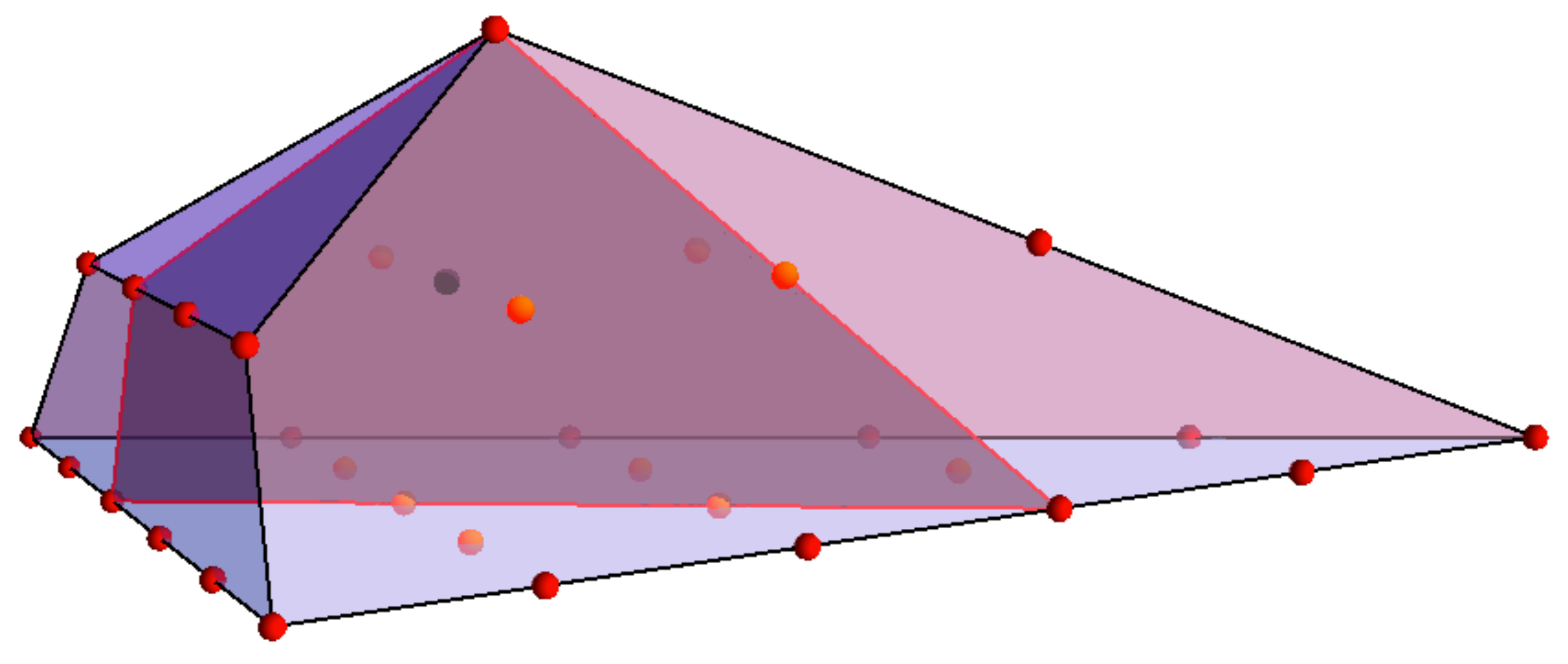}}
\hspace{20pt}
\vspace{10pt}
\end{minipage}
%\capt{5.7in}{fig:SecondK3}{The second $K3$ surface. Both polyhedra correspond to $\nabla$ and indicate two different elliptic fibration structures.}
\end{center}
\vskip -20pt
\end{figure}

\begin{figure}[h]
\begin{center}
\hskip-90pt
\begin{minipage}[t]{5.2in}
\vspace{10pt}
\raisebox{-1in}{\includegraphics[width=7.7cm]{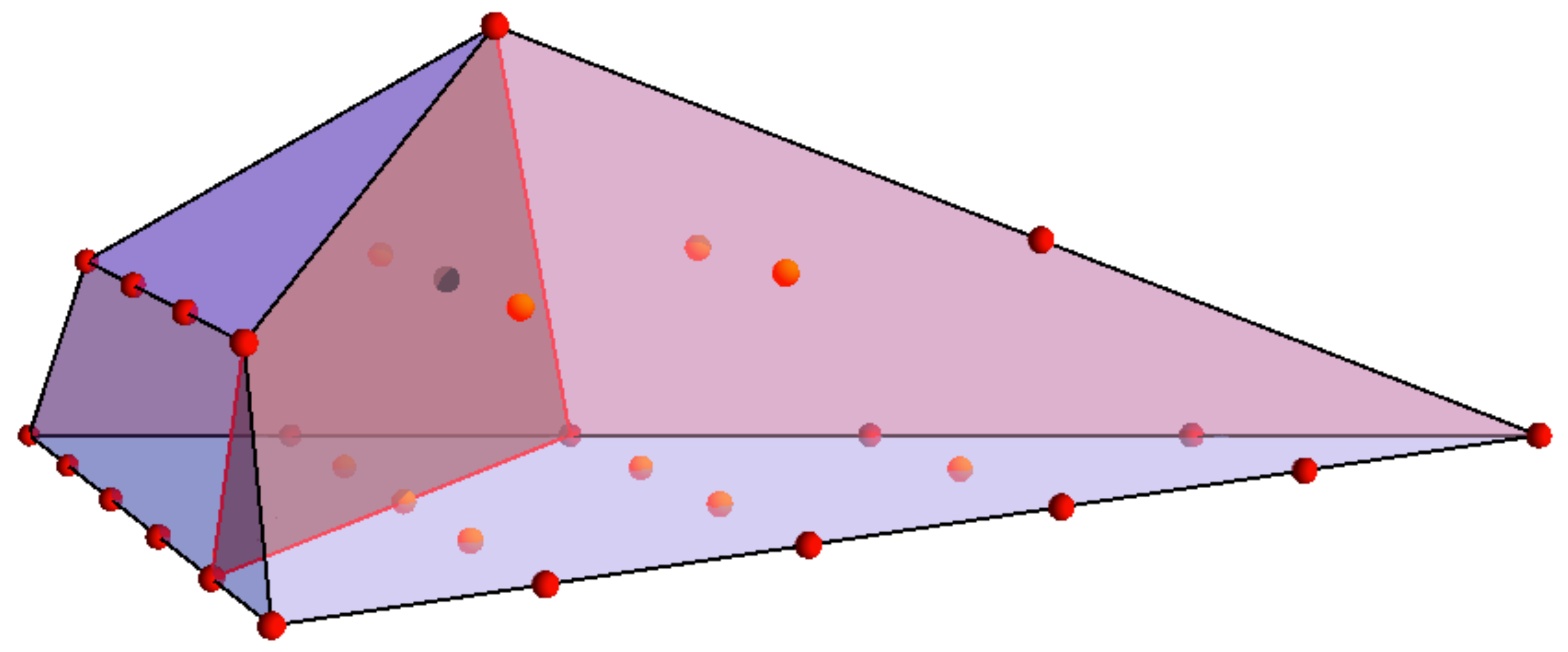}} 
\hfill \hspace*{21pt}
\raisebox{-0.95in}{\includegraphics[width=7.7cm]{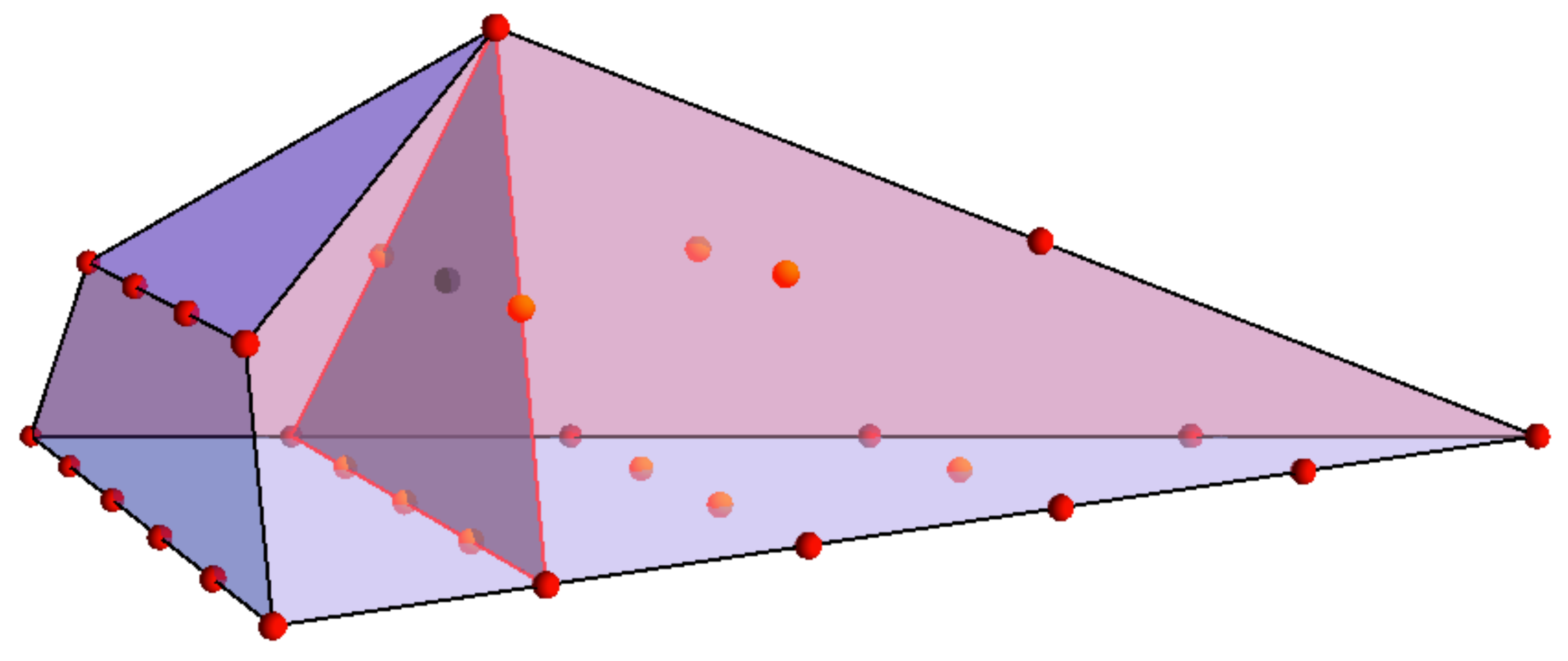}}
\hspace{20pt}
\vspace{10pt}
\end{minipage}
\capt{6.2in}{fig:ThirdK3}{The polyhedron $\nabla$ for the third $K3$ surface. This $K3$ admits four different elliptic fibration structures, of types $\{1\}\times SU(15)$, $E_6\times SO(18)$, $E_7\times E_8$ and $E_7\times SU(10)$, from top left to bottom right.}
\end{center}
\vskip -10pt
\end{figure}

\subsection*{The fifth $K3$}
The fifth $K3$ surface is given by the polyhedron $\nabla$ with vertices
\beq
w_i=\{(-1, -1, ~1), ~(-1, -1, -1), ~(-1, ~1, -1), ~(\,1,-1, ~3), ~(\,1, -1, -1),~ (\,1, ~3, -1)\}
\eeq
and its dual $\Delta$, with vertices
\beq
v_a=\{(-1, ~0,~ 0),~ (\,0,~ 0,~ 1),~ (\,0, ~1,~ 0), ~(\,1,~ 0,~ 0),~ (\,1, -1, -1)\}~.
\eeq
The pair of dual polyhedra $(\nabla, \Delta)$ leads, in this case, to the following defining polynomial:
\beq
f = -c_0\,z_1\, z_2\, z_3\, z_4\, z_5\,z_6+c_1\, z_1^2\, z_4^4 + c_2\,z_2^2\, z_5^4+c_3\,z_3^2\, z_6^4 + c_4\,z_1^2\, z_2^2\, z_3^2 + c_5\,z_4^2\, z_5^2\, z_6^2
\eeq
This $K3$ surface admits $6$ different elliptic fibration structures.

\begin{figure}[H]
\begin{center}
\hskip-70pt
\begin{minipage}[t]{5.2in}
\vspace{12pt}
\raisebox{-1in}{\includegraphics[width=7.3cm]{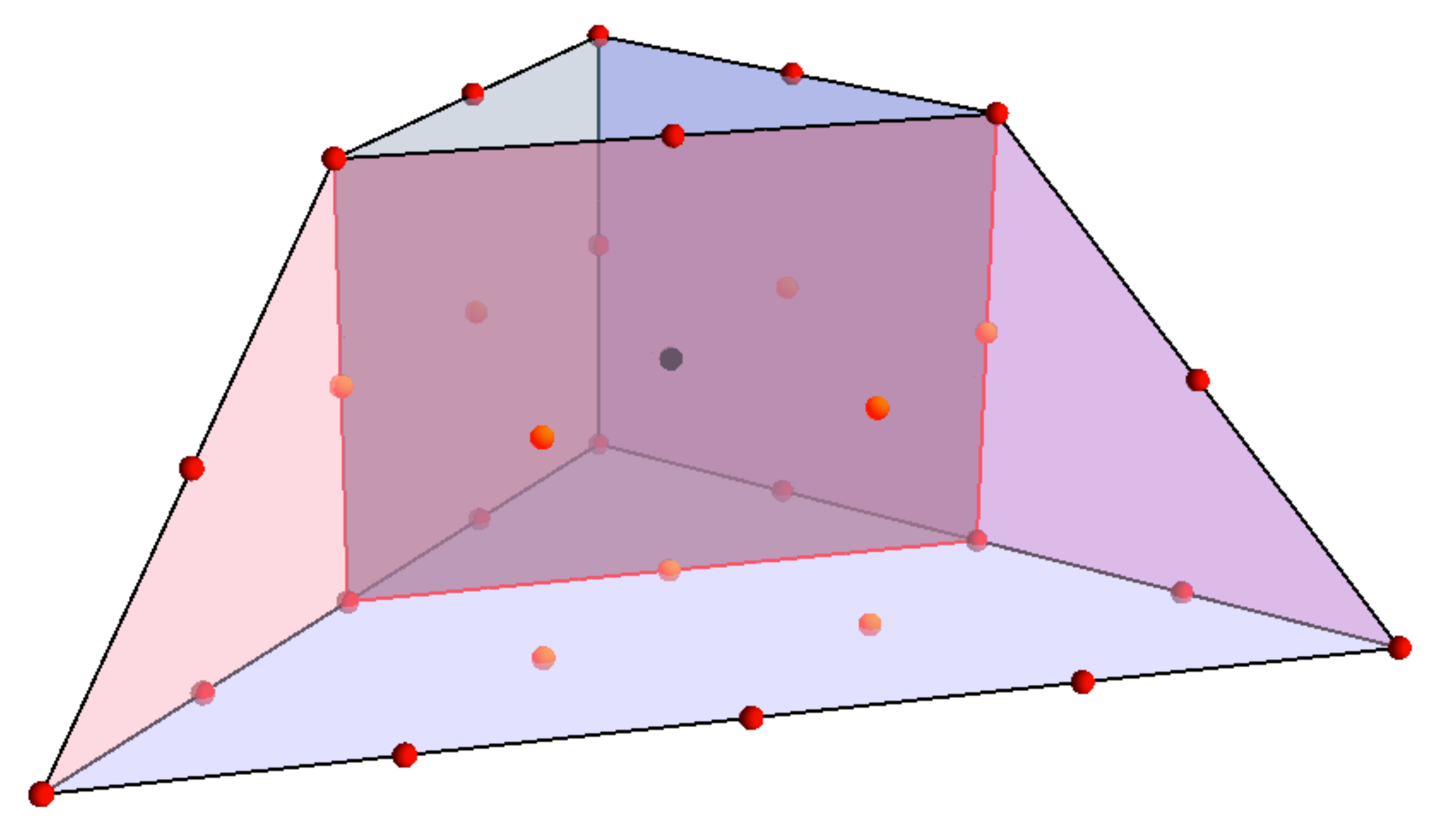}} 
\hfill \hspace*{21pt}
\raisebox{-0.95in}{\includegraphics[width=7.3cm]{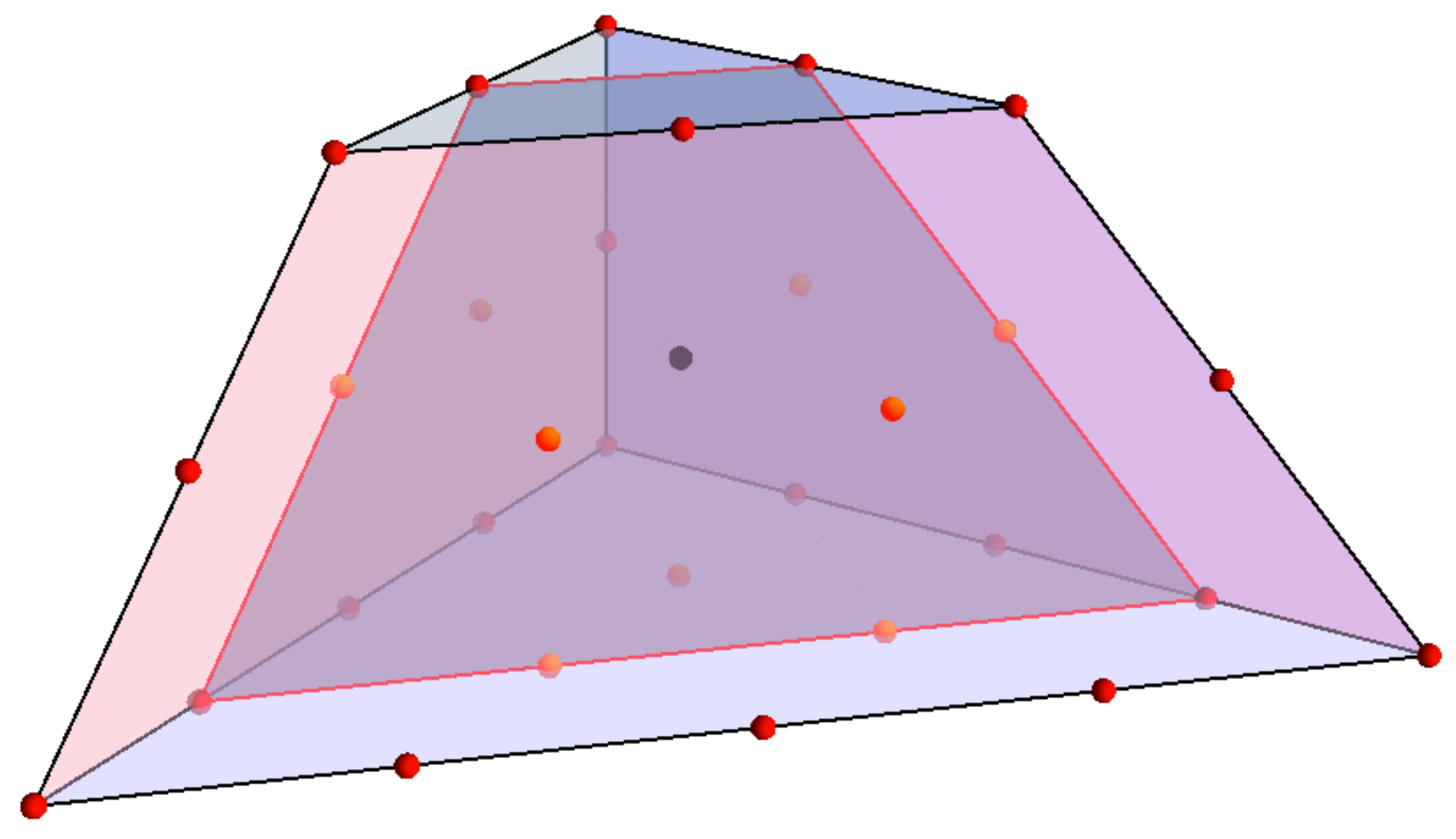}}
\hspace{20pt}
\vspace{10pt}
\end{minipage}
\end{center}
\vskip -24pt
\end{figure}

\begin{figure}[h]
\begin{center}
\hskip-70pt
\begin{minipage}[t]{5.2in}
\vspace{10pt}
\raisebox{-1in}{\includegraphics[width=7.3cm]{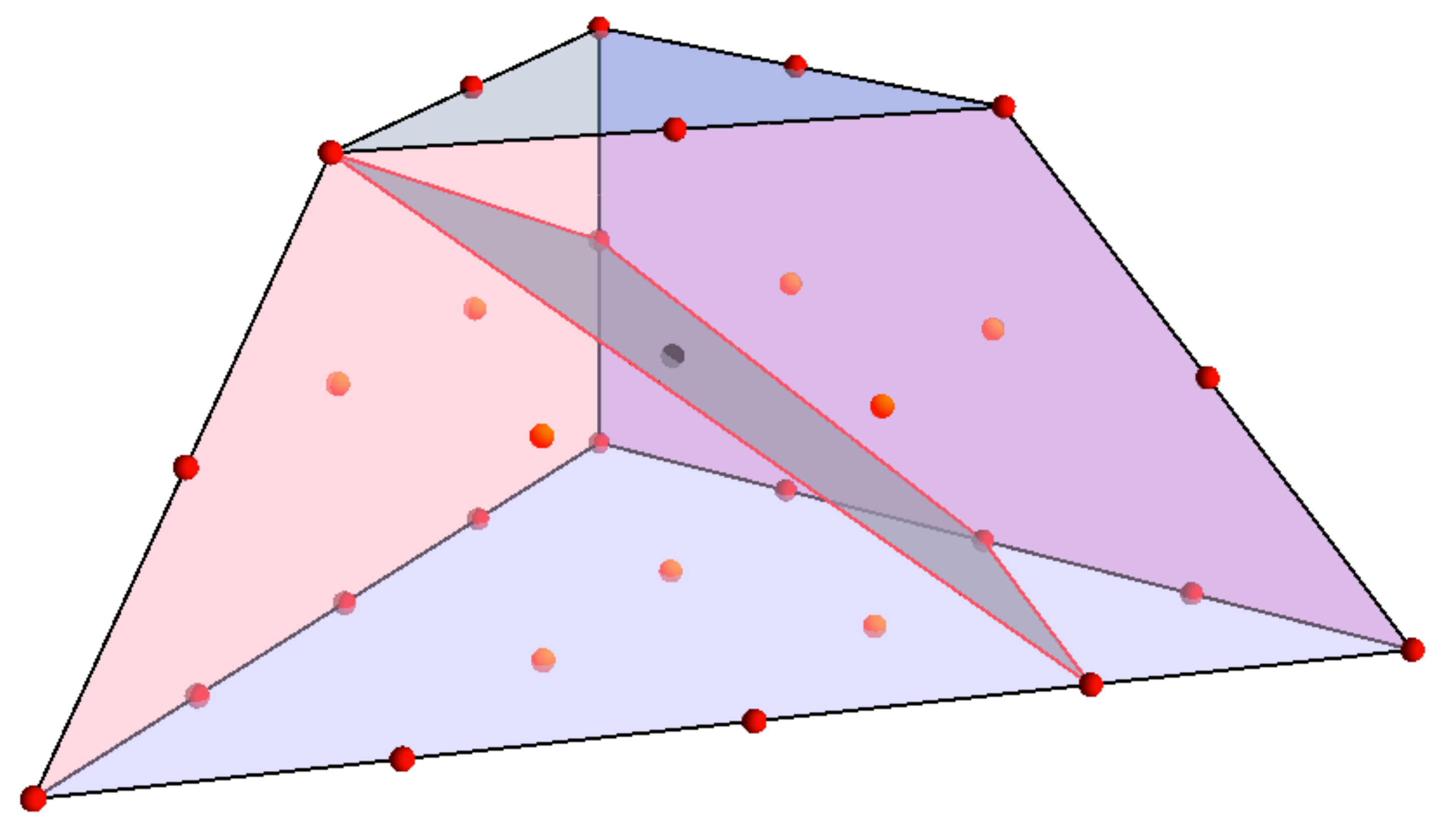}} 
\hfill \hspace*{21pt}
\raisebox{-0.95in}{\includegraphics[width=7.3cm]{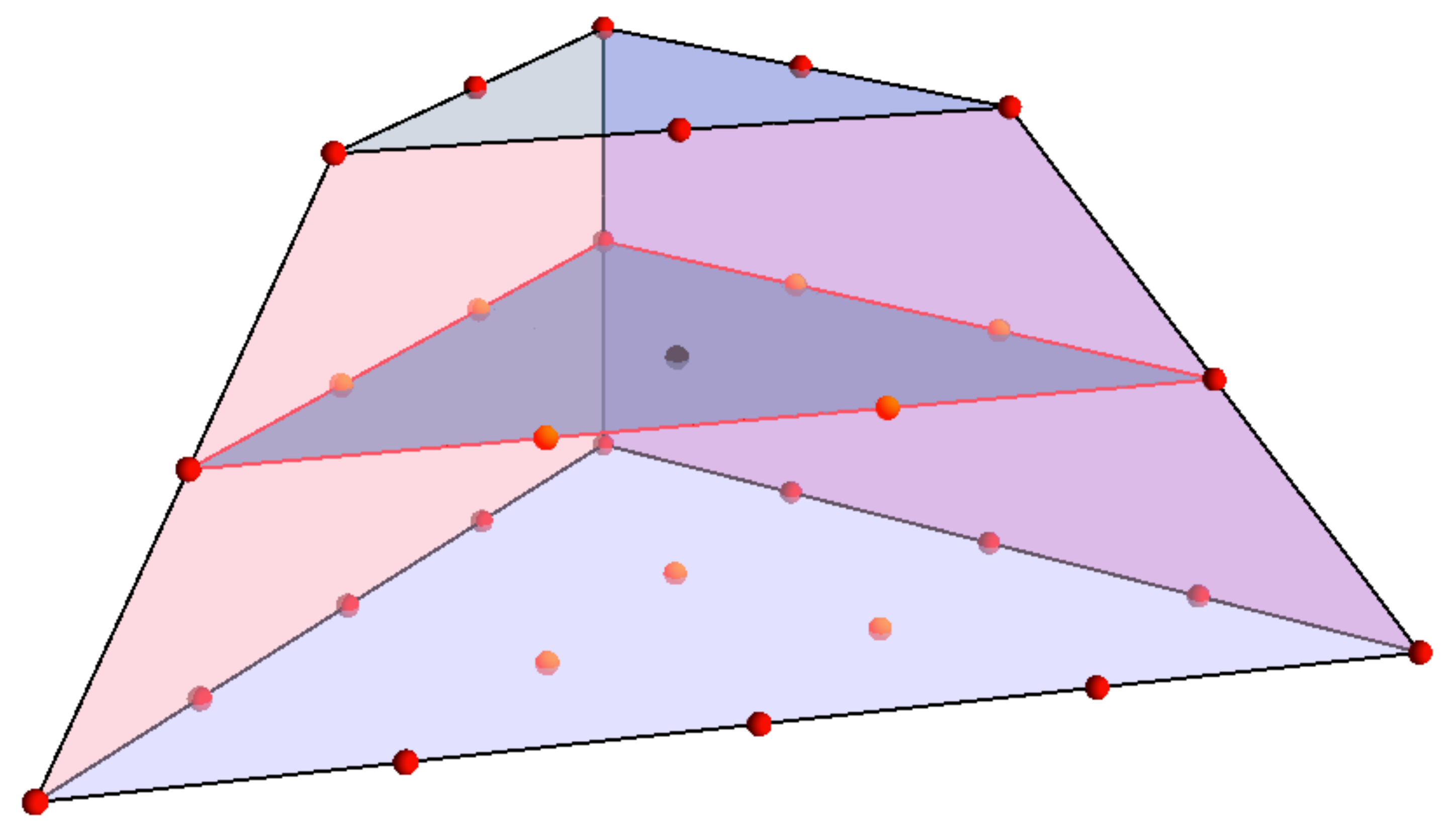}}
\hspace{20pt}
\vspace{10pt}
\end{minipage}
\end{center}
\vskip -24pt
\end{figure}

\begin{figure}[H]
\begin{center}
\hskip-70pt
\begin{minipage}[t]{5.2in}
\vspace{10pt}
\raisebox{-1in}{\includegraphics[width=7.3cm]{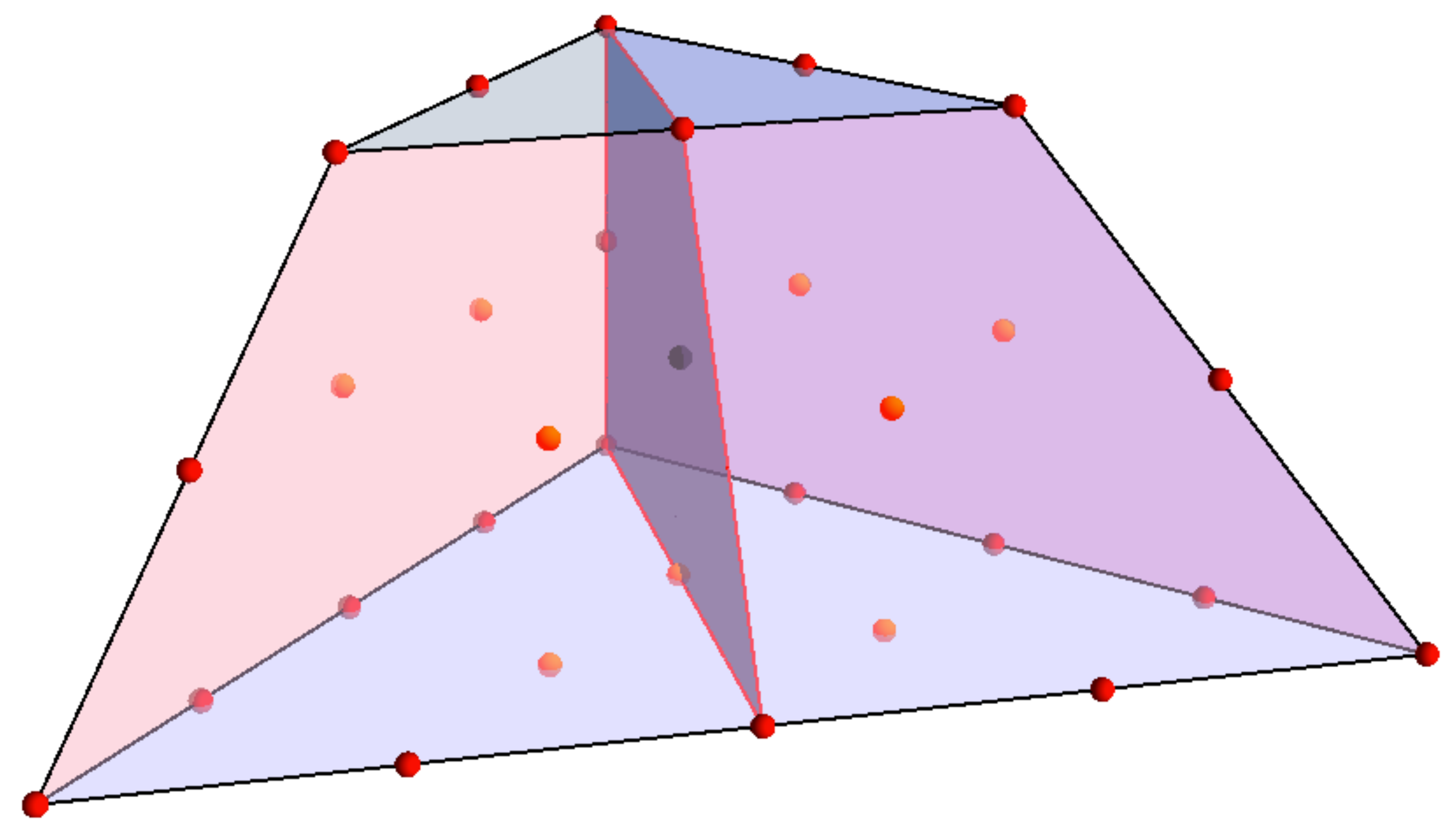}} 
\hfill \hspace*{21pt}
\raisebox{-0.95in}{\includegraphics[width=7.3cm]{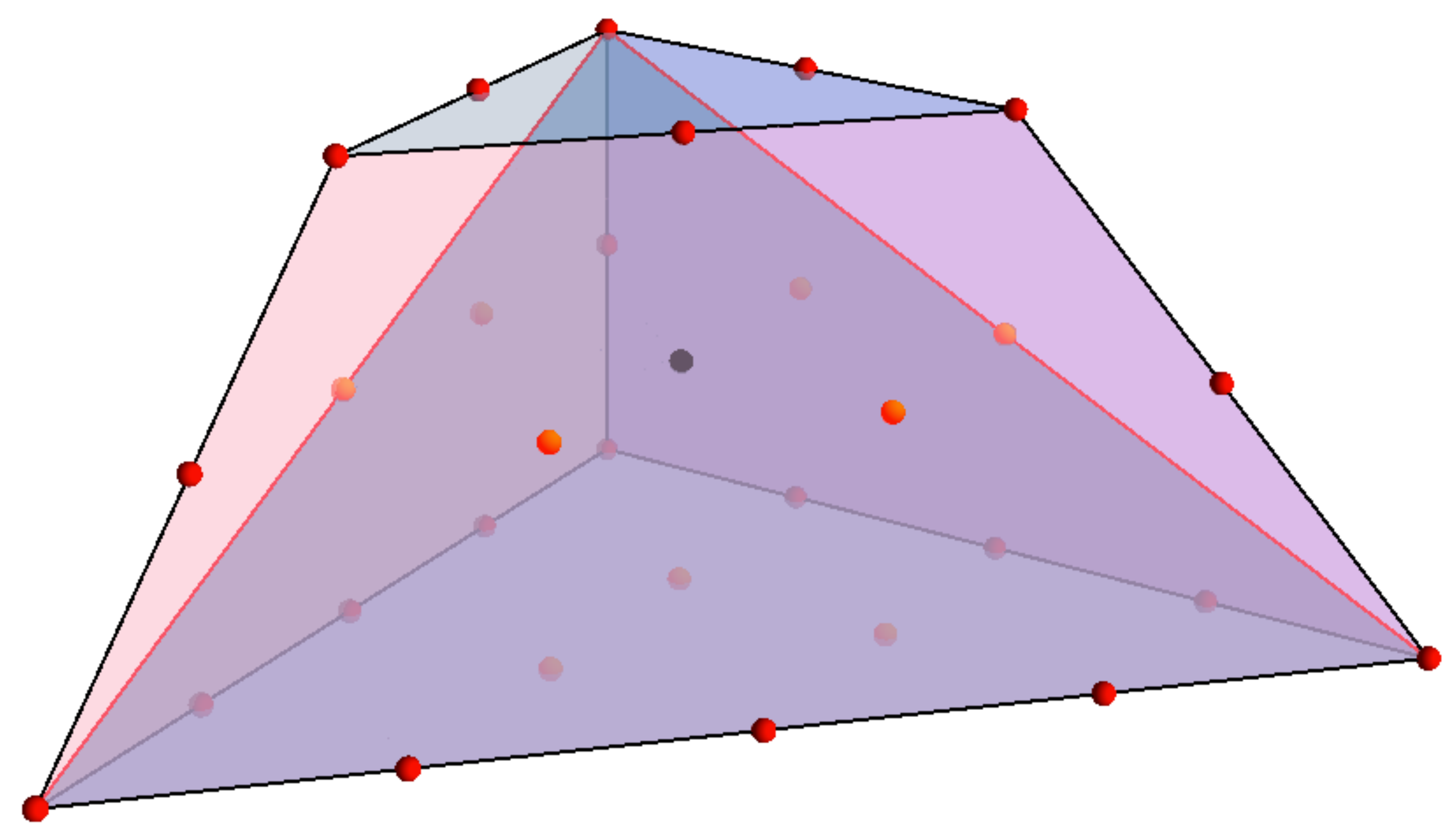}}
\hspace{20pt}
\vspace{10pt}
\end{minipage}
\capt{6.2in}{fig:FourthK3}{The polyhedron $\nabla$ for the fourth $K3$ surface. This $K3$ admits six different elliptic fibration structures, of types $SO(16)\times SO(12)$, $E_6\times SO(10)$, $E_8\times E_7$, $SU(12)\times SU(6)$, $E_7\times E_7$ and $E_7\times SO(12)$, from top left to bottom right.}
\end{center}
\vskip 0pt
\end{figure}

\subsection*{The sixth $K3$}
The sixth $K3$ surface is given by the polyhedron $\nabla$ with vertices
\beq
w_i=\{(-1, -1, ~0),~ (-1, -1, -1), ~(-1, ~2, ~0), ~(-1, ~2, -1), ~(\,2, -1,~ 3), ~(\,2, -1, -1)\}
\eeq
and its dual $\Delta$, with vertices
\beq
v_a=\{(-1, -1, ~0), ~(\,0, ~1, ~0), ~(\,1, ~0, -1), ~(\,1, ~0, ~0),~ (\,0, ~0, ~1)\}~.
\eeq
The pair of dual polyhedra $(\nabla, \Delta)$ leads, in this case, to the following defining polynomial:
\beq
f = -c_0\,z_1\, z_2\, z_3\, z_4\, z_5\,z_6+c_1\, z_1^3\, z_2^3 + c_2\,z_3^3\, z_4^3+c_3\,z_5^3\, z_6^3 + c_4\,z_1 \,z_3\, z_5^4 + c_5\, z_2\, z_4\, z_6^4
\eeq
This manifold admits $7$ different elliptic fibration structures, as displayed below.

\begin{figure}[h]
\begin{center}
\hskip80pt
\begin{minipage}[t]{4.4in}
\vspace{10pt}
\raisebox{-1in}{\includegraphics[width=7.3cm]{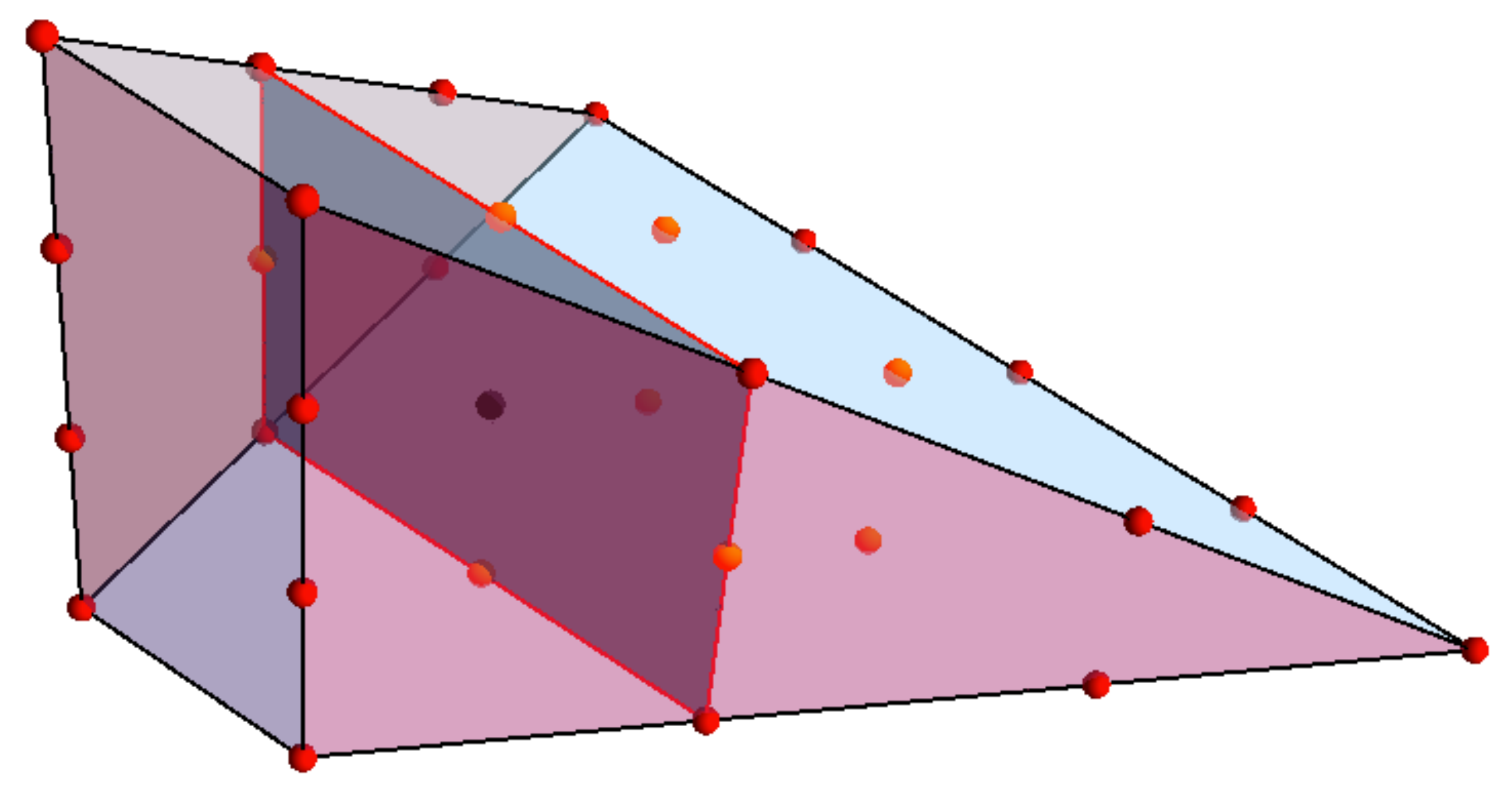}} 
\hspace{20pt}
\vspace{10pt}
\end{minipage}
\end{center}
\vskip -38pt
\end{figure}

\begin{figure}[h]
\begin{center}
\hskip-70pt
\begin{minipage}[t]{5.2in}
\vspace{10pt}
\raisebox{-1in}{\includegraphics[width=7.3cm]{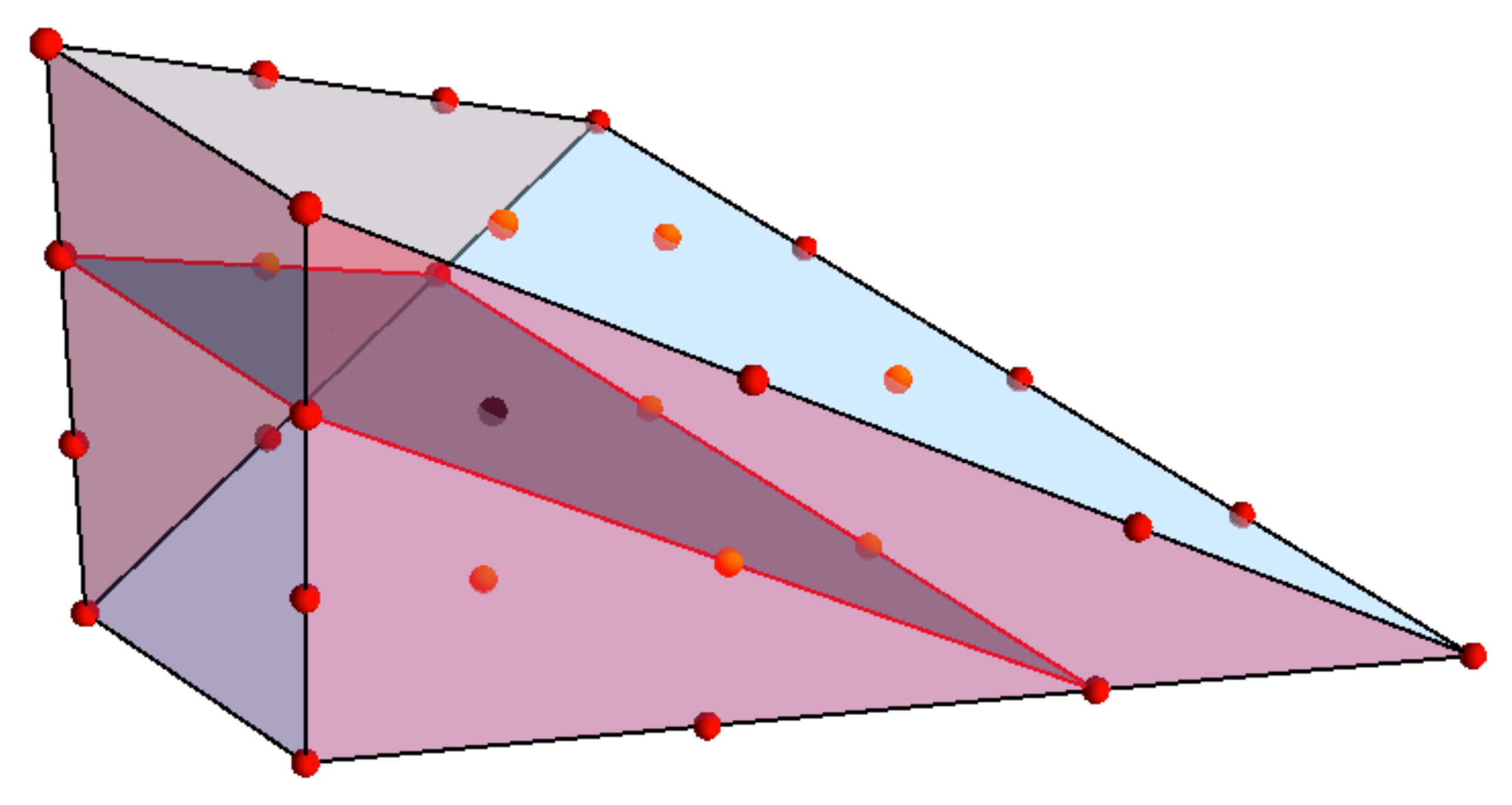}} 
\hfill \hspace*{24pt}
\raisebox{-0.95in}{\includegraphics[width=7.3cm]{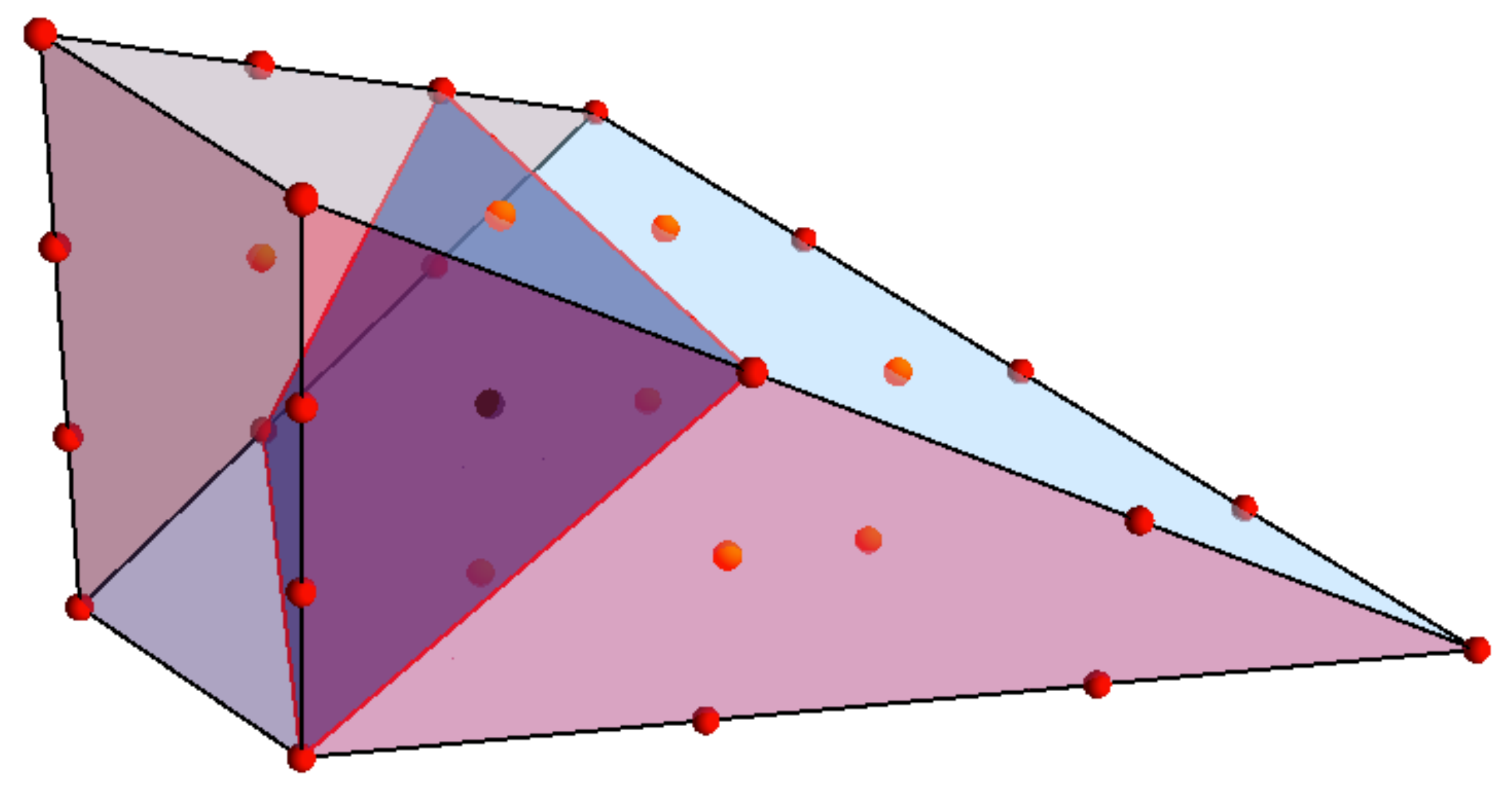}}
\hspace{20pt}
\vspace{10pt}
\end{minipage}
\end{center}
\vskip -30pt
\end{figure}

\begin{figure}[h]
\begin{center}
\hskip-70pt
\begin{minipage}[t]{5.2in}
\vspace{10pt}
\raisebox{-1in}{\includegraphics[width=7.3cm]{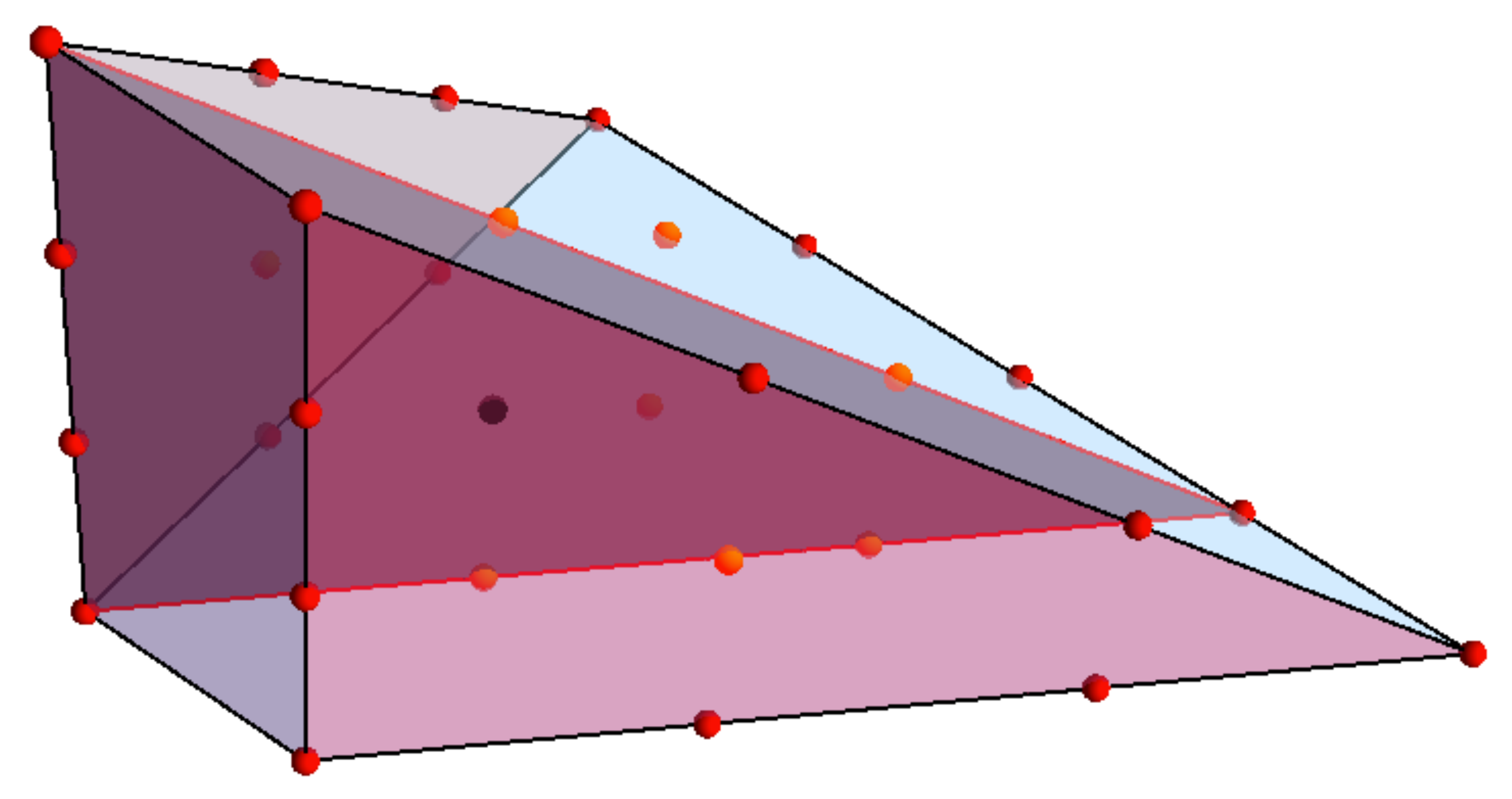}} 
\hfill \hspace*{24pt}
\raisebox{-0.95in}{\includegraphics[width=7.3cm]{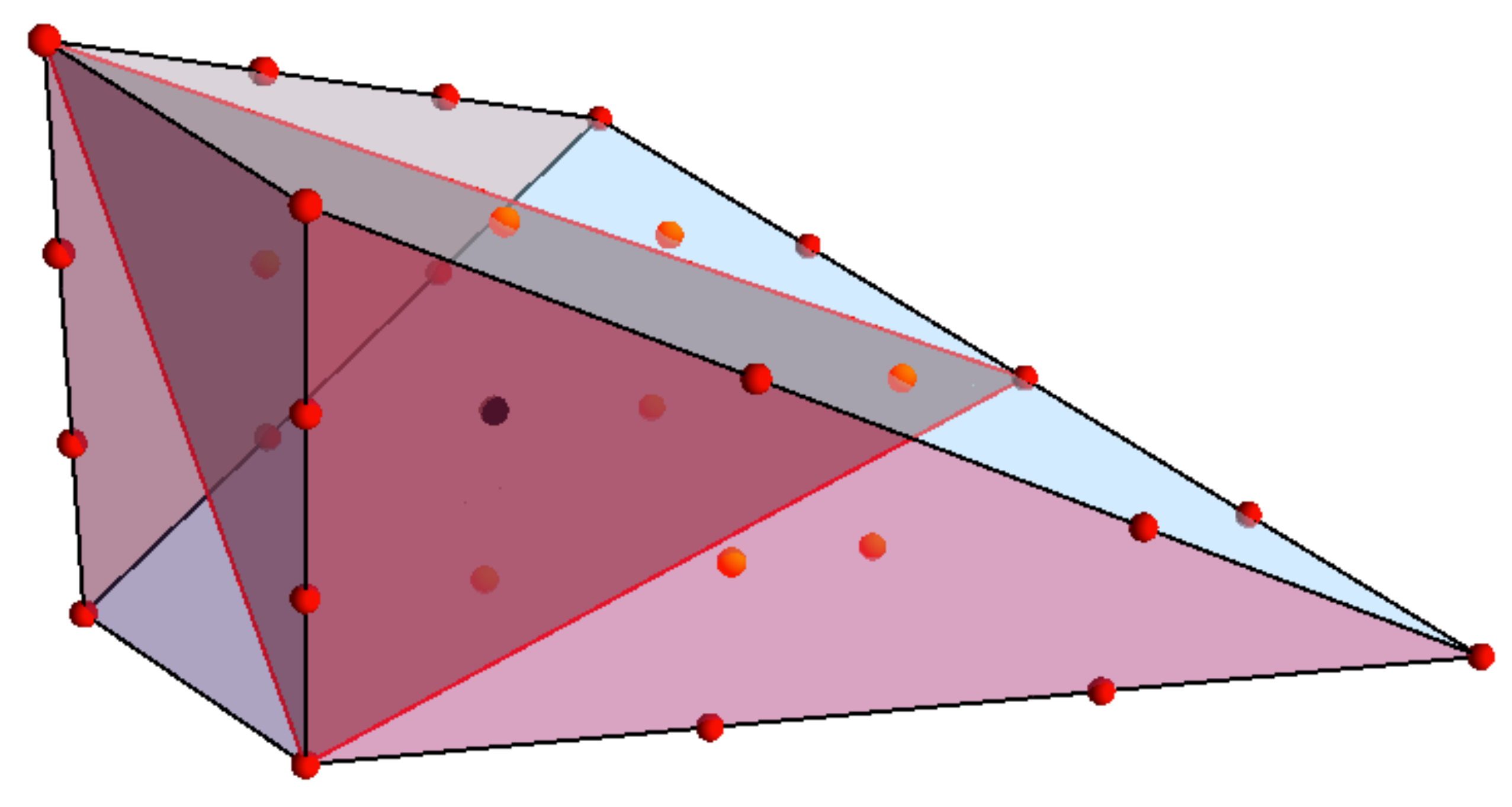}}
\hspace{20pt}
\vspace{10pt}
\end{minipage}
\end{center}
\vskip -30pt
\end{figure}

\begin{figure}[H]
\begin{center}
\hskip-70pt
\begin{minipage}[t]{5.2in}
\vspace{10pt}
\raisebox{-1in}{\includegraphics[width=7.3cm]{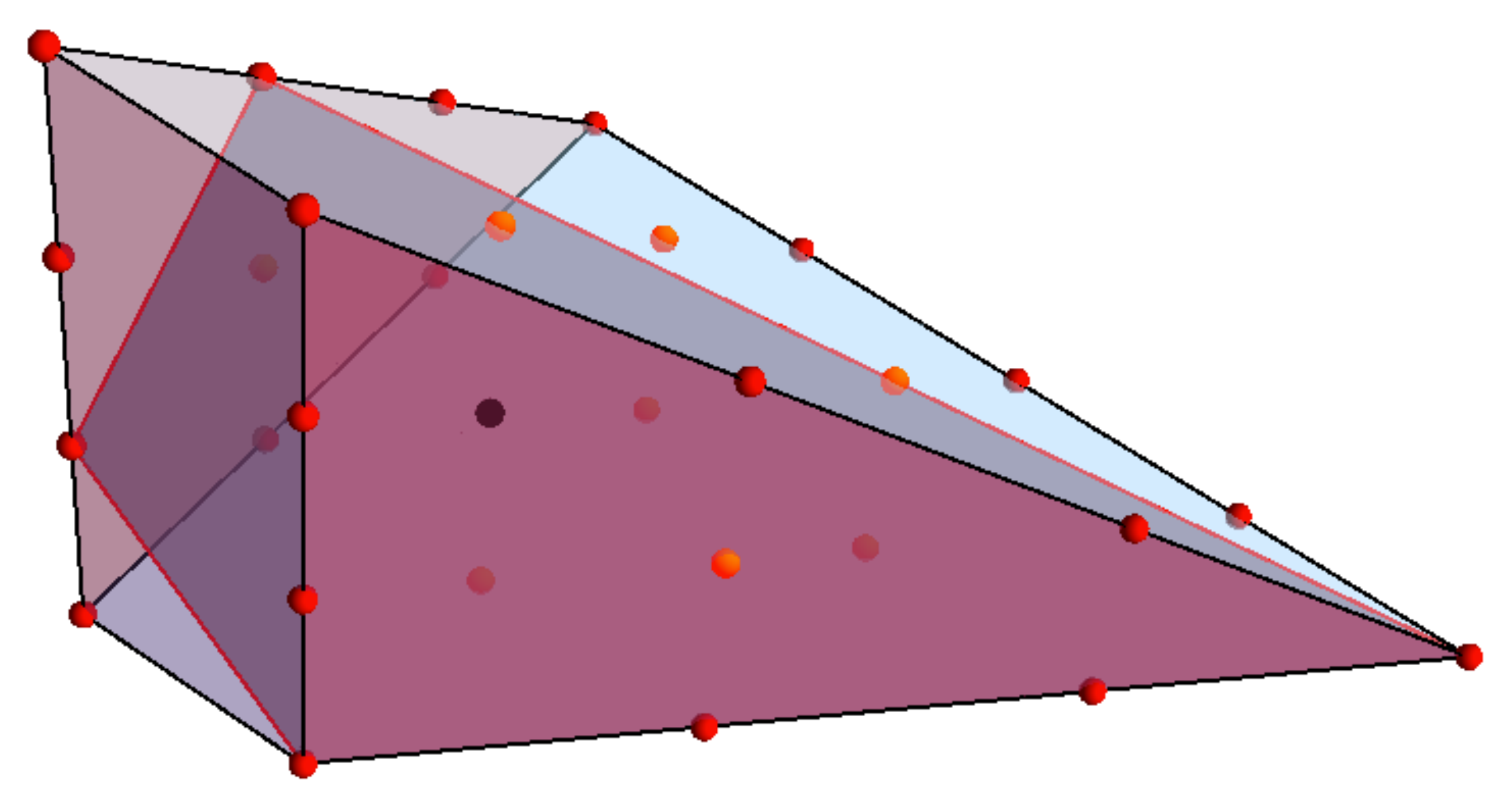}} 
\hfill \hspace*{24pt}
\raisebox{-0.95in}{\includegraphics[width=7.3cm]{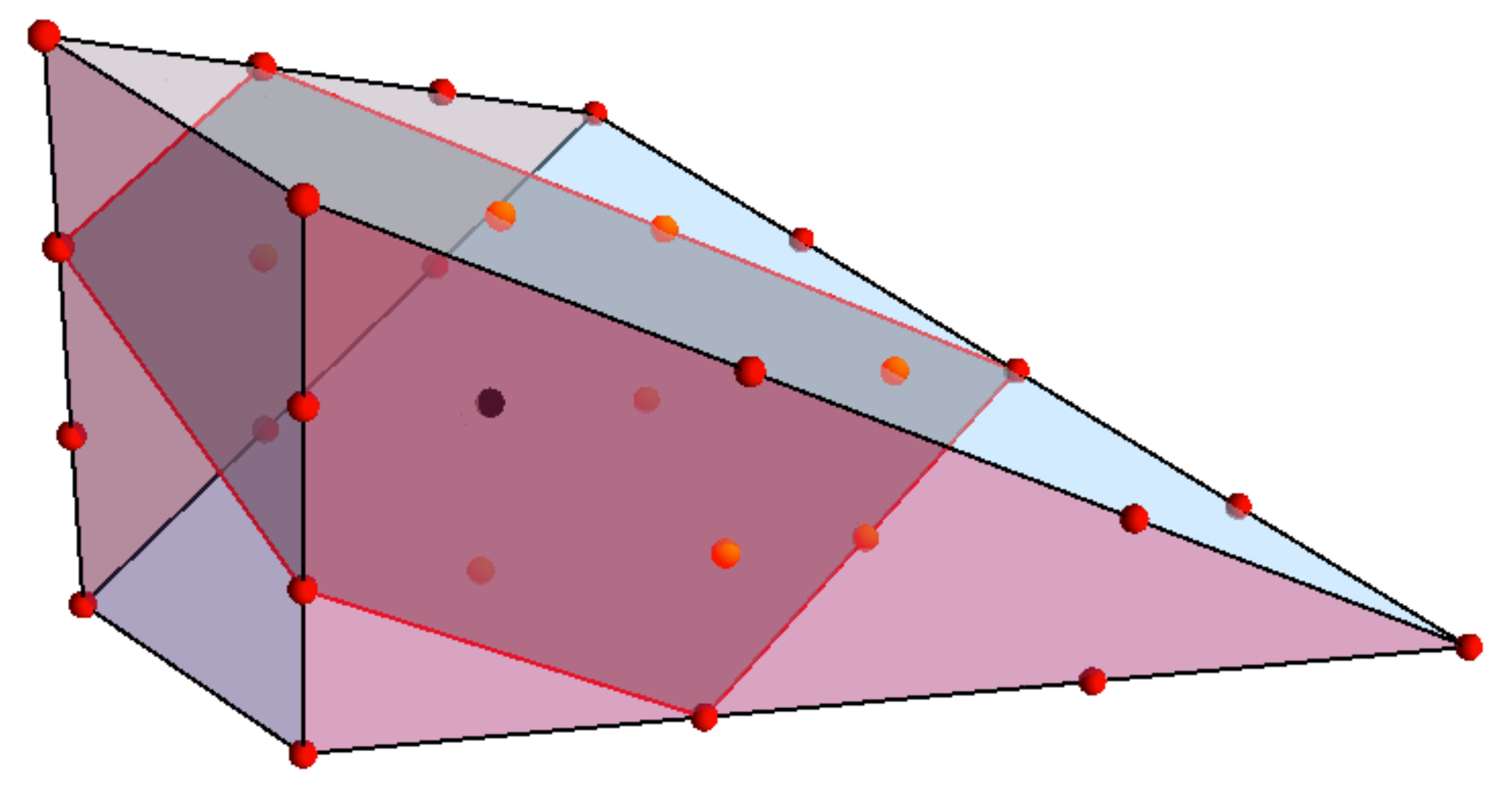}}
\hspace{20pt}
\vspace{8pt}
\end{minipage}
\capt{6.2in}{fig:SixthK3}{The polyhedron $\nabla$ for the fifth $K3$ surface. This $K3$ admits five different elliptic fibration structures, of types $SO(16)\times SU(8)$, $SO(10)\times SU(11)$, $E_8\times E_7$, $E_6\times SU(9)$, $E_8\times E_8$, $E_6\times E_7$ and $SO(14)\times SO(14)$ from top to bottom right.}
\end{center}
\vskip 21pt
\end{figure}

\subsection*{The seventh $K3$}
The seventh $K3$ surface is given by a polyhedron $\nabla$ with vertices
\beq
w_i=\{(-1, -1, ~0), ~(-1, ~0, -1), ~(-1, -1, -1),~ (\,1, -1, ~4),~ (\,1, ~4, -1), ~(\,1, -1, -1)\}
\eeq
and its dual $\Delta$, with vertices
\beq
v_a=\{(-1, ~0, ~0),~ (\,0, ~0, ~1),~ (\,0, ~1,~ 0),~ (\,2, -1, -1), ~(\,1, ~0, ~0)\}~.
\eeq
The pair $(\nabla, \Delta)$ leads, in this case, to the following defining polynomial:
\beq
f = -c_0\,z_1\, z_2\, z_3\, z_4\, z_5\,z_6+c_1\,  z_1\, z_4^5+ c_2\,z_2\, z_5^5+c_3\, z_3\, z_6^5 + c_4\,z_1^2 \,z_2^2\, z_3^2 + c_5\, z_4^2\, z_5^2\, z_6^2
\eeq
This $K3$ surface admits $4$ different elliptic fibration structures.

\begin{figure}[H]
\begin{center}
\hskip-90pt
\begin{minipage}[t]{5.2in}
\vspace{30pt}
\raisebox{-1in}{\includegraphics[width=8.cm]{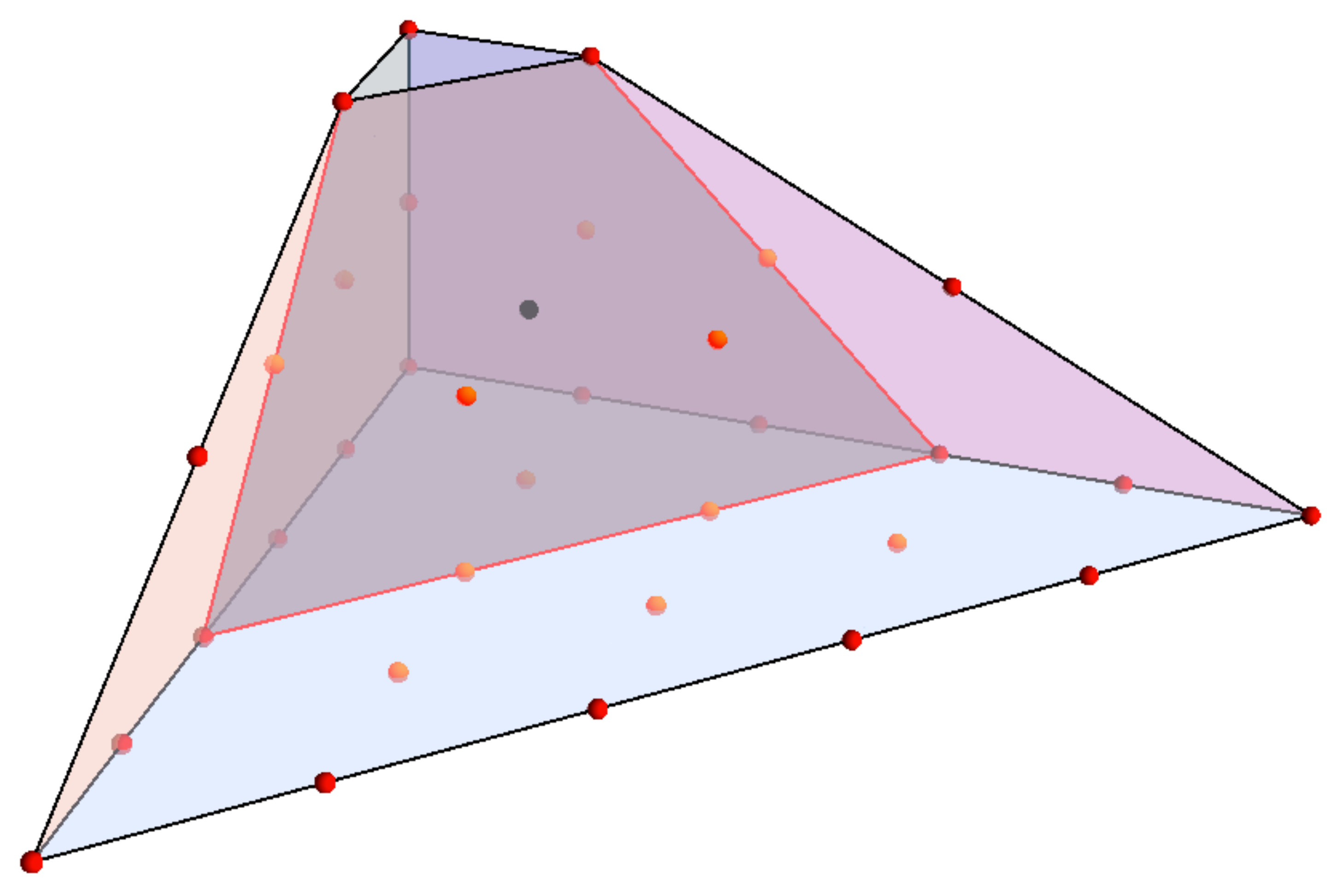}} 
\hfill \hspace*{4pt}
\raisebox{-0.95in}{\includegraphics[width=8.cm]{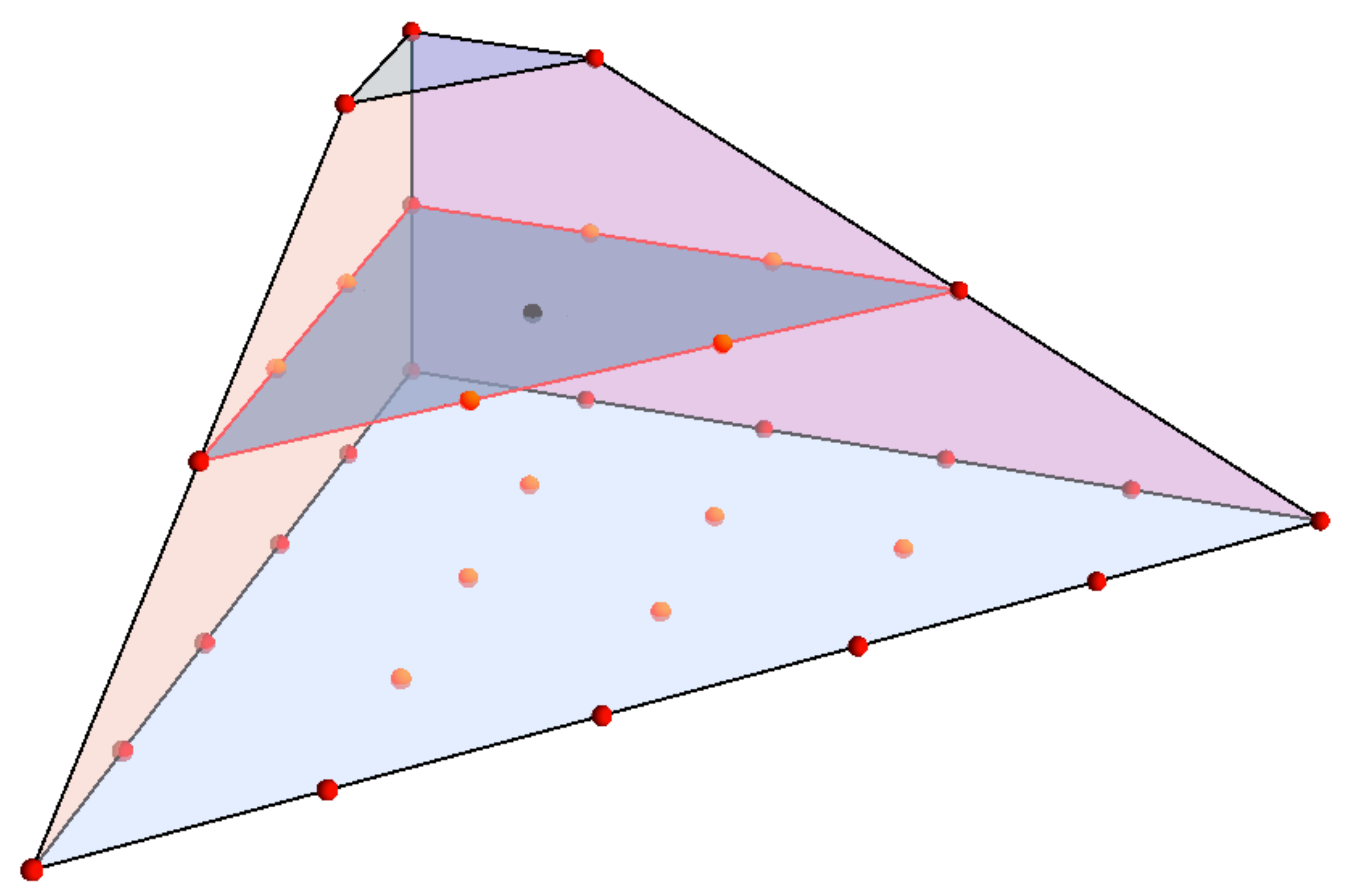}}
\hspace{20pt}
\vspace{10pt}
\end{minipage}
%\capt{5.7in}{fig:SecondK3}{The second $K3$ surface. Both polyhedra correspond to $\nabla$ and indicate two different elliptic fibration structures.}
\end{center}
\vskip -20pt
\end{figure}

\begin{figure}[h]
\begin{center}
\hskip-90pt
\begin{minipage}[t]{5.2in}
\vspace{10pt}
\raisebox{-1in}{\includegraphics[width=8.cm]{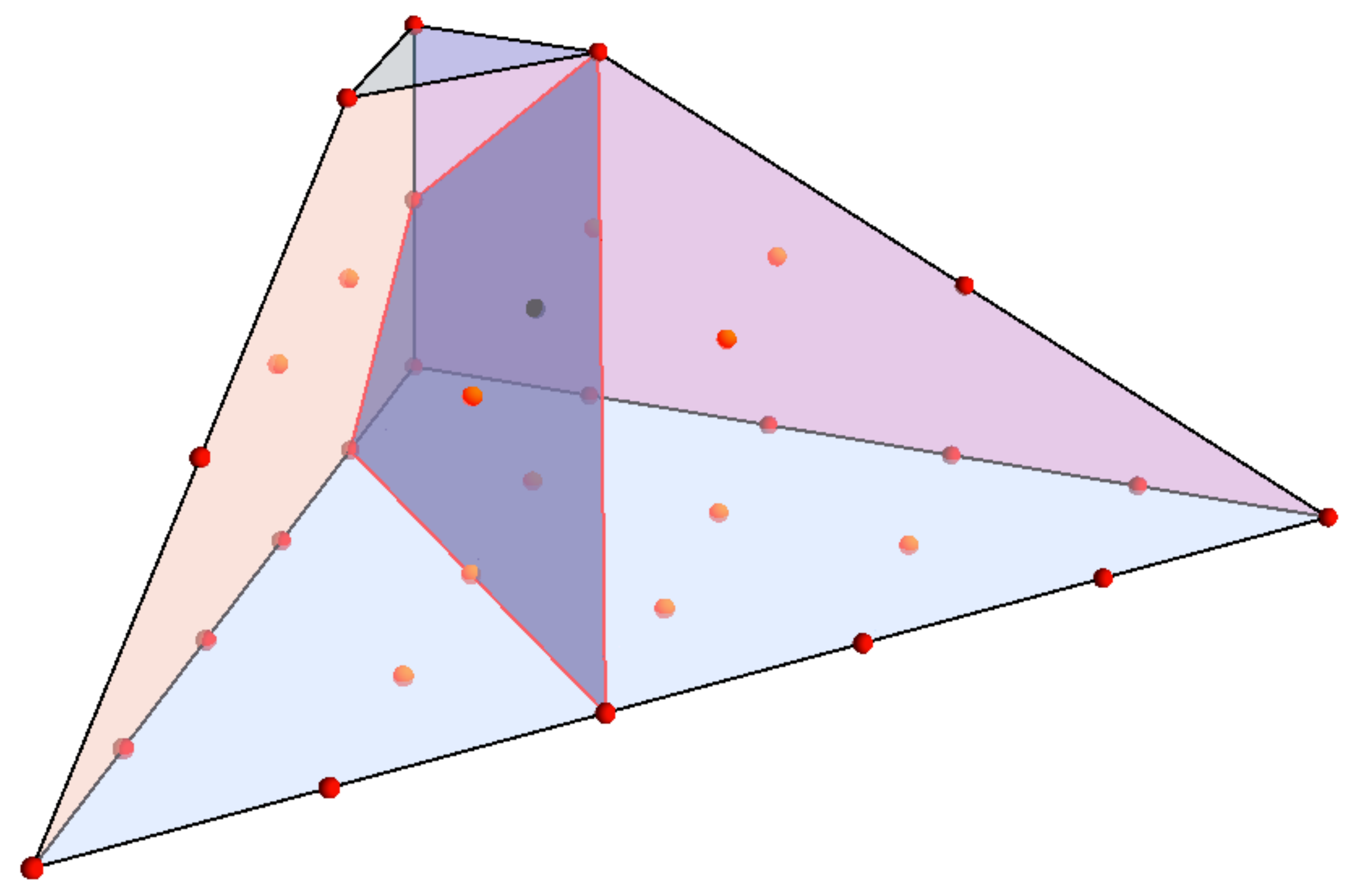}} 
\hfill \hspace*{4pt}
\raisebox{-0.95in}{\includegraphics[width=8.cm]{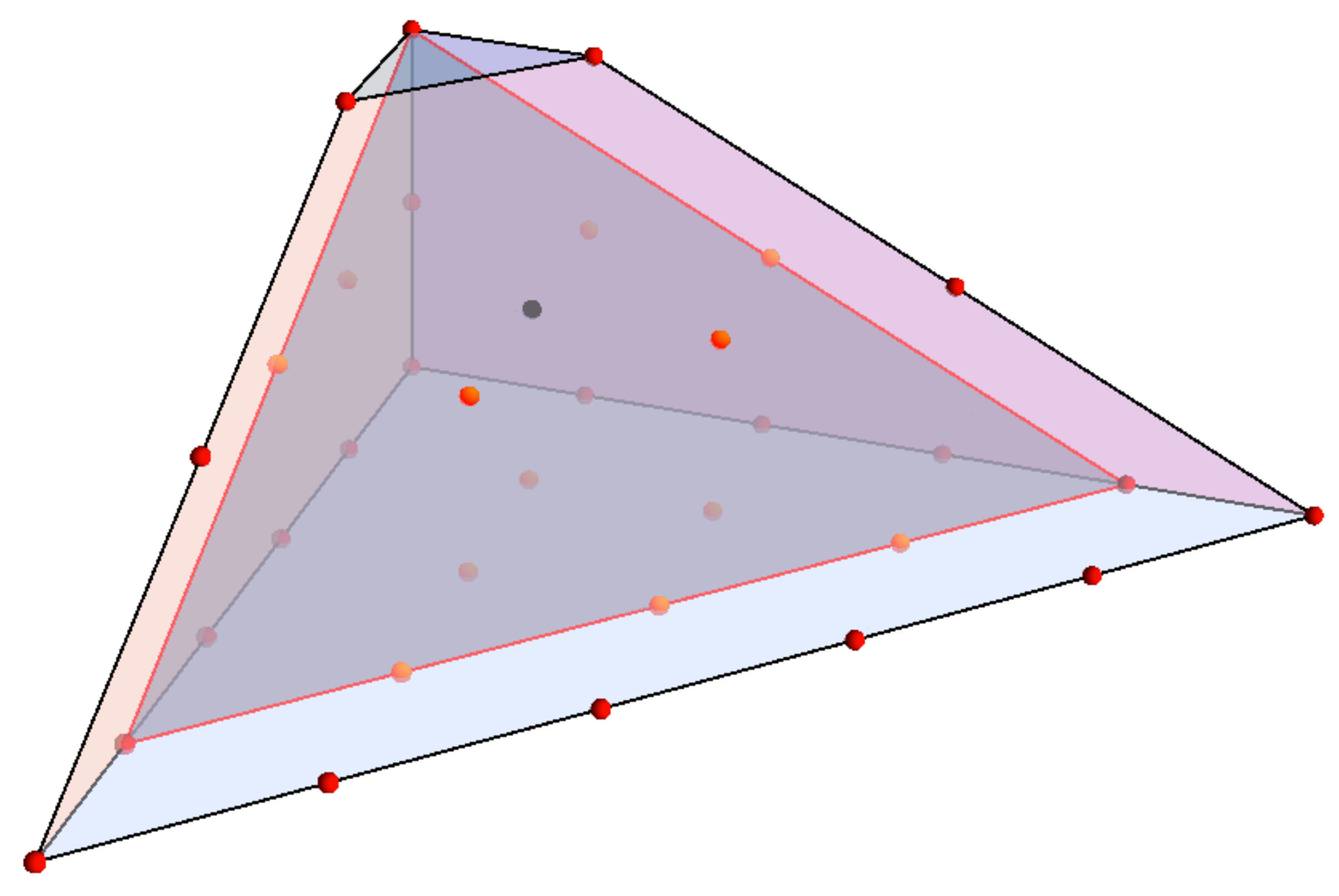}}
\hspace{20pt}
\vspace{0pt}
\end{minipage}
\capt{6.2in}{fig:ThirdK3}{The polyhedron $\nabla$ for the third $K3$ surface. This $K3$ admits four different elliptic fibration structures, of types $SO(18)\times E_6$, $SU(15)\times SU(3)$, $E_8\times E_7$ and $E_7\times SU(10)$, from top left to bottom right.}
\end{center}
\vskip 10pt
\end{figure}

\subsection*{The eighth $K3$}
The eighth $K3$ surface is given by a polyhedron $\nabla$ with vertices
\beq
w_i=\{(-1, -1, -1), ~(-1, ~3,~ 3), ~(\,1, -1, -1), ~(-1, -1, ~3),~ (-1, ~3, -1)\}
\eeq
and its dual $\Delta$, with vertices
\beq
v_a=\{(-2, -1, ~0), ~(-2, ~0, -1), ~(\,0,~ 0,~ 1),~ (\,0, ~1,~ 0), ~(\,1,~ 0,~ 0)\}~.
\eeq
The pair $(\nabla, \Delta)$ leads, in this case, to the following defining polynomial:
\beq
f = -c_0\,z_1\, z_2\, z_3\, z_4\, z_5 + c_1\,  z_1^4\, z_4^4+ c_2\,z_1^4\, z_5^4+c_3\, z_2^4\, z_4^4 + c_4\,z_2^4 \,z_5^4 + c_5\, z_3^2
\eeq
This $K3$ surface admits $3$ different elliptic fibration structures.

\begin{figure}[H]
\begin{center}
\hskip3cm
\begin{minipage}[t]{4.3in}
\vspace{20pt}
\raisebox{-1in}{\includegraphics[width=8.cm]{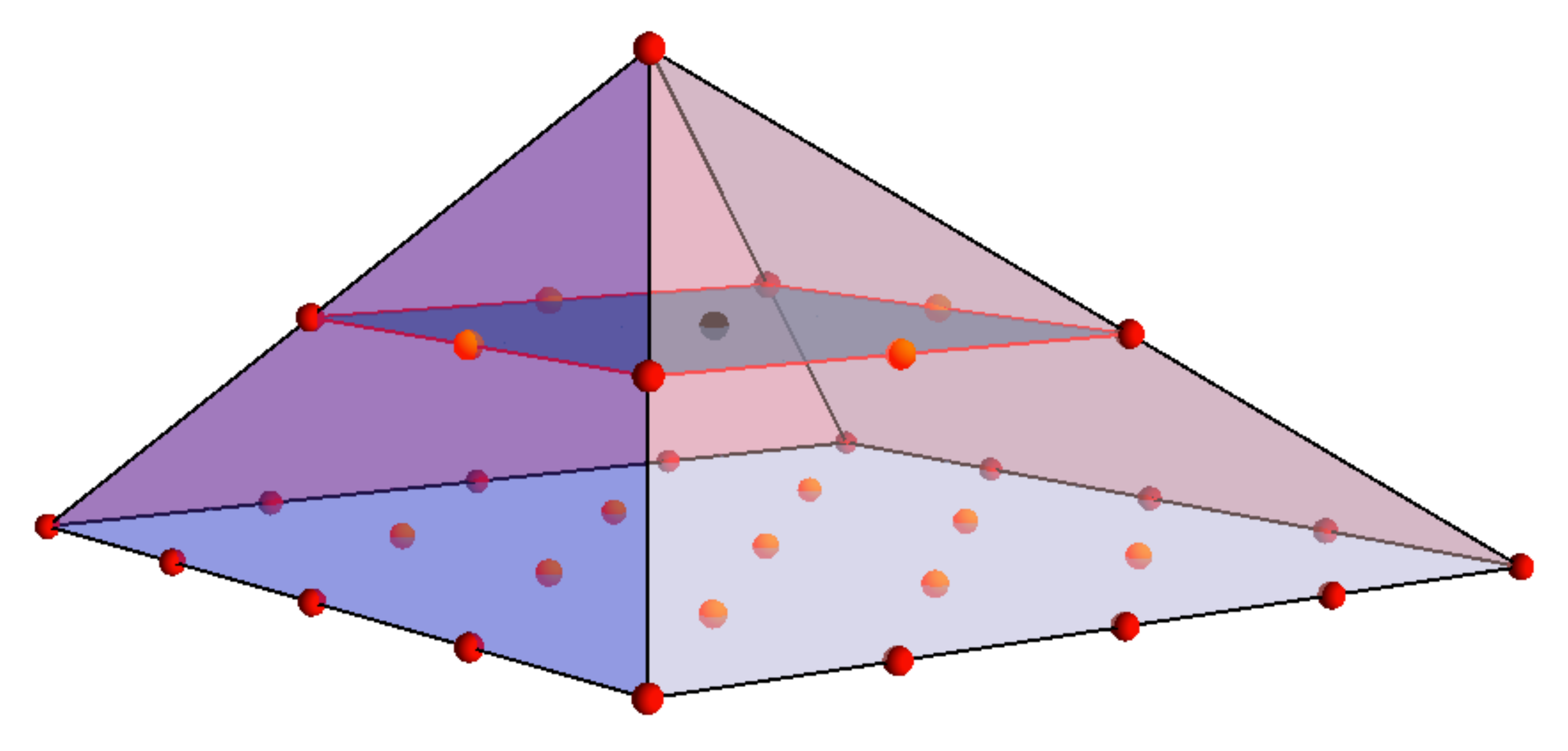}} 
%\hfill \hspace*{0pt}
%\raisebox{-0.95in}{\includegraphics[width=7.cm]{Case2NablaFib4.pdf}}
\hspace{20pt}
\vspace{10pt}
\end{minipage}
\end{center}
\vskip 0pt
\end{figure}

\begin{figure}[h]
\begin{center}
\hskip-90pt
\begin{minipage}[t]{5.2in}
\vspace{20pt}
\raisebox{-1in}{\includegraphics[width=8.cm]{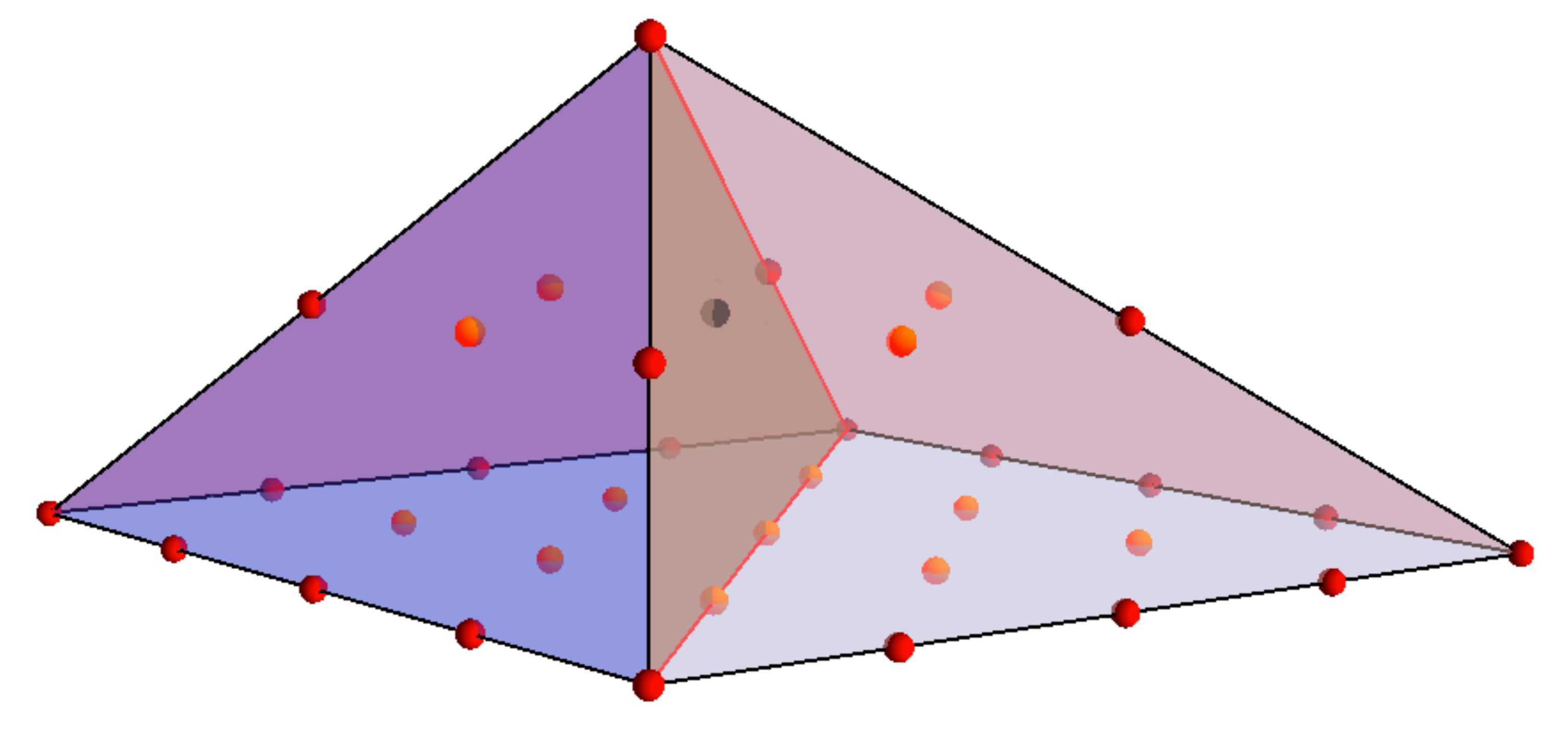}} 
\hfill \hspace*{10pt}
\raisebox{-0.95in}{\includegraphics[width=8.cm]{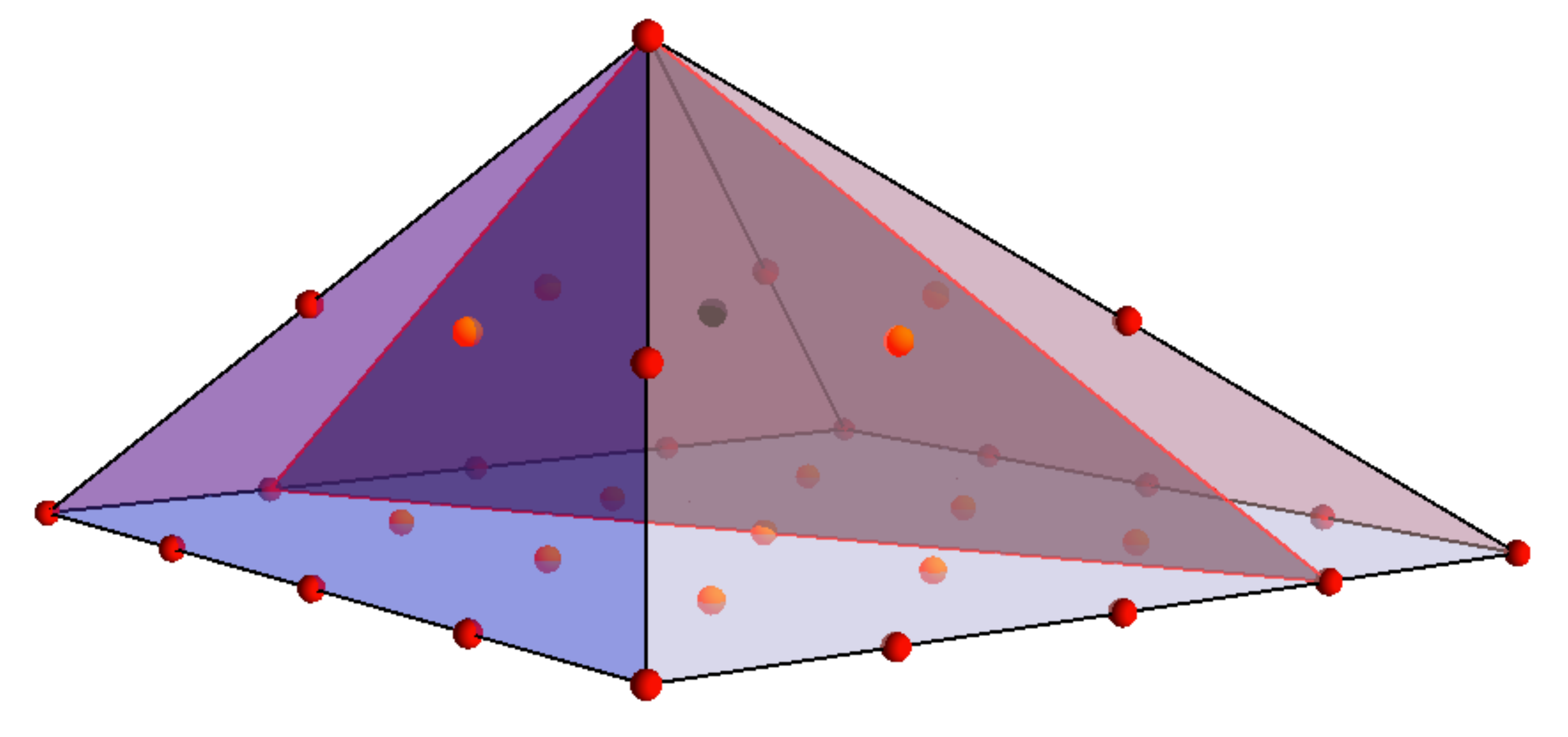}}
\hspace{20pt}
\vspace{10pt}
\end{minipage}
\capt{6.2in}{fig:ThirdK3}{The polyhedron $\nabla$ for the third $K3$ surface. This $K3$ admits four different elliptic fibration structures, of types $\{1\}\times SU(16)$, $E_7\times E_7$ and $E_8\times E_8$, from top to bottom right.}
\end{center}
\vskip -10pt
\end{figure}

\subsection*{The ninth $K3$}
The ninth $K3$ surface is given by a polyhedron $\nabla$ with vertices
\beq
w_i=\{(-1, -1, -1), ~(-1, -1, ~1), (\,1, -1 -1), ~(-1, ~3, ~5),~ (-1, ~3,-1)\}
\eeq
and its dual $\Delta$, with vertices
\beq
v_a=\{(-2, -1, ~0), ~(-1,~ 1,-1),~ (\,0, ~1,~ 0), (\,1,~ 0,~ 0), (\,0, ~0, ~1)\}~.
\eeq
The pair $(\nabla, \Delta)$ leads, in this case, to the following defining polynomial:
\beq
f = -c_0\,z_1\, z_2\, z_3\, z_4\, z_5 + c_1\,  z_1^4\, z_2^4 + c_2\,z_4^4\, z_5^4+c_3\, z_1^2\, z_5^6 + c_4\,z_2^2 \,z_4^6 + c_5\, z_3^2
\vspace{-4pt}
\eeq

%This $K3$ surface admits $4$ different elliptic fibration structures.

\begin{figure}[h]
\begin{center}
\hskip-45pt
\begin{minipage}[t]{5.2in}
\vspace{0pt}
\raisebox{-1in}{\includegraphics[width=7.cm]{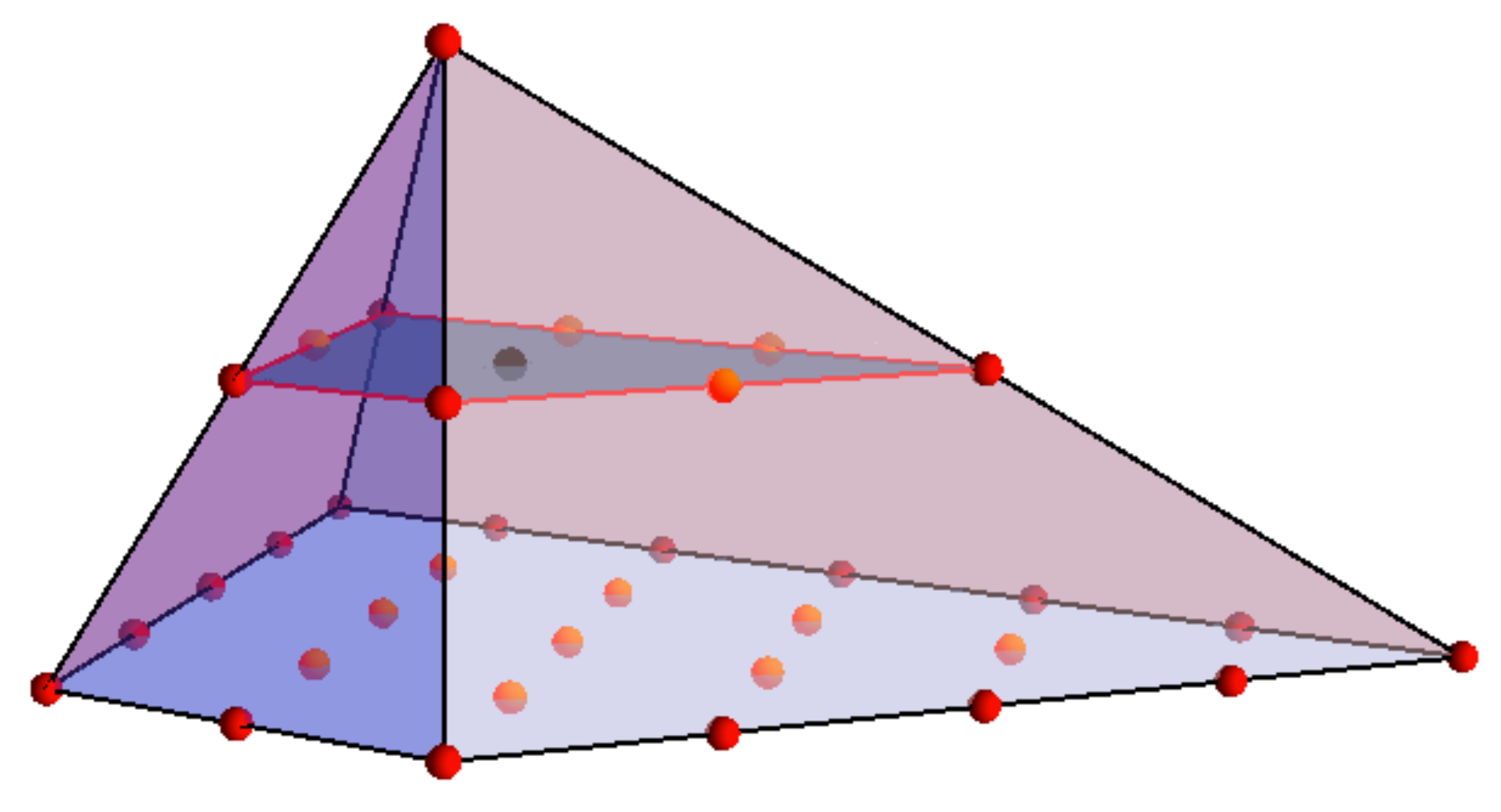}} 
\hfill \hspace*{30pt}
\raisebox{-0.95in}{\includegraphics[width=7.cm]{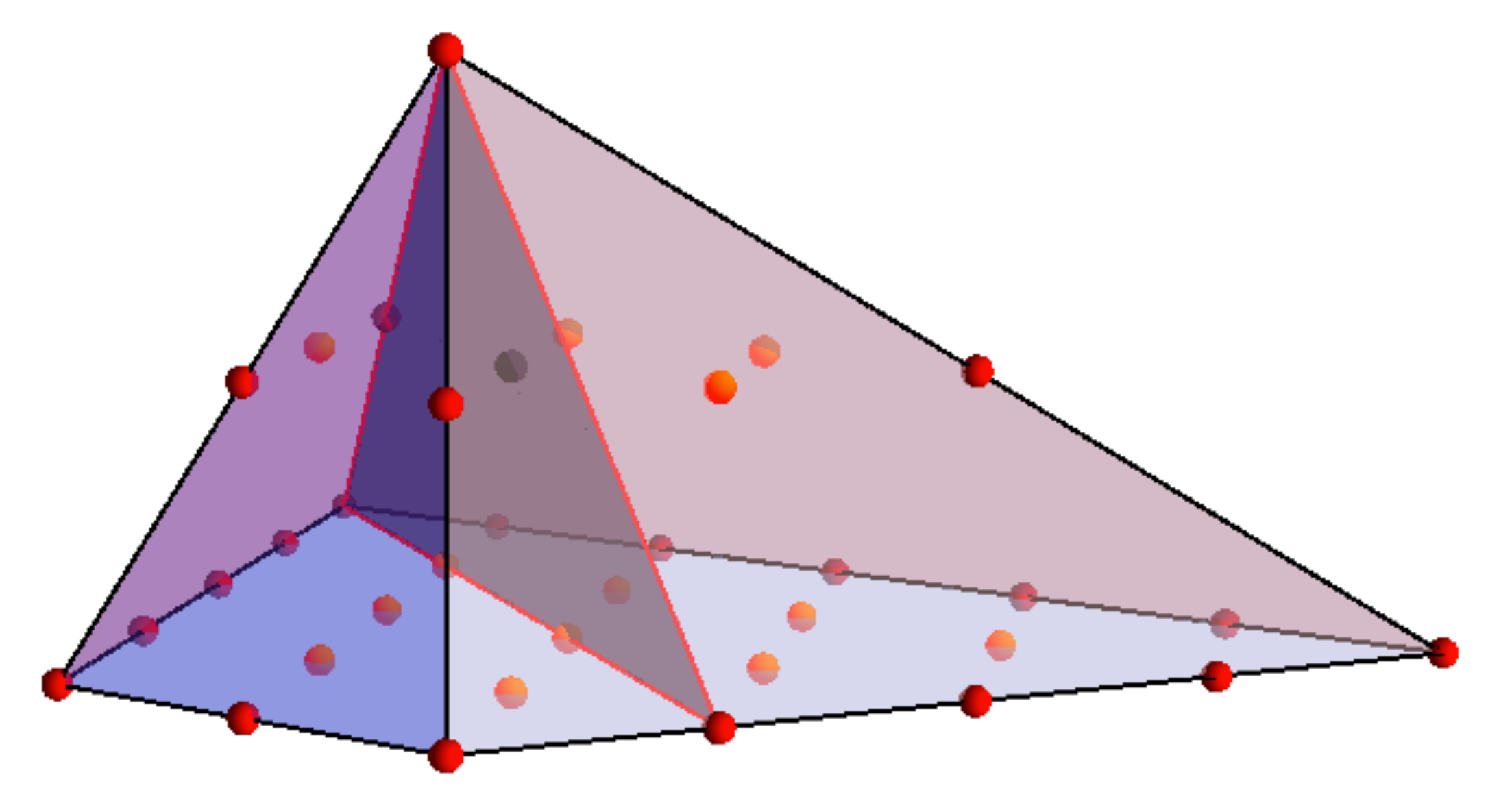}}
\hspace{20pt}
\vspace{10pt}
\end{minipage}
%\capt{5.7in}{fig:SecondK3}{The second $K3$ surface. Both polyhedra correspond to $\nabla$ and indicate two different elliptic fibration structures.}
\end{center}
\vskip -50pt
\end{figure}

\begin{figure}[H]
\begin{center}
\hskip-45pt
\begin{minipage}[t]{5.2in}
\vspace{10pt}
\raisebox{-1in}{\includegraphics[width=7.cm]{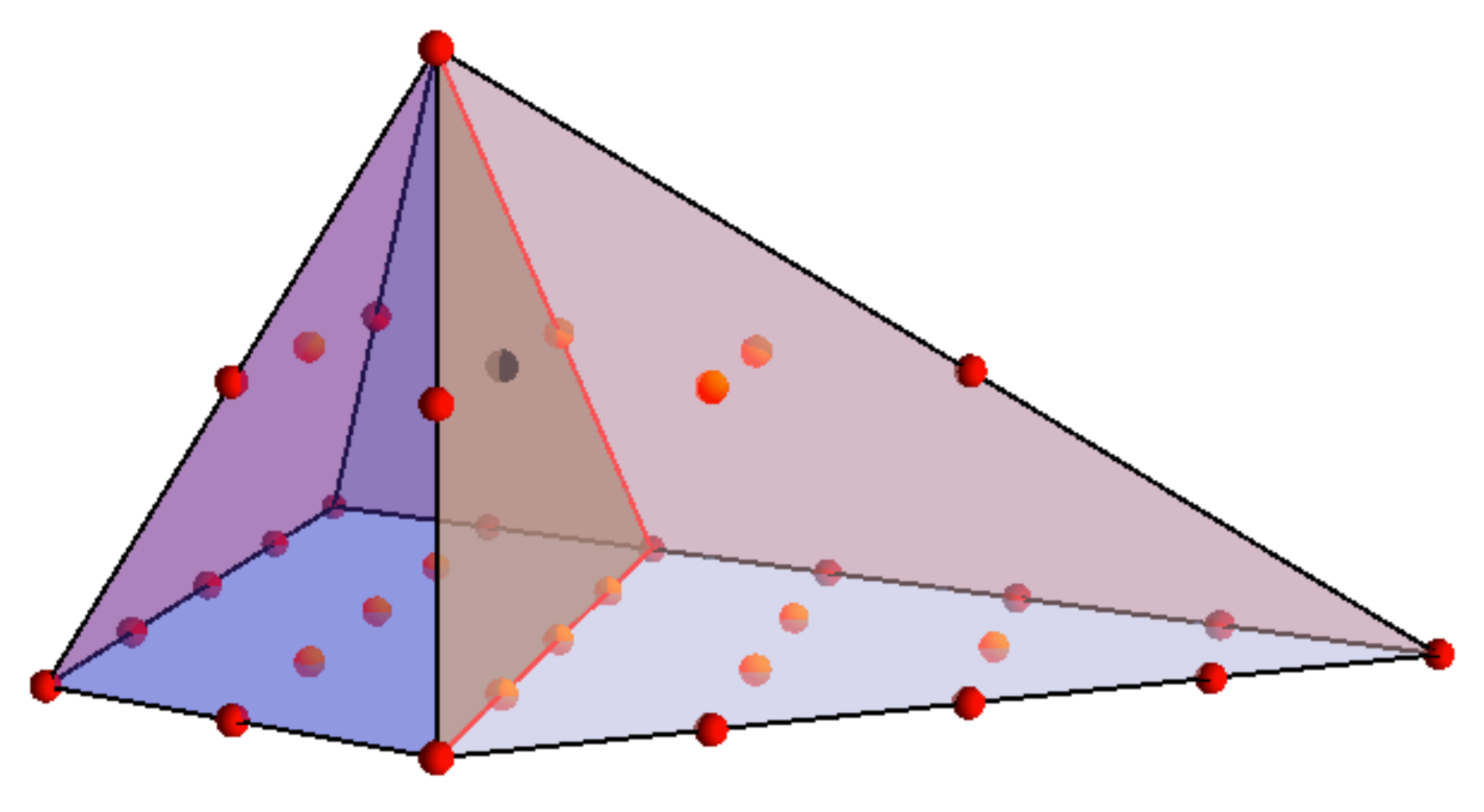}} 
\hfill \hspace*{30pt}
\raisebox{-0.95in}{\includegraphics[width=7.cm]{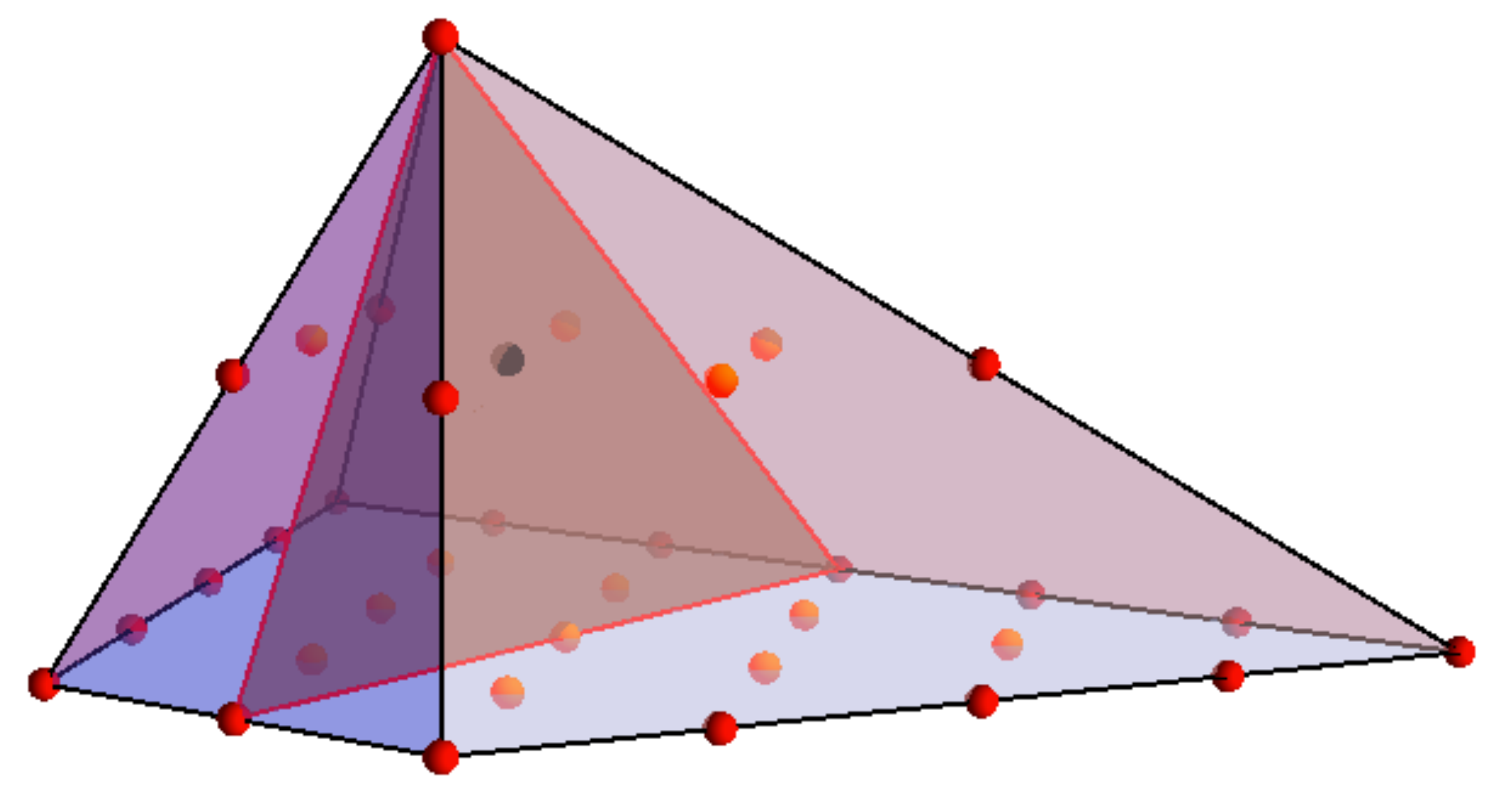}}
\hspace{20pt}
\vspace{0pt}
\end{minipage}
\capt{6.2in}{fig:ThirdK3}{The polyhedron $\nabla$ for the third $K3$ surface. This $K3$ admits four different elliptic fibration structures, of types $\{1\}\times SU(16)$, $E_8\times E_7$, $E_7\times S0(16)$ and $E_8\times E_8$, from top left to bottom right.}
\end{center}
\vskip 0pt
\end{figure}

\end{appendix} 

%\newpage
%\bibliographystyle{JHEP}
%\bibliography{Bibliography3}

\begin{thebibliography}{10}

\bibitem{paper:local}
P.~Candelas, A.~Constantin, C.~Damian, M.~Larfors, and J.~F. Morales, {\it
  {Type IIB flux vacua from G-theory II}},
  \href{http://xxx.lanl.gov/abs/1411.4786}{{\tt arXiv:1411.4786}}.

\bibitem{Grana:2004bg}
M.~Gra\~na, R.~Minasian, M.~Petrini, and A.~Tomasiello, {\it {Supersymmetric
  backgrounds from generalized Calabi-Yau manifolds}},  {\em JHEP} {\bf 0408}
  (2004) 046, [\href{http://xxx.lanl.gov/abs/hep-th/0406137}{{\tt
  hep-th/0406137}}].

\bibitem{Maldacena:2000mw}
J.~M. Maldacena and C.~Nunez, {\it {Supergravity description of field theories
  on curved manifolds and a no go theorem}},  {\em Int.J.Mod.Phys.} {\bf A16}
  (2001) 822--855, [\href{http://xxx.lanl.gov/abs/hep-th/0007018}{{\tt
  hep-th/0007018}}].

\bibitem{Giddings:2001yu}
S.~B. Giddings, S.~Kachru, and J.~Polchinski, {\it {Hierarchies from fluxes in
  string compactifications}},  {\em Phys.Rev.} {\bf D66} (2002) 106006,
  [\href{http://xxx.lanl.gov/abs/hep-th/0105097}{{\tt hep-th/0105097}}].

\bibitem{Martucci:2012jk}
L.~Martucci, J.~F. Morales, and D.~R. Pacifici, {\it {Branes, U-folds and
  hyperelliptic fibrations}},  {\em JHEP} {\bf 1301} (2013) 145,
  [\href{http://xxx.lanl.gov/abs/1207.6120}{{\tt arXiv:1207.6120}}].

\bibitem{Braun:2013yla}
A.~P. Braun, F.~Fucito, and J.~F. Morales, {\it {U-folds as K3 fibrations}},
  {\em JHEP} {\bf 1310} (2013) 154,
  [\href{http://xxx.lanl.gov/abs/1308.0553}{{\tt arXiv:1308.0553}}].

\bibitem{Vafa:1996xn}
C.~Vafa, {\it {Evidence for F theory}},  {\em Nucl.Phys.} {\bf B469} (1996)
  403--418, [\href{http://xxx.lanl.gov/abs/hep-th/9602022}{{\tt
  hep-th/9602022}}].

\bibitem{Kumar:1996zx}
A.~Kumar and C.~Vafa, {\it {U manifolds}},  {\em Phys.Lett.} {\bf B396} (1997)
  85--90, [\href{http://xxx.lanl.gov/abs/hep-th/9611007}{{\tt
  hep-th/9611007}}].

\bibitem{Liu:1997mb}
J.~T. Liu and R.~Minasian, {\it {U-branes and T**3 fibrations}},  {\em
  Nucl.Phys.} {\bf B510} (1998) 538--554,
  [\href{http://xxx.lanl.gov/abs/hep-th/9707125}{{\tt hep-th/9707125}}].

\bibitem{Hellerman:2002ax}
S.~Hellerman, J.~McGreevy, and B.~Williams, {\it {Geometric constructions of
  nongeometric string theories}},  {\em JHEP} {\bf 0401} (2004) 024,
  [\href{http://xxx.lanl.gov/abs/hep-th/0208174}{{\tt hep-th/0208174}}].

\bibitem{Hull:2004in}
C.~Hull, {\it {A Geometry for non-geometric string backgrounds}},  {\em JHEP}
  {\bf 0510} (2005) 065, [\href{http://xxx.lanl.gov/abs/hep-th/0406102}{{\tt
  hep-th/0406102}}].

\bibitem{Flournoy:2004vn}
A.~Flournoy, B.~Wecht, and B.~Williams, {\it {Constructing nongeometric vacua
  in string theory}},  {\em Nucl.Phys.} {\bf B706} (2005) 127--149,
  [\href{http://xxx.lanl.gov/abs/hep-th/0404217}{{\tt hep-th/0404217}}].

\bibitem{Dabholkar:2005ve}
A.~Dabholkar and C.~Hull, {\it {Generalised T-duality and non-geometric
  backgrounds}},  {\em JHEP} {\bf 0605} (2006) 009,
  [\href{http://xxx.lanl.gov/abs/hep-th/0512005}{{\tt hep-th/0512005}}].

\bibitem{Gray:2005ea}
J.~Gray and E.~J. Hackett-Jones, {\it {On T-folds, G-structures and
  supersymmetry}},  {\em JHEP} {\bf 0605} (2006) 071,
  [\href{http://xxx.lanl.gov/abs/hep-th/0506092}{{\tt hep-th/0506092}}].

\bibitem{Hull:2007zu}
C.~Hull, {\it {Generalised Geometry for M-Theory}},  {\em JHEP} {\bf 0707}
  (2007) 079, [\href{http://xxx.lanl.gov/abs/hep-th/0701203}{{\tt
  hep-th/0701203}}].

\bibitem{Vegh:2008jn}
D.~Vegh and J.~McGreevy, {\it {Semi-Flatland}},  {\em JHEP} {\bf 0810} (2008)
  068, [\href{http://xxx.lanl.gov/abs/0808.1569}{{\tt arXiv:0808.1569}}].

\bibitem{Pacheco:2008ps}
P.~P. Pacheco and D.~Waldram, {\it {M-theory, exceptional generalised geometry
  and superpotentials}},  {\em JHEP} {\bf 0809} (2008) 123,
  [\href{http://xxx.lanl.gov/abs/0804.1362}{{\tt arXiv:0804.1362}}].

\bibitem{Grana:2008yw}
M.~Gra\~na, R.~Minasian, M.~Petrini, and D.~Waldram, {\it {T-duality,
  Generalized Geometry and Non-Geometric Backgrounds}},  {\em JHEP} {\bf 0904}
  (2009) 075, [\href{http://xxx.lanl.gov/abs/0807.4527}{{\tt
  arXiv:0807.4527}}].

\bibitem{McOrist:2010jw}
J.~McOrist, D.~R. Morrison, and S.~Sethi, {\it {Geometries, Non-Geometries, and
  Fluxes}},  {\em Adv.Theor.Math.Phys.} {\bf 14} (2010)
  [\href{http://xxx.lanl.gov/abs/1004.5447}{{\tt arXiv:1004.5447}}].

\bibitem{Andriot:2011uh}
D.~Andriot, M.~Larfors, D.~L\"ust, and P.~Patalong, {\it {A ten-dimensional
  action for non-geometric fluxes}},  {\em JHEP} {\bf 1109} (2011) 134,
  [\href{http://xxx.lanl.gov/abs/1106.4015}{{\tt arXiv:1106.4015}}].

\bibitem{Coimbra:2011nw}
A.~Coimbra, C.~Strickland-Constable, and D.~Waldram, {\it {Supergravity as
  Generalised Geometry I: Type II Theories}},  {\em JHEP} {\bf 1111} (2011)
  091, [\href{http://xxx.lanl.gov/abs/1107.1733}{{\tt arXiv:1107.1733}}].

\bibitem{Berman:2011cg}
D.~S. Berman, H.~Godazgar, M.~Godazgar, and M.~J. Perry, {\it {The Local
  symmetries of M-theory and their formulation in generalised geometry}},  {\em
  JHEP} {\bf 1201} (2012) 012, [\href{http://xxx.lanl.gov/abs/1110.3930}{{\tt
  arXiv:1110.3930}}].

\bibitem{Berman:2011jh}
D.~S. Berman, H.~Godazgar, M.~J. Perry, and P.~West, {\it {Duality Invariant
  Actions and Generalised Geometry}},  {\em JHEP} {\bf 1202} (2012) 108,
  [\href{http://xxx.lanl.gov/abs/1111.0459}{{\tt arXiv:1111.0459}}].

\bibitem{Coimbra:2011ky}
A.~Coimbra, C.~Strickland-Constable, and D.~Waldram, {\it {$E_{d(d)} \times
  \mathbb{R}^+$ generalised geometry, connections and M theory}},  {\em JHEP}
  {\bf 1402} (2014) 054, [\href{http://xxx.lanl.gov/abs/1112.3989}{{\tt
  arXiv:1112.3989}}].

\bibitem{Hohm:2011si}
O.~Hohm and B.~Zwiebach, {\it {On the Riemann Tensor in Double Field Theory}},
  {\em JHEP} {\bf 1205} (2012) 126,
  [\href{http://xxx.lanl.gov/abs/1112.5296}{{\tt arXiv:1112.5296}}].

\bibitem{Andriot:2012wx}
D.~Andriot, O.~Hohm, M.~Larfors, D.~L\"ust, and P.~Patalong, {\it {A geometric
  action for non-geometric fluxes}},  {\em Phys.Rev.Lett.} {\bf 108} (2012)
  261602, [\href{http://xxx.lanl.gov/abs/1202.3060}{{\tt arXiv:1202.3060}}].

\bibitem{Andriot:2012an}
D.~Andriot, O.~Hohm, M.~Larfors, D.~L\"ust, and P.~Patalong, {\it
  {Non-Geometric Fluxes in Supergravity and Double Field Theory}},  {\em
  Fortsch.Phys.} {\bf 60} (2012) 1150--1186,
  [\href{http://xxx.lanl.gov/abs/1204.1979}{{\tt arXiv:1204.1979}}].

\bibitem{Blumenhagen:2012nt}
R.~Blumenhagen, A.~Deser, E.~Plauschinn, and F.~Rennecke, {\it {Non-geometric
  strings, symplectic gravity and differential geometry of Lie algebroids}},
  {\em JHEP} {\bf 1302} (2013) 122,
  [\href{http://xxx.lanl.gov/abs/1211.0030}{{\tt arXiv:1211.0030}}].

\bibitem{Coimbra:2012af}
A.~Coimbra, C.~Strickland-Constable, and D.~Waldram, {\it {Supergravity as
  Generalised Geometry II: $E_{d(d)} \times \mathbb{R}^+$ and M theory}},  {\em
  JHEP} {\bf 1403} (2014) 019, [\href{http://xxx.lanl.gov/abs/1212.1586}{{\tt
  arXiv:1212.1586}}].

\bibitem{Aldazabal:2013mya}
G.~Aldazabal, M.~Graña, D.~Marqués, and J.~Rosabal, {\it {Extended geometry and
  gauged maximal supergravity}},  {\em JHEP} {\bf 1306} (2013) 046,
  [\href{http://xxx.lanl.gov/abs/1302.5419}{{\tt arXiv:1302.5419}}].

\bibitem{Cederwall:2013naa}
M.~Cederwall, J.~Edlund, and A.~Karlsson, {\it {Exceptional geometry and tensor
  fields}},  {\em JHEP} {\bf 1307} (2013) 028,
  [\href{http://xxx.lanl.gov/abs/1302.6736}{{\tt arXiv:1302.6736}}].

\bibitem{Blumenhagen:2013aia}
R.~Blumenhagen, A.~Deser, E.~Plauschinn, F.~Rennecke, and C.~Schmid, {\it {The
  Intriguing Structure of Non-geometric Frames in String Theory}},  {\em
  Fortsch.Phys.} {\bf 61} (2013) 893--925,
  [\href{http://xxx.lanl.gov/abs/1304.2784}{{\tt arXiv:1304.2784}}].

\bibitem{Andriot:2013xca}
D.~Andriot and A.~Betz, {\it {$\beta$-supergravity: a ten-dimensional theory
  with non-geometric fluxes, and its geometric framework}},  {\em JHEP} {\bf
  1312} (2013) 083, [\href{http://xxx.lanl.gov/abs/1306.4381}{{\tt
  arXiv:1306.4381}}].

\bibitem{Cederwall:2014opa}
M.~Cederwall, {\it {T-duality and non-geometric solutions from double
  geometry}},  \href{http://xxx.lanl.gov/abs/1409.4463}{{\tt arXiv:1409.4463}}.

\bibitem{deBoer:2012ma}
J.~de~Boer and M.~Shigemori, {\it {Exotic Branes in String Theory}},  {\em
  Phys.Rept.} {\bf 532} (2013) 65--118,
  [\href{http://xxx.lanl.gov/abs/1209.6056}{{\tt arXiv:1209.6056}}].

\bibitem{Hitchin:2000jd}
N.~J. Hitchin, {\it {The geometry of three-forms in six and seven dimensions}},
   \href{http://xxx.lanl.gov/abs/math/0010054}{{\tt math/0010054}}.

\bibitem{Larfors:2010wb}
M.~Larfors, D.~L\"ust, and D.~Tsimpis, {\it {Flux compactification on smooth,
  compact three-dimensional toric varieties}},  {\em JHEP} {\bf 1007} (2010)
  073, [\href{http://xxx.lanl.gov/abs/1005.2194}{{\tt arXiv:1005.2194}}].

\bibitem{b87}
T.~Buscher, {\it {A Symmetry of the String Background Field Equations}},  {\em
  Phys.Lett.} {\bf B194} (1987) 59.

\bibitem{b88}
T.~Buscher, {\it {Path Integral Derivation of Quantum Duality in Nonlinear
  Sigma Models}},  {\em Phys.Lett.} {\bf B201} (1988) 466.

\bibitem{Lunin:2001fv}
O.~Lunin and S.~D. Mathur, {\it {Metric of the multiply wound rotating
  string}},  {\em Nucl.Phys.} {\bf B610} (2001) 49--76,
  [\href{http://xxx.lanl.gov/abs/hep-th/0105136}{{\tt hep-th/0105136}}].

\bibitem{Aspinwall:1996mn}
P.~S. Aspinwall, {\it {K3 surfaces and string duality}},
  \href{http://xxx.lanl.gov/abs/hep-th/9611137}{{\tt hep-th/9611137}}.

\bibitem{Greene:1989ya}
B.~R. Greene, A.~D. Shapere, C.~Vafa, and S.-T. Yau, {\it {Stringy Cosmic
  Strings and Noncompact Calabi-Yau Manifolds}},  {\em Nucl.Phys.} {\bf B337}
  (1990) 1.

\bibitem{Batyrev:1993dm}
V.~V. Batyrev, {\it {Dual Polyhedra and Mirror Symmetry for Calabi-Yau
  Hypersurfaces in Toric Varieties}},  {\em J.Alg.Geom.} {\bf 3} (1994)
  [\href{http://xxx.lanl.gov/abs/alg-geom/9310003}{{\tt alg-geom/9310003}}].

\bibitem{Kreuzer:1998vb}
M.~Kreuzer and H.~Skarke, {\it {Classification of reflexive polyhedra in
  three-dimensions}},  {\em Adv.Theor.Math.Phys.} {\bf 2} (1998) 847--864,
  [\href{http://xxx.lanl.gov/abs/hep-th/9805190}{{\tt hep-th/9805190}}].

\bibitem{Berglund:1993ax}
P.~Berglund, P.~Candelas, X.~De~La~Ossa, A.~Font, T.~Hubsch, {\em et.~al.},
  {\it {Periods for Calabi-Yau and Landau-Ginzburg vacua}},  {\em Nucl.Phys.}
  {\bf B419} (1994) 352--403,
  [\href{http://xxx.lanl.gov/abs/hep-th/9308005}{{\tt hep-th/9308005}}].

\bibitem{Griffihs}
P.~Griffiths, {\it {On the periods of certain rational integrals. I, II}},
  {\em Ann.~of Maths.~(2)} {\bf 90} (1969) 460--495; 466--541.

\bibitem{Morrison:1991cd}
D.~R. Morrison, {\it {Picard-Fuchs equations and mirror maps for
  hypersurfaces}},  \href{http://xxx.lanl.gov/abs/hep-th/9111025}{{\tt
  hep-th/9111025}}.

\bibitem{Candelas:1993dm}
P.~Candelas, X.~De~La~Ossa, A.~Font, S.~H. Katz, and D.~R. Morrison, {\it
  {Mirror symmetry for two parameter models. 1.}},  {\em Nucl.Phys.} {\bf B416}
  (1994) 481--538, [\href{http://xxx.lanl.gov/abs/hep-th/9308083}{{\tt
  hep-th/9308083}}].

\bibitem{Candelas:1994hw}
P.~Candelas, A.~Font, S.~H. Katz, and D.~R. Morrison, {\it {Mirror symmetry for
  two parameter models. 2.}},  {\em Nucl.Phys.} {\bf B429} (1994) 626--674,
  [\href{http://xxx.lanl.gov/abs/hep-th/9403187}{{\tt hep-th/9403187}}].

\bibitem{Hosono:1994ax}
S.~Hosono, A.~Klemm, S.~Theisen, and S.-T. Yau, {\it {Mirror symmetry, mirror
  map and applications to complete intersection Calabi-Yau spaces}},  {\em
  Nucl.Phys.} {\bf B433} (1995) 501--554,
  [\href{http://xxx.lanl.gov/abs/hep-th/9406055}{{\tt hep-th/9406055}}].

\bibitem{Hosono:1993qy}
S.~Hosono, A.~Klemm, S.~Theisen, and S.-T. Yau, {\it {Mirror symmetry, mirror
  map and applications to Calabi-Yau hypersurfaces}},  {\em Commun.Math.Phys.}
  {\bf 167} (1995) 301--350,
  [\href{http://xxx.lanl.gov/abs/hep-th/9308122}{{\tt hep-th/9308122}}].

\bibitem{Bianchi:1999uq}
M.~Bianchi, J.~F. Morales, and G.~Pradisi, {\it {Discrete torsion in
  nongeometric orbifolds and their open string descendants}},  {\em Nucl.Phys.}
  {\bf B573} (2000) 314--334,
  [\href{http://xxx.lanl.gov/abs/hep-th/9910228}{{\tt hep-th/9910228}}].

\bibitem{Braun:2011ux}
V.~Braun, {\it {Toric Elliptic Fibrations and F-Theory Compactifications}},
  {\em JHEP} {\bf 1301} (2013) 016,
  [\href{http://xxx.lanl.gov/abs/1110.4883}{{\tt arXiv:1110.4883}}].

\bibitem{Malmendier:2014uka}
A.~Malmendier and D.~R. Morrison, {\it {K3 surfaces, modular forms, and
  non-geometric heterotic compactifications}},
  \href{http://xxx.lanl.gov/abs/1406.4873}{{\tt arXiv:1406.4873}}.

\bibitem{Billo:2012st}
M.~Billo, M.~Frau, F.~Fucito, L.~Giacone, A.~Lerda, {\em et.~al.}, {\it
  {Non-perturbative gauge/gravity correspondence in N=2 theories}},  {\em JHEP}
  {\bf 1208} (2012) 166, [\href{http://xxx.lanl.gov/abs/1206.3914}{{\tt
  arXiv:1206.3914}}].

\bibitem{0813.14039}
W.~Fulton, {\em {Introduction to toric varieties. The 1989 William H. Roever
  lectures in geometry.}}
\newblock {Annals of Mathematics Studies. 131. Princeton, NJ: Princeton
  University Press. xi, 157 p.}, 1993.

\bibitem{1223.14001}
D.~A. Cox, J.~B. Little, and H.~K. Schenck, {\em {Toric varieties.}}
\newblock {Graduate Studies in Mathematics 124. Providence, RI: American
  Mathematical Society (AMS). xxiv, 841~p. }, 2011.

\bibitem{Skarke:1998yk}
H.~Skarke, {\it {String dualities and toric geometry: An Introduction}},  {\em
  Chaos Solitons Fractals} (1998)
  [\href{http://xxx.lanl.gov/abs/hep-th/9806059}{{\tt hep-th/9806059}}].

\bibitem{Avram:1996pj}
A.~Avram, M.~Kreuzer, M.~Mandelberg, and H.~Skarke, {\it {Searching for K3
  fibrations}},  {\em Nucl.Phys.} {\bf B494} (1997) 567--589,
  [\href{http://xxx.lanl.gov/abs/hep-th/9610154}{{\tt hep-th/9610154}}].

\bibitem{Candelas:1996su}
P.~Candelas and A.~Font, {\it {Duality between the webs of heterotic and type
  II vacua}},  {\em Nucl.Phys.} {\bf B511} (1998) 295--325,
  [\href{http://xxx.lanl.gov/abs/hep-th/9603170}{{\tt hep-th/9603170}}].

\bibitem{Candelas:1997pq}
P.~Candelas and H.~Skarke, {\it {F theory, SO(32) and toric geometry}},  {\em
  Phys.Lett.} {\bf B413} (1997) 63--69,
  [\href{http://xxx.lanl.gov/abs/hep-th/9706226}{{\tt hep-th/9706226}}].

\bibitem{Candelas:2012uu}
P.~Candelas, A.~Constantin, and H.~Skarke, {\it {An Abundance of K3 Fibrations
  from Polyhedra with Interchangeable Parts}},  {\em Commun. Math. Phys.} {\bf
  324} (2013) 937--959, [\href{http://xxx.lanl.gov/abs/1207.4792}{{\tt
  arXiv:1207.4792}}].

\bibitem{Kreuzer:1997zg}
M.~Kreuzer and H.~Skarke, {\it {Calabi-Yau four folds and toric fibrations}},
  {\em J.Geom.Phys.} {\bf 26} (1998) 272--290,
  [\href{http://xxx.lanl.gov/abs/hep-th/9701175}{{\tt hep-th/9701175}}].

\bibitem{Chialva:2007sv}
D.~Chialva, U.~H. Danielsson, N.~Johansson, M.~Larfors, and M.~Vonk, {\it
  {Deforming, revolving and resolving - New paths in the string theory
  landscape}},  {\em JHEP} {\bf 0802} (2008) 016,
  [\href{http://xxx.lanl.gov/abs/0710.0620}{{\tt arXiv:0710.0620}}].

\end{thebibliography}
%\end{document}

% ---------------------------------------- Bibliography

\newpage
\providecommand{\href}[2]{#2}\begingroup\raggedright\endgroup

\end{document}